\documentclass[a4paper,11pt]{article}
\usepackage{jheppub} 
\usepackage{comment}
\usepackage{lineno}
\usepackage[dvipsnames]{xcolor}
\usepackage[utf8]{inputenc}  
\usepackage[inkscapelatex=false]{svg}
\usepackage{fix-cm}  
\usepackage{tikz}    
\usepackage{float}
\usepackage{longtable}
\usepackage{array}
\usetikzlibrary{arrows}
\usepackage{tikz-cd} 
\usepackage{tkz-euclide}
\usepackage{circuitikz}
\usepgfmodule{shapes}
\usetikzlibrary{decorations.pathmorphing}
\usetikzlibrary{backgrounds, arrows,calc,shapes,decorations.pathreplacing, automata,positioning}
\usepackage{cancel} 
\usepackage{graphicx}
\usepackage{mathdots}
\usepackage{slashed}
\usetikzlibrary{decorations.markings,backgrounds}
\usetikzlibrary{matrix}
\usetikzlibrary{shapes.misc}

\definecolor{CScolor}{rgb}{1.0, 0.0, 0.0} 
\definecolor{BFcolor} {RGB} {0, 158, 96} 
\definecolor{FIcolor}{RGB}{250, 95, 85} 

\definecolor{bcolor}{RGB}{4, 55, 242} 
\definecolor{fcolor}{RGB}{255, 36, 0} 
\definecolor{labelcolor}{HTML}{808080} 

\def\elephi{\varphi}
\def\elepsi{\psi}
\def\magphi{\phi}
\def\magpsi{\psi}

\def\mon{\mathfrak{M}}

\newcommand{\zz}{\mathbb{Z}}

\def\dualto{
	\qquad\leftrightarrow\qquad
}
\def\CSg{
	\text{CS$_g$}
}

\def\identityN[#1,#2]{
	_{#1}\mathbb{I}_{#2}^{[N]}
}

\def\identityOne[#1,#2]{
	_{#1}\mathbb{I}_{#2}^{[1^N]}
}

\def\DeltaN[#1]{
	\Delta^{[N]}\left( #1 \right)
}

\def\DeltaOne[#1]{
	\Delta^{[1^N]}\left( #1 \right)
}

\newcommand{\tikznode}[2]{\relax
\ifmmode%
  \tikz[remember picture,baseline=(#1.base),inner sep=0pt]\node(#1){$#2$};
\else
  \tikz[remember picture,baseline=(#1.base),inner sep=0pt]\node(#1){#2};%
\fi}

\tikzstyle{BFline}=[dashed,black,draw]
\tikzstyle{farrowstyle}=[fcolor,thick,draw]
\tikzstyle{barrowstyle}=[bcolor,thick,draw]

\tikzstyle{flavor}=[rectangle,draw=black,thick,inner sep = 0pt, minimum size = 6mm]
\tikzstyle{manifest}=[rectangle,draw=blue!50,thick,inner sep = 0pt, minimum size = 6mm]
\tikzstyle{gauge}=[circle,draw=black,thick,inner sep = 0pt, minimum size = 6mm] 
\tikzstyle{gauge2}=[circle,draw=black!50,thick,inner sep = 0pt, minimum size = 4.5mm] 
\tikzstyle{gauge3}=[rounded rectangle, draw=black!100, thick, minimum size=5mm] 
\tikzset{->-/.style={decoration={
  markings,
  mark=at position .5 with {\arrow{>}}},postaction={decorate}}}
\tikzset{-<-/.style={decoration={
  markings,
  mark=at position .5 with {\arrow{<}}},postaction={decorate}}}
\tikzstyle{BFline}=[dashed]
  
\def\nodeCS(#1,#2)(#3,#4,#5){ 	
	\node at (#1,#2) (#3) [gauge,black] {#4};
	\draw[CScolor] (#3) ++(6pt,-7pt) node[anchor=west] {\tiny#5};
}

\def\flavorCS(#1,#2)(#3,#4,#5){ 	
	\node at (#1,#2) (#3) [flavor,black] {#4};
	\draw[CScolor] (#3) ++(6pt,-8pt) node[anchor=west] {\tiny#5};
}

\def\arrowBF(#1,#2)(#3){ 
	\begin{scope}[every node/.style={auto, outer sep=-1pt}]
	\path (#1) edge[->-] node[BFcolor,midway] {\tiny#3} (#2);
	\end{scope}
}

\def\arrowBFlr(#1,#2)(#3,#4){ 
	\begin{scope}[every node/.style={auto=#4, outer sep=-1pt}]
	\path (#1) edge[->-] node[BFcolor,midway] {\tiny#3} (#2);
	\end{scope}
}

\def\dottedBF(#1,#2)(#3){ 
	\begin{scope}[every node/.style={auto, outer sep=-1pt}]
	\path (#1) edge[BFline] node[BFcolor,midway] {\tiny#3} (#2);
	\end{scope}
}

\def\farrow(#1,#2,#3){
	\begin{scope}[every node/.style={auto=#3, outer sep=-1pt}]
	\path (#1) edge[->-,fcolor] node[fcolor,midway] {\tiny$\psi$} (#2);
	\end{scope}
}

\def\barrow(#1,#2,#3){
	\begin{scope}[every node/.style={auto=#3, outer sep=-1pt}]
	\path (#1) edge[->-,bcolor] node[bcolor,midway] {\tiny$\phi$} (#2);
	\end{scope}
}

\def\barrowBF(#1,#2,#3,#4){
	\begin{scope}[every node/.style={auto=#3, outer sep=-1pt}]
	\path (#1) edge[->-,bcolor] node[bcolor,midway] {\tiny$\phi$}  node[BFcolor,midway,swap] {\tiny#4} (#2);
	\end{scope}
}

\def\farrowBF(#1,#2,#3,#4){
	\begin{scope}[every node/.style={auto=#3, outer sep=-1pt}]
	\path (#1) edge[->-,fcolor] node[fcolor,midway] {\tiny$\psi$}  node[BFcolor,midway,swap] {\tiny#4} (#2);
	\end{scope}
}

\def\fQED(#1,#2){ 
\begin{tikzpicture}{baseline=(current bounding box).center}
	\node at (0,0) (g) [gauge,black] {$1$};
	\draw[CScolor] (g.south east) ++(-3pt,3pt) node[anchor=north west] {\tiny$#2$};
	\node at (1.5,0) (f) [flavor,black] {$#1$};
	\farrow(g,f,left)
\end{tikzpicture}
}

\def\sQED(#1,#2){ 
\begin{tikzpicture}{baseline=(current bounding box).center}
	\node at (0,0) (g) [gauge,black] {$1$};
	\draw[CScolor] (g.south east) ++(-3pt,3pt) node[anchor=north west] {\tiny$#2$};
	\node at (1.5,0) (f) [flavor,black] {$#1$};
	\barrow(g,f,left)
\end{tikzpicture}
}
\def\bQED(#1,#2){ \sQED(#1,#2) }

\tikzset{cross/.style={cross out, draw=black, minimum size=5*(#1-\pgflinewidth), inner sep=0pt, outer sep=0pt},
cross/.default={2pt}}

\tikzset{snake it/.style={decorate, decoration=snake}}

\tikzset{mid arrow/.style={postaction={decorate,decoration={
        markings,
        mark = at position .55 with {\arrow[#1]{Straight Barb[width=5pt]}}
      }}}}
      
\tikzset{mid arrowsm/.style={postaction={decorate,decoration={
        markings,
        mark = at position .55 with {\arrow[#1]{Straight Barb[width=3pt]}}
      }}}}
      
\tikzset{middx arrowsm/.style={postaction={decorate,decoration={
        markings,
        mark = at position .7 with {\arrow[#1]{Straight Barb[width=3pt]}}
      }}}}
      
\tikzset{midsx arrowsm/.style={postaction={decorate,decoration={
        markings,
        mark = at position .4 with {\arrow[#1]{Straight Barb[width=3pt]}}
      }}}}

\newcommand{\SB}[1]{{\textcolor{Red} {[SB: #1]}}}
\newcommand{\RC}[1]{{\textcolor{Purple} {[RC: #1]}}}
\newcommand{\GP}[1]{{\color{blue} {[GP: #1]}}}
\newcommand{\SR}[1]{{\color{orange} {[SR: #1]}}}
\newcommand{\SP}[1]{{\color{brown} {[SP: #1]}}}
\newcommand{\AS}[1]{{\color{Blue} {[\textbf{AS}: #1]}}}

\usepackage{amsmath} 
\usepackage{amssymb}
\usepackage{graphicx}
\graphicspath{ {images/} }
\usepackage{caption}
\usepackage{subcaption}

\usepackage[most]{tcolorbox}
\usepackage{hyperref}

\title{\boldmath Planar Abelian Duals of Chern-Simons QCD}

\author[a]{Sergio Benvenuti}
\author[b,c]{Riccardo Comi}
\author[b,c]{Sara Pasquetti}
\author[a,d]{Gabriel Pedde Ungureanu}
\author[a,d]{Simone Rota}
\author[a,d]{Anant Shri}

\affiliation[a]{INFN, Sezione di Trieste, Via Valerio 2, I-34127 Trieste, Italy}
\affiliation[b]{Dipartimento di Fisica, Università di Milano-Bicocca, Piazza della Scienza 3, I-20126 Milano, Italy}
\affiliation[c]{INFN, Sezione di Milano-Bicocca, Piazza della Scienza 3, I-20126 Milano, Italy}
\affiliation[d]{SISSA, Via Bonomea 265, I-34136 Trieste, Italy}

\emailAdd{benve79@gmail.com} \emailAdd{r.comi2@campus.unimib.it} \emailAdd{sara.pasquetti@gmail.com} \emailAdd{gpeddeun@sissa.it} \emailAdd{srota@sissa.it}  \emailAdd{ashri@sissa.it}

\abstract{
We propose novel infrared dualities connecting 2+1 dimensional non-Abelian gauge theories (with unitary or special unitary gauge groups) to Abelian gauge theories. The dual Abelian theories are characterized by a planar quiver structure, where interactions are fully encoded in the quiver diagram. These dualities are rooted in supersymmetric mirror symmetry and display the characteristic exchange of mesonic and monopole operators. Furthermore, our proposed dualities exhibit features of bosonization: the addition of a fermionic (bosonic) flavor to the non-Abelian side corresponds to the addition of a bosonic (fermionic) column in the dual planar quiver.
}

\begin{document}
\maketitle
\flushbottom

\newpage
\section{Introduction and Summary}
\label{sec: introduction}

Over the past fifteen years, the study of infrared (IR) \textit{bosonization} dualities in three-dimensional Chern-Simons (CS) gauge theories has become a vibrant and rapidly developing area of theoretical physics. These developments have extended the framework of supersymmetric dualities deep into the more challenging and intricate domain of non-supersymmetric quantum field theories (QFTs)
\cite{Giombi:2011kc, Aharony:2011jz, Minwalla:2015sca, Aharony:2015mjs, Son:2015xqa, Metlitski:2015eka, Seiberg:2016gmd, Benini:2017aed, Senthil:2018cru, Turner:2019wnh}.

A particularly fruitful idea has been to derive non-supersymmetric dualities as mass deformations of known supersymmetric ones \cite{Jain:2013gza,Gur-Ari:2015pca, Karch:2016sxi, Kachru:2016rui,Kachru:2016aon, Karch:2018mer}. Thus, non-supersymmetric 3D Chern-Simons dualities have become a central tenet within contemporary research on QFTs, with broad implications across particle physics, condensed matter physics, and mathematics.

\vspace{.3cm}

This paper proposes a new class of dualities between non-Abelian theories such as $SU(N)$ CS-QCD$_3$ with bosonic and fermionic matter, and Abelian theories featuring multiple $U(1)$ gauge groups.
A hallmark of the Abelian duals is that gauge and matter content are encoded in a \textit{quiver} - a diagrammatic representation where circular (square) nodes denote gauge (flavor) groups\footnote{We use a circle for unitary groups and a double circle for special unitary groups.}, and arrows represent matter fields. Scalar fields are depicted in blue, while fermionic fields are shown in red. Additionally, our quivers are \emph{planar}, meaning that the diagram can be embedded in the plane such that each face corresponds to an interaction term in the action. The concept of planar quivers originates in the study of the $\text{AdS}_5/\text{CFT}_4$ correspondence with minimal supersymmetry \cite{Hanany:2005ve, Franco:2005rj, Benvenuti:2005cz, Benvenuti:2005ja, Franco:2005sm, Butti:2005sw}; here, we generalize this to the non-supersymmetric context.

The dualities we propose bear a strong resemblance to the recently established $\mathcal{N}=2$ mirror dualities\footnote{We actually expect that our dualities arise from supersymmetry breaking mass deformations of the $\mathcal{N}=2$ dualities of \cite{Benvenuti:2024seb, Benvenuti:2025a, Benvenuti:2025b}, but an in-depth study of the RG flow from the $\mathcal{N}=2$ to the non-supersymmetric dualities goes beyond the scope of this paper.} of \cite{Benvenuti:2024seb, Benvenuti:2025a, Benvenuti:2025b}, which themselves arise from suitably chosen mass deformations of $\mathcal{N}=4$ mirror pairs \cite{Intriligator:1996ex, Hanany:1996ie}. As an illustrative example, Figure~\eqref{fig:SU(3)_w8sa} presents the dual description of $SU(3)_{-2}$ QCD$_3$ with eight fundamental scalar fields.

\begin{equation}
    \label{fig:SU(3)_w8sa}
    \begin{tikzpicture}
   \begin{scope}[yshift=-5.5cm]
    \begin{scope}[xshift=9cm]
    \node at (-7.5,1) (g100) [gauge, black] {$3$};
    \node at (-7.5,1) [gauge2,black] {$ $};
    \node at (-5.5,1) (f100) [flavor,black] {$8$};
    \draw[bcolor,thick,->-] (g100)-- (f100);
    \draw[bcolor] node at (-6.5,1.3) {\tiny$\varphi$};
    \draw [red] (g100)+(0.3,-0.5) node {\tiny$_{-2}$};
    \draw (-3.5, 1) node {\large $\Longleftrightarrow$};

    \begin{scope}[yshift=-1cm,xshift=-.85cm]

    \node at (-2,4.5) (q01) [gauge,black] {$1$};
    \node at (-.5,4.5) (q02) [gauge,black] {$1$};
    \node at (1,4.5) (q03) [gauge,black] {$1$};
    \node at (2.5,4.5) (q04) [gauge,black] {$1$};
    \node at (4,4.5) (q05) [gauge,black] {$1$};
    \node at (5.5,4.5) (q0f) [flavor,black] {$1$};
    
    \node at (-.5,2.5) (q11) [gauge,black] {$1$};
    \node at (1,2.5) (q12) [gauge,black] {$1$};
    \node at (2.5,2.5) (q13) [gauge,black] {$1$};
    \node at (4,2.5) (q14) [gauge,black] {$1$};
    \node at (5.5,2.5) (q15) [gauge,black] {$1$};
    
    \node at (-.5,.5) (q21) [gauge,black] {$1$};
    \node at (1,.5) (q22) [gauge,black] {$1$};
    \node at (2.5,.5) (q23) [gauge,black] {$1$};
    \node at (4,.5) (q24) [gauge,black] {$1$};
    \node at (5.5,.5) (q25) [gauge,black] {$1$};
    \node at (7,.5) (q26) [gauge,black] {$1$};

    \draw[fcolor,thick,->-] (q11)--(q12); 
    \draw[fcolor,thick,->-] (q12)--(q13);
    \draw[fcolor,thick,->-] (q13)--(q14);
    \draw[fcolor,thick,->-] (q01)--(q02);
    \draw[fcolor,thick,->-] (q02)--(q03);
    \draw[fcolor,thick,->-] (q04)--(q05); 
    \draw[fcolor,thick,->-] (q05)--(q0f); 
    \draw[bcolor,thick,->-] (q05)--(q14);
    \draw[bcolor,thick,->-] (q02)--(q11);
    \draw[bcolor,thick,->-] (q03)--(q12); 
    \draw[fcolor,thick,->-] (q03)--(q04);
    \draw[bcolor,thick,->-] (q04) -- (q13); 
    \draw[fcolor,thick,->-] (q11)--(q01); 
    \draw[fcolor,thick,->-] (q14)--(q15);
    \draw[fcolor,thick,->-] (q15)--(q05);
    \draw[bcolor,thick,->-] (q15)--(q25);
    \draw[fcolor,thick,->-] (q25)--(q26);
    \draw[fcolor,->-,thick] (q26)--(q15);
    \draw[fcolor,thick,->-] (q12)--(q02); 
    \draw[fcolor,thick,->-] (q13)--(q03); 
    \draw[fcolor,thick,->-] (q14)--(q04);
    \draw[bcolor,thick,->-] (q12)--(q22);
    \draw[bcolor,thick,->-] (q13)--(q23);
    \draw[fcolor,thick,->-] (q21)--(q22); 
    \draw[fcolor,thick,->-] (q22)--(q23);
    \draw[fcolor,thick,->-] (q23)--(q24);
    \draw[fcolor,thick,->-] (q22)--(q11); 
    \draw[fcolor,thick,->-] (q23)--(q12); 
    \draw[fcolor,thick,->-] (q24)--(q13); 
    \draw[fcolor,thick,->-] (q24)--(q25);
    \draw[bcolor,thick,->-] (q14)--(q24); 
    \draw[fcolor,thick,->-] (q25)--(q14);

\end{scope}

\end{scope}
\end{scope}

\end{tikzpicture}
\end{equation}

The corresponding planar quiver incorporates additional CS couplings and potential terms not shown in the figure. Section~\ref{sec: non-Abelian} presents the general proposal for the planar Abelian dual of $SU(N)_{2N-N_s-k}$ QCD with $N_s$ bosonic flavors and expands on these features\footnote{Throughout this paper, we focus on the regime $N_s \geq 2N - 2$ and $k \geq 0$.}. 

The duality displays characteristic features of mirror dualities. The width of the quiver scales with the number of QCD flavors $N_s$. More precisely, the number of vertical columns of gauge nodes is equal to $N_s$.  On the other hand, the maximum height of these columns is the number of colors $N$. On the planar side, the UV global symmetry $U(1)^{N_s}$, with each vertical column associated to a topological $U(1)$ symmetry (as a consequence of the interaction term, which we write down in the main text and includes monopole potentials) enhances in the IR to $U(N_s)$.

We further generalize these dualities to theories with both bosonic and fermionic matter, proposing a duality between $SU(N)$ CS-QCD with $N_s$ scalars and $N_f$ fermions,\footnote{The CS level is $2N - N_s - \frac{N_f}{2} - k$. In this work, we consider $N_s \geq 2N - 2$, $N_f \geq 0$, and $k \geq 0$.} as detailed in Section~\ref{sec: qcd_b&f}. Figure~\eqref{fig: SU(2)_w5s3fa} displays the dual for $SU(2)_{-\frac{5}{2}}$ QCD$_3$ with five scalars and three fermions.

\begin{equation}
    \label{fig: SU(2)_w5s3fa}
    \begin{tikzpicture}
    \begin{scope}[yshift=-5.5cm]
    \begin{scope}[xshift=9cm] 
    \node at (-7.5,1) (g100) [gauge, black] {$2$};
    \node at (-7.5,1) [gauge2,black] {$ $};
    \node at (-6,2) (f100) [flavor,black] {$5$};
    \node at (-6,0) (f110) [flavor,black] {$3$};
    \draw[bcolor,thick,thick,->-] (g100)-- (f100);
    \draw[fcolor,thick,thick,->-] (g100)--(f110);
    \draw [red] (g100)+(.3,-0.6) node {\tiny$_{-\frac{5}{2}}$};
    \draw[bcolor,thick,thick] node at (-6.8,1.75) {\tiny$^{\varphi}$};
    \draw[fcolor,thick,thick] node at (-6.8,.25) {\tiny$^{\psi}$};
    \draw (-3.5, 1) node {\large $\Longleftrightarrow$};

    \begin{scope}[yshift=-.5cm,xshift=-1.5cm]

    \node at (-.5,2.5) (q11) [gauge,black] {$1$};
    \node at (1,2.5) (q12) [gauge,black] {$1$};
    \node at (2.5,2.5) (q13) [gauge,black] {$1$};
    \node at (4,2.5) (q14) [gauge,black] {$1$};
    \node at (5.5,2.5) (q15) [gauge,black] {$1$};
    \node at (7,2.5) (q16) [gauge,black] {$1$};
    \node at (8.5,2.5) (q1f) [flavor,black] {$1$};
    
    \node at (-.5,.5) (q21) [gauge,black] {$1$};
    \node at (1,.5) (q22) [gauge,black] {$1$};
    \node at (2.5,.5) (q23) [gauge,black] {$1$};
    \node at (4,.5) (q24) [gauge,black] {$1$};
    \node at (5.5,.5) (q25) [gauge,black] {$1$};
    \node at (7,.5) (q26) [gauge,black] {$1$};
    \node at (8.5,.5) (q27) [gauge,black] {$1$};

    \draw[thick, fcolor,thick,->-] (q11)--(q12); 
    \draw[thick, fcolor,thick,->-] (q12)--(q13);
    \draw[thick, bcolor,thick,->-] (q13)--(q14);
    \draw[thick, bcolor,thick,->-] (q14)--(q15);
    \draw[thick, bcolor,thick,->-] (q15)--(q16);
    \draw[thick, fcolor,thick,->-] (q16)--(q1f);
    
    \draw[thick, bcolor,thick,->-] (q12)--(q22);
    \draw[thick, bcolor,thick,->-] (q13)--(q23);
    \draw[thick, bcolor,thick,->-] (q14)--(q24);
    \draw[thick, bcolor,thick,->-] (q15)--(q25);
    \draw[thick, bcolor,thick,->-] (q16)--(q26);

    \draw[thick, fcolor,thick,->-] (q21)--(q22); 
    \draw[thick, fcolor,thick,->-] (q22)--(q23);
    \draw[thick, bcolor,thick,->-] (q23)--(q24);
    \draw[thick, bcolor,thick,->-] (q24)--(q25);
    \draw[thick, bcolor,thick,->-] (q25)--(q26);
    \draw[thick, fcolor,thick,->-] (q26)--(q27);

    \draw[thick,fcolor,->-] (q22)--(q11);
    \draw[thick,fcolor,->-] (q23)--(q12);
    \draw[thick,bcolor,->-] (q24)--(q13);
    \draw[thick,bcolor,->-] (q25)--(q14);
    \draw[thick,bcolor,->-] (q26)--(q15);
    \draw[thick,fcolor,->-] (q27)--(q16);

    


\end{scope}

\end{scope}
\end{scope}

\end{tikzpicture}
\end{equation}

A simple rule governs the quiver structure: adding a bosonic flavor on the QCD side corresponds to inserting a fermionic column in the quiver (red), while adding a fermionic flavor adds a bosonic column (blue). In this way, our dualities can be viewed as generalizations of bosonization dualities \cite{Shaji:1990is,Chen:1993cd, Fradkin:1994tt}, now extended to non-Abelian theories with richer matter content.

Although we focus extensively on $SU(N)$ gauge theories here, we also demonstrate that these dualities can be extended to $U(N)$ gauge theories easily by gauging the baryonic symmetry and its image under the duality. By setting $N = 1$, one recovers the bosonization dualities of \cite{Karch:2016aux}, which relate $U(1)$ gauge theories with $F$ bosons (fermions) to a linear $U(1)^{F-1}$ quiver with fermionic (bosonic) matter. \\

We present several nontrivial checks supporting the proposed dualities.
We establish a dictionary mapping gauge-invariant mesons, baryons, and conserved currents on the non-Abelian QCD side to suitably dressed disorder operators (monopoles) of the mirror theory.
We analyze the effect of adding mass terms for individual fundamental fields in the QCD theory and trace the RG flow in the dual quiver. These deformations yield new dual pairs consistent with our general framework, with precise outcomes depending on the sign and type (bosonic or fermionic) of the mass.


In addition, as mentioned before, we observe that our dualities may arise as mass deformations of the $\mathcal{N}=2$ supersymmetric dualities developed in~\cite{Benvenuti:2024seb, Benvenuti:2025a, Benvenuti:2025b}, which relate $\mathcal{N}=2$ CS-QCD to Abelian planar quivers. The diagrams used here have the same topology as those in the supersymmetric case, with the distinction that each edge now corresponds to a single boson or fermion. Along the RG flow, precisely one field per edge acquires a mass, breaking supersymmetry. 
Although a detailed study of this flow is beyond the
scope of this paper, the existence of such a flow could provide further checks supporting the dualities proposed here.

Driven by this analogy one can think of our dualities as \textit{non-Abelian mirror symmetry in the absence of supersymmetry}.

\subsubsection*{Future Directions}
We restrict ourselves to the study of $SU(N)$ QCD$_3$ with at least $N_s=2N-2$ bosonic fundamental flavors and any number of fermionic fundamental flavors. Extending our proposal to study duals for smaller values of $N_s$ would be the next step. This can be achieved by further mass deformations of the present dualities, as discussed in \cite{Benvenuti:2025b} for the $\mathcal{N}=2$ supersymmetric case. 

\vspace{0.2cm}

Another promising direction of research would be to develop an algorithmic paradigm to construct these bosonization dualities, in a vein similar to their $\mathcal{N}=2$ \cite{Benvenuti:2025a, Benvenuti:2023qtv} and $\mathcal{N}=4$ \cite{Comi:2022aqo} counterparts. Such an approach would help construct planar Abelian duals for broader classes of non-Abelian gauge theories.

\vspace{0.3cm}

\subsubsection*{Organization of the Paper}
\begin{itemize}
    \item Section~\ref{sec: Abelian} discusses dualities between $U(1)_k$ QED with $N_f$ bosonic plus $N_f$ fermionic flavors and $U(1)^{N_s + N_f + k - 1}$ Abelian quivers.
    \item Section~\ref{sec: non-Abelian} turns to bosonic CS-QCD, analyzing three representative examples and detailing the mapping of operators across the dualities.
    \item Section~\ref{sec: qcd_b&f} generalizes to CS-QCD with both bosonic and fermionic matter, analyzing two representative examples.
    \item Section \ref{sec: U(n)_b&f} extends our framework to unitary CS-QCD with bosonic and fermionic matter.
    \item Section~\ref{sec:massdef} investigates mass deformations, offering some consistency checks of our proposal.
    \item Section \ref{sec: ns<2n} extends the dualities into the $N_s < 2N$ regime, focusing on two representative examples.
\end{itemize}

\section{Linear Abelian Duals of CS-QED\texorpdfstring{$_3$}{3} with Bosons and Fermions}
\label{sec: Abelian}

In this Section we propose a duality for QED$_3$ with bosonic and fermionic matter and a Chern-Simons interaction, which is a mild generalization of two \emph{bosonization} dualities of \cite{Karch:2016aux}. 
This duality relates the $U(1)_{-N_{s}-N_{f}/2-k}$ QED theory with $N_s$ scalars and $N_f$ fermions to an Abelian linear quiver theory with $N_s+N_f+k-1$ $U(1)$ nodes. This duality is the $N=1$ case of the general duality for $U(N)$ or $SU(N)$ QCD with bosons and fermions discussed in this paper, we discuss the Abelian case as a warm up.

The relevant matter content are complex fermions, usually denoted as $\psi$,  complex scalars, denoted as $\phi$ and  real scalars denoted as $\sigma$.

The simplest instances of the duality are the well known bosonization and fermionization dualities, which can be schematically written down as follows:\footnote{In this paper we normalize a fermion charged under a $U(1)$ gauge field $A$ as:
\begin{equation}
Z_{fermion}=\left|\operatorname{det} \cancel D_{{A}}\right| e^{-\frac{i \pi}{2} \eta({A})} .
\end{equation}
therefore a mass deformation $m$ for a fermion produces:
\begin{equation}
i \bar \psi \cancel D_{{A}} \psi
\quad\xrightarrow[]{m\bar\elepsi \elepsi}\quad
\left\{
\begin{array}{ll}
0 & \quad m>0
\\
-\frac{1}{4\pi}AdA -2\CSg & \quad m<0
\end{array}
\right.
\end{equation}
When we present dualities in compact notation, such as in \eqref{eq:bosonization_compact}, \eqref{eq:fermionization_compact} and in the quiver-like notation described below we use the slightly imprecise but convenient notation where $\frac{i \pi}{2} \eta({A})$ is identified with a $\tfrac{1}{2}$ shift in the CS level of $A$, and we do not keep track of background gravitational CS terms.
}
\begin{equation}    \label{eq:bosonization_compact}
U(1)_{-\frac{1}{2}} \; + \; 1 \psi
\qquad \leftrightarrow \qquad
1 \phi
\end{equation} 

\begin{equation}    \label{eq:fermionization_compact}
U(1)_{-1} \; + \; 1 \phi
\qquad \leftrightarrow \qquad
1 \psi
\end{equation}

It is convenient to write the dualities \eqref{eq:bosonization_compact}, \eqref{eq:fermionization_compact} in a quiver notation in view of the generalization discussed below:
\begin{equation}
\text{bosonization:} \qquad 
\fQED(1,-\frac{1}{2}) \dualto
\begin{tikzpicture}[baseline=(current bounding box).center]
    \flavorCS(0,0)(f1, 1, )
    \flavorCS(1.5,0)(f2, 1, )
    \barrow(f1,f2,left)
    \node (V) at (0.75,-1)  {$\mathcal{V} = |\magphi|^4$};
\end{tikzpicture}
\end{equation}

\begin{equation}
\text{fermionization:} \qquad 
\begin{tikzpicture}[baseline=(current bounding box).center]
    \nodeCS(0,0)(g,1,$-1$)
    \node[flavor] (f) at (1.5,0)  {$1$};
    \barrow(g,f,left)
    \node (V) at (0.75,-1)  {$\mathcal{V} = |\magphi |^4$};
\end{tikzpicture}
\dualto
\begin{tikzpicture}[baseline=(current bounding box).center]
    \flavorCS(0,0)(f1, 1, )
    \flavorCS(1.5,0)(f2, 1, )
    \farrow(f1,f2,left)
\end{tikzpicture}
\end{equation}

\begin{figure}[ht]
    \centering
    \includegraphics[width=1.1\linewidth]{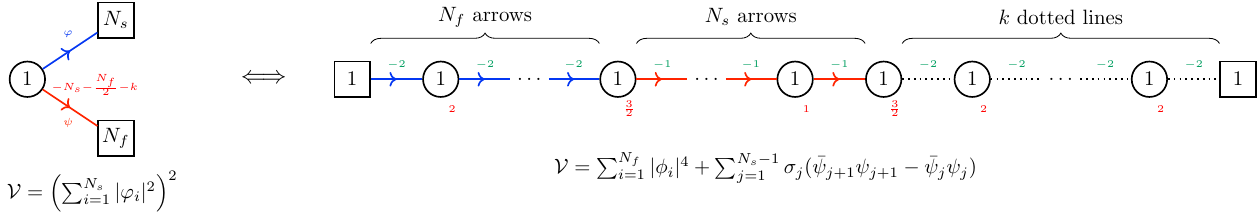}
    \caption{The proposed dual for QED$_3$ with $N_s$ scalars, $N_f$ fermions and CS level $-N_s-\tfrac{N_f}{2}-k$ with $k>0$. Round nodes denote $U(1)$ gauge groups, square node denote flavor groups. Red and blue arrows denote fermions and scalars, respectively, charged under the nodes that they connect. Red labels denote CS levels and green labels denote mixed CS levels.}
    \label{fig:Abelian_dual_bosons_fermions}
\end{figure}

We propose a generalization for the case of QED with $N_f$ fermions $\psi$, $N_s$ complex scalars $\phi$ and CS level $-N_s - \frac{N_f}{2} - k$ with $k\geq0$, shown in Figure \ref{fig:Abelian_dual_bosons_fermions}.
This proposal is a refinement of the dualities already considered in \cite{Karch:2016sxi,Karch:2016aux}.
The dual is a $U(1)^{N_s + N_f + k -1}$ gauge theory with $N_f$ complex scalars and $N_s$ complex fermions charged under two $U(1)$ gauge groups with opposite charge. Furthermore for every $U(1)$ gauge group with at least two charged fermions there is a real scalar $\sigma_j$ that interacts with the two fermions via the Yukawa coupling. 

The off-diagonal mesons of the QED theory are mapped to monopole operators in the quiver theory. In particular, consider thes spin-0 and spin-1 mesons $\bar\elepsi_i \elepsi_j$ with $i < j$. 
The corresponding operator is a monopole with GNO flux $+1$ under all the gauge groups from the $i$-th to the $(j-1)$-th gauge group. 
As reviewed in Appendix \ref{app:monopoles}, this monopole is not gauge invariant and has gauge charges:
\begin{equation}
    Q\left[\mon^{(0, \dots, 0, \overset{i}{+}, \dots, \overset{j-1}{+},0 \dots)}\right] 
    =
    (0,\dots, -1,\overset{i}{+1},0 \dots, 0,\overset{j-1}{+1},-1,0, \dots)
\end{equation}
and can be made gauge invariant by dressing with the scalar $\magphi_{i-1}$ connecting the $(i-1)$-th and the $i$-th node and the complex conjugate scalar $\magphi^*_{j}$. 
Therefore we have the mapping:
\\

\begin{equation}
\bar\elepsi_i \elepsi_j \dualto
\mon^{
\begin{pmatrix}
        0& \dots& \tikznode{g0}{0}  \mkern 30mu  & \tikznode{g1}{+} & \dots & \tikznode{g2}{+}  \mkern 30mu& \tikznode{g3}{0} &\dots
    \end{pmatrix}
   } \quad ,
   \qquad \qquad 
   1\leq i<j \leq N_f
\end{equation}
\begin{tikzpicture}[overlay,remember picture,line width=.7pt,transform canvas={yshift=0mm}]
	\draw[bcolor,->-] (g0)--(g1);
        \draw[bcolor,->-] (g3)--(g2);
        \draw (g1)++(0,0.5) node {\tiny$\scriptstyle i$};
        \draw (g2)++(0,0.5) node {\tiny$\scriptstyle j-1$};
\end{tikzpicture}
where the two arrows denote the dressing with the corresponding scalars and the left-pointing arrow indicates that the dressing is performed with the complex conjugate scalar.
Each boson in the monopole background “sees" a GNO flux $+1$ and therefore has spin-$\tfrac{1}{2}$, resulting in a spin-0 and a spin-1 operator. Similarly, the mesons $\elephi^*_i \elephi_j$ and $\bar\elepsi_i \elephi_j$ are mapped to:
\begin{equation}
\elephi^*_i \elephi_j \dualto
\mon^{
\begin{pmatrix}
        0& \dots& \tikznode{g0}{0}  \mkern 30mu  & \tikznode{g1}{+} & \dots & \tikznode{g2}{+}  \mkern 30mu& \tikznode{g3}{0} &\dots
    \end{pmatrix}
   } \quad ,
    \qquad \qquad 
   1\leq i<j \leq N_s
\end{equation}
\begin{tikzpicture}[overlay,remember picture,line width=.7pt,transform canvas={yshift=0mm}]
	\draw[fcolor,->-] (g0)--(g1);
        \draw[fcolor,->-] (g3)--(g2);
        \draw (g1)++(0,0.5) node {\tiny$\scriptstyle i+N_f$};
        \draw (g2)++(0,0.5) node {\tiny$\scriptstyle j-1+N_f$};
\end{tikzpicture}

\begin{equation}
\bar\elepsi_i \elephi_j \dualto
\mon^{
\begin{pmatrix}
        0& \dots& \tikznode{g0}{0}  \mkern 30mu  & \tikznode{g1}{+} & \dots & \tikznode{g2}{+}  \mkern 30mu& \tikznode{g3}{0} &\dots
    \end{pmatrix}
   }
\end{equation}
\begin{tikzpicture}[overlay,remember picture,line width=.7pt,transform canvas={yshift=0mm}]
	\draw[bcolor,->-] (g0)--(g1);
        \draw[fcolor,->-] (g3)--(g2);
        \draw (g1)++(0,0.5) node {\tiny$\scriptstyle i$};
        \draw (g2)++(0,0.5) node {\tiny$\scriptstyle{j-1+N_f}$};
\end{tikzpicture}
which are operators with spin-0 and spin-$\tfrac{1}{2}$, respectively.

For generic $k$ the monopole operators in the QED transform in non-trivial representations of the flavor symmetry $S(U(N_s)\times U(N_f))$. These monopole operators are expected to map to monopole operators of the dual theory. We will not discuss in detail the mapping of such operators.
When $k=0$ the QED theory has some monopole operators which are neutral under the flavor symmetry $S(U(N_s)\times U(N_f))$.
These are obtained by dressing the bare monopole $\mon^{+}$, which has gauge charge $-N_s$, with $N_s$ (derivatives of) the scalars $\elephi$.
The scalars are antisymmetrized in flavor indices, resulting in an operator that is charged under the topological symmetry but is neutral under $S(U(N_s)\times U(N_f))$.

On the mirror side, only for $k=0$, the quiver has long mesonic operators connecting the two flavor nodes, built out of (derivatives of) all the fields. The long mesons are mapped to the monopoles described above:
\begin{equation}
\mon^{+} \;
\overbrace{\partial_{\bullet} \elephi_{\{ i_1} \;\cdots\;  \partial_{\bullet} \elephi_{ i_{N_s}\} } }
	^{N_s}
\dualto
\partial_{\star}\magphi_1  \;\cdots\; \partial_{\star} \magphi_{N_f} 
\partial_{\star}\magpsi_1  \;\cdots\; \partial_{\star} \magpsi_{N_s} 
\end{equation}

where $\partial_{\bullet}$ and $\partial_{\star}$ indicate generic sets of derivatives. On the LHS Lorentz indices are contracted to obtain a non-vanishing operator. It would be interesting to precisely understand the mapping between the $\partial_{\bullet}$ derivatives on the right and the $\partial_{\star}$ derivatives on the left, but we will not attempt to do so in the present paper. 
While the spin of these operators depend on the sets of derivatives $\partial_{\bullet}$ and $\partial_{\star}$, respectively, the integrality of the spin is given by $\tfrac{N_s}{2} \text{ mod }1$ on both sides. 
\\

For particular values of the three parameters $N_f,N_s$ and $k$, which are all non-negative, our proposal \ref{fig:Abelian_dual_bosons_fermions} reproduces known dualities.
For $k=0$ and a single matter species we recover the Abelian dualities for fermionic or bosonic QED discussed in \cite{Karch:2016aux}:

\begin{equation}	\label{eq:Abelian_mirror_fermion}
\fQED(N_f,-\frac{N_f}{2})
\dualto
\begin{tikzpicture}[baseline=(current bounding box).center]
	\flavorCS(0,0)(f1, 1,)
	\nodeCS(1.5,0)(g1,1, 2)
	\node (dots) at (3,0) {$\dots$};
	\nodeCS(4.5,0)(g2,1, 2)
	\flavorCS(6,0)(f2, 1,)
	
	\barrowBF(f1,g1,left,)
	\barrowBF(g1,dots,left,-2)
	\barrowBF(dots,g2,left,-2)
	\barrowBF(g2,f2,left,)
	
	\node (V) at (3,-1) {$V=\sum_{i=1}^{N_f} |\phi_i|^4 $};
\end{tikzpicture}
\end{equation}

\begin{equation}	\label{eq:Abelian_mirror_boson}
\begin{tikzpicture}[baseline=(current bounding box).center]
    \nodeCS(0,0)(g,1,$-N_s$)
    \node[flavor] (f) at (1.5,0)  {$N_s$};
    \path[bcolor,draw,->-] (g) -- node[midway,above] {$_\elephi$} (f);
    \node (V) at (0.75,-1)  {$\mathcal{V} = \left(\sum_{i=1}^{N_s}| \elephi_i |^2\right)^2$};
\end{tikzpicture}
\dualto
\begin{tikzpicture}[baseline=(current bounding box).center]
	\flavorCS(0,0)(f1, 1,)
	\nodeCS(1.5,0)(g1,1, 1)
	\node (dots) at (3,0) {$\dots$};
	\nodeCS(4.5,0)(g2,1, 1)
	\flavorCS(6,0)(f2, 1,)
	
	\farrowBF(f1,g1,left,)
	\farrowBF(g1,dots,left,-1)
	\farrowBF(dots,g2,left,-1)
	\farrowBF(g2,f2,left,)
	
	\node (V) at (3,-1) {$V=\sum_{i=1}^{N_s-1} \sigma_i( \bar{\psi}_{i+1} \psi_{i+1}- \bar{\psi}_{i} \psi_{i}) $};
\end{tikzpicture}
\end{equation}

In particular, for $k=0$ and $N_f+N_s=1$ we reproduce the original bosonization dualities \eqref{eq:bosonization_compact}, \eqref{eq:fermionization_compact}.
For $k=0$ and $N_f+N_s=2$ the duality \ref{fig:Abelian_dual_bosons_fermions} reduces to dualities between QED theories. For $N_f=N_s=1$ we recover a duality analyzed in \cite{Benini:2017aed} for $U(1)_{-\frac{3}{2}}$ with one scalar and one fermion.
For $N_f=2$, $N_s=0$ or $N_s=2$, $N_f=2$ we recover dualities between scalar and fermionic QED. We discuss these dualities in details in Appendix \ref{sec: Abelian-technical}. 
In the bosonic QED in \eqref{eq:Abelian_mirror_boson} one can turn on a negative mass for all the scalars that preserves the $SU(N_s)$ global symmetry. The scalars condense, Higgsing the gauge group and the theory reduces to a $\mathbb{CP}^{N_s-1}$ NLSM. On the dual side the fermions take negative mass, resulting in a $(U(1)_0)^{N_s-1}$ gauge theory. We conjecture that the $N_s-1$ dual photons together with the $N_s-1$ real scalars $\sigma_i$ combine into a $N_s-1$ complex dimensional NLSM, reproducing the dual side.

Let us discuss some features of the duality \ref{fig:Abelian_dual_bosons_fermions}. 
The symmetry of the QED theory is, up to discrete factors, $U(1)_{top} \times S(U(N_s) \times U(N_f))$. On the quiver side for $k>0$ there are $N_f + N_s$ topological symmetries, while for $k=0$ there are $N_f + N_s-1$ topological symmetries and a $U(1)$ flavor symmetry rotating the fields. In both cases the rank of the global symmetry matches between the two theories. The duality implies that the $U(1)^{N_f+N_f}$ global symmetry of the quiver theory exhibit non-Abelian enhancement in the IR, matching the global symmetry of the QED theory.

Mass deformations for the matter fields of the QED theory are mapped to masses of the quiver theory, schematically:
\begin{equation}
\text{LHS} \qquad
\begin{gathered}
|\phi|^2 \\ \bar\psi \psi
\end{gathered}
\dualto
\begin{gathered}
\bar\psi \psi \\ - |\phi|^2
\end{gathered}
\qquad \text{RHS}
\end{equation}
One can check that turning on masses for the fermions and positive masses for the scalars one recovers the same duality \ref{fig:Abelian_dual_bosons_fermions} with different values of $k, N_f$ and $N_s$. This provides a non-trivial consistency check of our proposal.
Turning on a negative mass for a scalar Higgses the QED theory. In the dual side one of the fermions is integrated out with a positive mass, 
triggering a sequential confinement that results in a theory of free fields, matching the electric theory.

\section{Planar Abelian Duals of Bosonic CS-QCD\texorpdfstring{$_3$}{3}}
\label{sec: non-Abelian}
In this section, we construct a quiver gauge theory dual to non-Abelian QCD$_3$ with fundamental matter. Our proposal builds upon recent results \cite{Benvenuti:2024seb, Benvenuti:2025a, Benvenuti:2025b} that established a duality between $\mathcal{N}=2$ supersymmetric non-Abelian linear CS quiver gauge theories and planar Abelian quiver gauge theories with mixed CS interactions. Motivated by these supersymmetric dualities, we conjecture an analogous correspondence in the non-supersymmetric case: non-Abelian CS QCD$_3$ with fundamental fermionic and bosonic matter should be dual to a planar Abelian quiver theory containing both bosons and fermions. This expectation follows from the observation that our proposed non-supersymmetric dualities should emerge as mass deformations of their supersymmetric counterparts.

\subsection{General Proposal}\label{sec: gen_proposal_bqcd}
\begin{figure}[ht]
    \centerline{\includegraphics[width=1.3\textwidth]{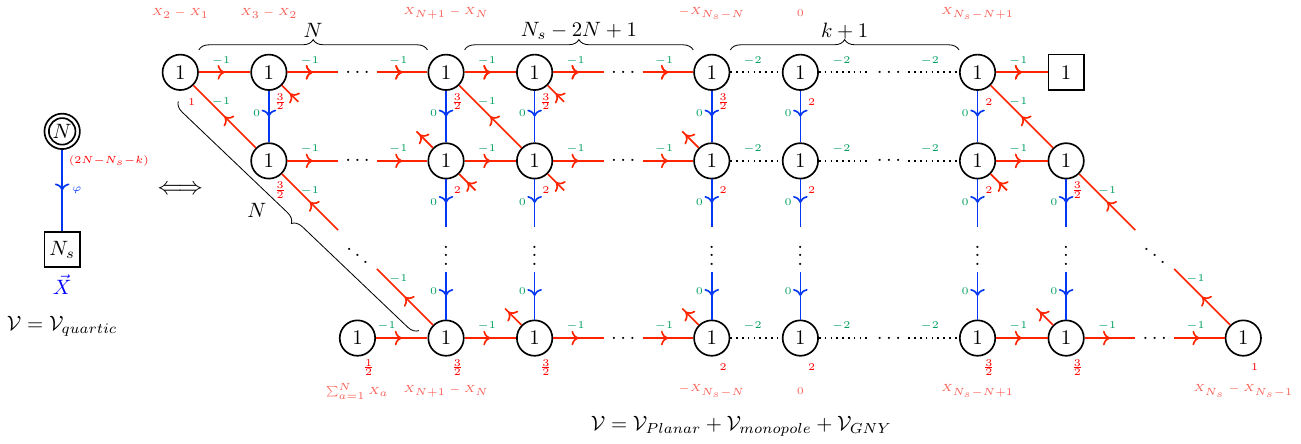}}
    \caption{The proposed dual for $SU(N)$ QCD$_3$ with $N_s$ scalars at Chern-Simons level $2N-N_s-k$ is shown here. Single/double circles (squares) correspond to $U/SU$ gauge (flavor) symmetries. We describe scalar (fermion) bifundamental fields in blue (red). Chern-Simons levels of gauge nodes are indicated in maroon, while mixed CS couplings between gauge nodes are indicated in green. Dotted lines indicate the absence of any bifundamental matter fields and only the presence of BF interactions. 
    BF couplings between adjacent gauge nodes connected by a fermionic field are $-1$, BF couplings between gauge nodes connected by a bosonic field are $0$, while BF couplings between adjacent gauge nodes connected by dotted lines alone are $-2$. We also indicate the fugacities of the topological symmetries of each column in orange. The orientation of arrows indicates the representation of the field under the symmetry group, i.e.: an arrow pointing out of (into) a node transforms in the (anti-)fundamental representation of the symmetry group.
    }
    \label{fig:scalar_Qcd_general}
\end{figure}

We propose the planar Abelian dual for $SU(N)$ QCD$_3$ with $N_s$ critical scalars at Chern-Simons level $2N-N_s-k$ ($k\geq 0$) shown in Figure \ref{fig:scalar_Qcd_general}. This duality maps
flavor symmetries to topological symmetries and many
baryonic and mesonic operators of one theory are mapped to gauge invariant disorder operators (monopoles) of its mirror dual. In this section, we discuss some simple examples explicitly to justify our proposal.
The proposed duality is believed to hold in the following range of parameters:
\begin{equation*}
    N_s  \geq 2N, \qquad k\geq 0.
\end{equation*}
In the planar quiver shown in Figure~\ref{fig:scalar_Qcd_general}, the parameter $k$ counts the number of ``empty” columns — columns that are connected to the rest of the quiver solely through horizontal BF couplings.

We now comment on the potential $\mathcal{V}$ shown in Figure \ref{fig:scalar_Qcd_general} for the planar mirror theory:
\begin{itemize}
    \item $\mathcal{V}_{Planar}$ contains a cubic Yukawa interaction ($\phi_{i}\psi_{j}\psi_k$ + $c.c.$) for every closed planar triangle. There is a term with a coefficient of +1 (-1) for each triangle closed clockwise (anticlockwise). 
    \item $\mathcal{V}_{monopole}$ captures the monopole terms in the potential.  For each vertical bosonic bifundamental field, there is a monopole potential with 
    GNO flux $+1$ and $-1$ under the nodes connected by the arrow, from top to bottom ($\mathfrak{M}^{\binom{+}{-}}$+$c.c.$). 
    As explained in Appendix \ref{app:monopoles} these monopoles can be dressed to obtain a gauge invariant spin-0 operator.
    This breaks the topological symmetry of the gauge nodes in a column of the quiver to a single, diagonal combination. We label the unbroken $U(1)$ symmetry with the orange label at the top (or bottom) of the corresponding column. The label encodes the mapping of the $U(1)$ group to a subgroup $U(1)\subset U(N_s)$ of the electric flavor symmetry, as explained below.
    
    \item $\mathcal{V}_{GNY}$ captures additional cubic interactions between a real scalar field associated with the $\alpha^{\text{th}}$ gauge node in the $I^{\text{th}}$ column, and the fermions charged under this gauge group (an arrow pointing out of (into) a node has charge +1 (-1) under the gauge symmetry). For example, consider a gauge theory represented by the quiver diagram in \eqref{eq: GNY potential}.

    \begin{center}
  \begin{equation}  \label{eq: GNY potential}
    \begin{tikzpicture}
        \node at (0,0) (a1) [gauge,black] {$1$};
        \node at (1.5,0) (a2) [gauge,black] {$1$};
        \draw[fcolor,->-] (-1,0) -- (a1); \draw[fcolor,->-] (a1)--(a2); \draw[fcolor,->-] (a2)--(2.5,0); 
        \draw[fcolor] node at (-.5,.2) {$^{\psi_1}$};
        \draw[fcolor] node at (.75,.2) {$^{\psi_2}$};
        \draw[fcolor] node at (2,.2) {$^{\psi_3}$};
        \draw[fcolor,->-] (a1)++(1,-1) -- node[midway,right] {$^{\psi_4}$} (a1); 
        \draw[fcolor,->-] (a2)++(1,-1) -- node[midway,right] {$^{\psi_5}$} (a2); 
        \draw[bcolor,->-] (a1) -- ++ (0,-1);
        \draw[bcolor,->-] (a2) -- ++ (0,-1);
    \end{tikzpicture}
    \end{equation}
    \end{center}
    $\mathcal{V}_{GNY}$ for this theory has the form:
    \begin{equation}
        \mathcal{V}_{GNY}=\sigma_1(\bar{\psi}_2\psi_2-\bar{\psi}_1\psi_1-\bar{\psi}_4\psi_4) + \sigma_2(\bar{\psi}_3\psi_3-\bar{\psi}_2\psi_2-\bar{\psi}_5\psi_5).
    \end{equation}
    Each gauge node \textbf{except} for the bottom-most gauge node has a GNY potential accompanying the Yukawa interactions. 

    By contrast, if we consider the following gauge theory (the dashed line indicates the absence of any fermionic bifundamental field, and the two nodes may share a bosonic bifundamental field or only mixed CS interactions):
    \begin{center}
    \begin{equation}  \label{eq: GNY potential_BF}
    \begin{tikzpicture}
        \node at (0,0) (a1) [gauge,black] {$1$};
        \node at (1.5,0) (a2) [gauge,black] {$1$};
        \draw[fcolor,->-] (-1,0) -- (a1); \draw[dashed] (a1)--(a2); \draw[fcolor,->-] (a2)--(2.5,0); 
        \draw[fcolor] node at (-.5,.2) {$^{\psi_1}$};
        \draw[fcolor] node at (2,.2) {$^{\psi_2}$};
        \draw[fcolor,->-] (a2)++(1,-1) -- node[midway,right] {$^{\psi_3}$} (a2); 
        \draw[bcolor,->-] (a1) -- ++ (0,-1);
        \draw[bcolor,->-] (a2) -- ++ (0,-1);
    \end{tikzpicture}
    \end{equation}
    \end{center}
    we expect the identification of $\sigma_1$ and $\sigma_2$ so the resulting $\mathcal{V}_{GNY}$ has the form:
    \begin{equation}
        \mathcal{V}_{GNY}=\sigma(\bar{\psi}_2\psi_2-\bar{\psi}_1\psi_1-\bar{\psi}_3\psi_3). 
    \end{equation}

    In the quiver theory in Figure \ref{fig:scalar_Qcd_general} there are $N(N_s-N)$ reals scalars $\sigma$.
    
\end{itemize}

In the QCD$_3$ theory, the quartic scalar potential is $\mathcal{V}_{quartic}$\footnote{N.B.: Latin indices $i,j=1,\ldots,F$ run over flavor indices, and Greek indices $\alpha,\beta=1,\ldots,N$ run over gauge indices.}:
\begin{equation}\label{eq:quarticV}
    \mathcal{V}_{quartic} = \sum_{\alpha,\beta=1}^N\sum_{I,J=1}^{N_s}\big[\varphi^{\alpha I}\bar{\varphi}_{\alpha J}\varphi^{\beta J}\bar{\varphi}_{\beta I}+(\varphi^{\alpha I}\bar{\varphi}_{\alpha I})(\varphi^{\beta J}\bar{\varphi}_{\beta J})\big]
\end{equation}
which preserves the full $U(N_s)$ global symmetry.

In the planar mirror theory the presence of the  interactions $\mathcal{V}_{monopole}$ and $\mathcal{V}_{Planar}$  are essential to match the rank of the global symmetry on the electric side. In the presence of these terms, the UV symmetry of the quiver theory is $U(1)_{top}^{N_s}$, where each $U(1)$ factor corresponds to the topological symmetries of a column of the quiver, broken to a diagonal combination by $\mathcal{V}_{monopole}$\footnote{There are $N_s-1$ columns which contribute $N_s-1$ $U(1)$ factors, and the additional $U(1)$ factor is given by the topological symmetry of the bottom-most gauge node. The topological symmetries associated to the columns within the “empty" region of the quiver can be reabsorbed by a gauge transformation and do not contribute to the global symmetry.}.
We further claim that the symmetry enhances in the IR to:
\begin{equation}
    U(1)^{N_s}_{top}\to SU(N_s) \times U(1)
\end{equation}
reproducing the $SU(N_s)\times U(1)$ flavor and baryonic symmetries of the electric theory respectively. This is most transparent in our choice of parametrization of the fugacities of the topological symmetries of each column in Figure \ref{fig:scalar_Qcd_general}.
Explicitly, the diagonal combination of the topological symmetries in the $i$-th column, associated to $X_{i+1}-X_i$, is mapped to the combination $U(1)_{X_{i+1}}-U(1)_{X_i}$ in the maximal torus of $U(N_s)$. Similarly, the topological symmetry of the bottom left node, associated to $\sum_{a=1}^{N} X_a$, is mapped to the linear combination $\sum_{i=1}^{N} U(1)_{X_i}$.
Therefore the fugacities denoted with orange labels in \ref{fig:scalar_Qcd_general} encode the explicit embedding of $U(1)_{top}^{N_s}$ in the enhanced $U(N_s)$ global symmetry group.

Moreover, the faithful symmetry of the QCD theory is \cite{Benini:2017dus,Benini:2017aed}:
\begin{equation}
    \frac{U(N_s)}{\mathbb{Z}_N} \rtimes \mathbb{Z}_2^{\mathcal{C}}
\end{equation}
where $\mathbb{Z}_2^{\mathcal{C}}$ is charge conjugation, acting as complex conjugation on the scalars for $N\geq 2$\footnote{For $N=2$, as $SU(2) \cong USp(2)$, the global symmetry of the electric $SU(2)$ QCD$_3$ theory is:
\begin{equation}
    \frac{USp(2N_s)}{\mathbb{Z}_2};
\end{equation}
which is broken by $\mathcal{V}_{quartic}$ to $\frac{U(N_s)}{\mathbb{Z}_2}\rtimes \mathbb{Z}_2^{\mathcal{C}}$ \cite{Benini:2017aed}.}.
$\mathbb{Z}_N$ is generated by the following element of the maximal torus of $U(N_s)$:
\begin{equation}
    \left\{e^{\tfrac{2\pi i}{N}},\dots,e^{\tfrac{2\pi i}{N}}\right\}
    \in
    \left\{U(1)_{X_1},\dots,U(1)_{X_N}\right\}
\end{equation}
All the gauge invariant dressed monopoles in the dual quiver transform trivially under this element, and are therefore compatible with being components of representations of the enhanced $\tfrac{U(N_s)}{\zz_N}$.

The fact that there are no mesonic $U(1)$ symmetries in the planar theory is a non-trivial statement and can be checked as follows by looking at the planar quiver:
\begin{equation}
    \begin{split}
     \# \text{mesonic } ~ U(1)s &= ~ \# \text{lines}-\#\text{gauge nodes}- \#\text{triangles} = \\
     & = \{3N(N_s-N)-2N_s+3\}-\{N(N_s-N)+1\} = \\
     &\;\;\;\; -\{2N(N_s-N)-2(N_s-1)\} \\
     &=0
    \end{split}
\end{equation}
where we assume that there is a mesonic $U(1)$ rotating each boson/fermion in the mirror, and a cubic potential term associated to each triangle appearing in the quiver\footnote{Note that this straightforward computation holds when the parameter $k=0$. For $k>0$, matching the $U(1)$ symmetries becomes more subtle. Nonetheless, the matching can be verified in this case as well.}.

The presence of the Gross-Neveu-Yukawa potential $\mathcal{V}_{GNY}$ also plays an important role in the duality and can be motivated as follows. On adding a negative mass deformation for the $N_s$ scalars in the electric theory, the gauge group is completely Higgsed and the IR phase of the theory is a sigma model with the target manifold identified as the coset space \cite{Bando:1987br, Nakahara:2003nw}: 
\begin{equation}
    \frac{SU(N_s)}{SU(N_s-N)\times SU(N)}
\end{equation}
which has $2N N_s-2N^2 + 1$ real dimensions. Upon mapping the same deformation in the mirror theory, we expect to flow to $N(N_s-N)+1$ copies of $U(1)_0$, whose dual photons parameterize $(\mathbf{S}^1)^{N(N_s-N)+1}$ in the deep IR. This can be achieved by giving a negative mass to all the fermions and a positive mass to all the scalars in the quiver.
The dual photons are expected to combine with the $N(N_s-N)$ real scalars $\sigma$, resulting in a NLSM with real dimension $2N(N_s-N)+1$, consistent with the QCD theory.

In the rest of this section, we will make these ideas more precise by considering some explicit examples. Through these examples, we illustrate the effects of increasing the Chern-Simons level $k$ and the rank of the gauge symmetries on the respective planar duals.  
\subsection{Example: \texorpdfstring{$SU(2)_{-1}$  QCD$_3$ with $N_s=5$}{su(2)} scalars}
\label{sec: SU(2)_1w5s}
\begin{figure}[ht]
     \centerline{\includegraphics[width=.9\textwidth]{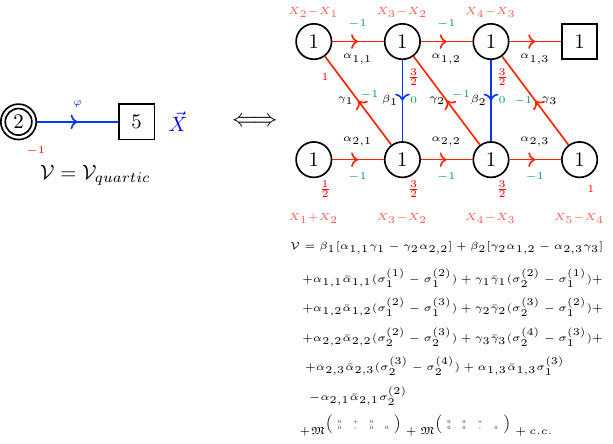}}
    \caption{The dual for $SU(2)_{-1}$ QCD$_3$ with 5 scalars is shown here. All labelling conventions follow those of Figure \ref{fig:scalar_Qcd_general}. The superscript (subscript) for the real scalar $\sigma$ indicates its column (position in the column), e.g.: $\sigma_{1}^{(3)}$ corresponds to the 1$^{\text{st}}$ gauge node in the 3$^{\text{rd}}$ column. 
    }
    \label{fig: SU(2)_1w5s}
\end{figure}

Let us begin with the example of $SU(2)_{-1}$ with $5$ scalars and its planar dual description. Since this is our first example, we are pedantic and give many details to illustrate our proposal. We propose the duality shown in Figure \ref{fig: SU(2)_1w5s}.

Henceforth, we refer to the SU(2)$_{-1}$ QCD$_3$ with 5 scalars as the ``electric" theory and the planar theory as its ``mirror". We are mainly concerned with the mapping of the following gauge-invariant operators in the electric theory:
\begin{enumerate}
    \item $25$ \textbf{mesons} of the form $\varphi_i\bar{\varphi}_j$ transforming in the adjoint representation of the flavor symmetry group $U(5)$. Hence, with respect to the maximal torus $U(1)_{X_i}^5 \subset U(5)$, they are charged under $U(1)_{X_i}-U(1)_{X_j}$.
    The off-diagonal mesons (those with $i \neq j$) should then be mapped to monopole operators of the dual quiver theory.
    The proposed mapping for the mesonic operators is shown in Table \ref{table: mesons of SU(2)_1 QCD3}.
    The diagonal mesons $|\varphi_i|^2$ are neutral under the $U(1)_{X_i}$ and are expected to map to mesons of the dual theory.
    \item $25-5 = 20$ \textbf{conserved currents} associated to the global symmetry enhancement $U(1)^5 \to U(5)$. 
    These are as spin-1 operators charged under $U(1)_{X_i}-U(1)_{X_j}$ with $i \neq j$ and are expected to be mapped to monopole operators, as illustrated below.
    \item $\binom{5}{2} = 10$ \textbf{baryons} constructed by anti-symmetrizing the color indices with the help of the Levi-Civita tensor $\epsilon_{\alpha\beta}\varphi^{\alpha}_i\varphi^{\beta}_j$. Together, these transform in the $2$-index antisymmetric representation of the flavor symmetry group $SU(5)$ and are charged under $U(1)_{X_i}+U(1)_{X_j}$, with $i \neq j$. 
    The proposed mapping for the baryonic operators is shown in Table \ref{table: baryons of SU(2)_1 QCD3}.
\end{enumerate}
The off-diagonal mesons, off-diagonal currents and the baryons are charged under the maximal torus $U(1)_{X_i}^5\subset U(5)$, therefore they should be mapped to monopole operators in the dual quiver theory. 
In what follows, we propose a mapping for these operators, starting from the off-diagonal mesons and the baryons.
Thus, the task at hand is to determine gauge invariant monopole operators that are \textbf{Lorentz scalars} and are charged under 
$U(1)_{X_i}\mp U(1)_{X_j}$, which are mapped to the topological symmetries of the quiver theory as described above. We make this statement precise in the next paragraphs by considering some examples. 
The techniques needed to compute the gauge charges of monopole operators, the dressing needed to construct gauge invariant monopoles and the spin of the resulting operator are reviewed in Appendix \ref{app:monopoles}.

\paragraph{``Mesonic" Monopole Operators}
We study monopole operators with the right quantum numbers to be mapped to the mesons of the electric theory here. As an example, we consider the following monopole operator:
\begin{equation}
    \mathfrak{M}^{\left(\:\hspace{-2pt}
        \resizebox{50pt}{!}{%
        \begin{tabular}{cccc} +&+&0& \\0&0&0&0
        \end{tabular}}\right)}
\end{equation}
This operator has charge $+1$ under the topological symmetries of the first two gauge nodes; hence, it is charged under $U(1)_{X_3}-U(1)_{X_1}$, and transforms as a component of the adjoint representation of $SU(5)$. We can calculate the gauge charge of this operator:
\begin{equation}
    \begin{pmatrix}
        2 & -1 & 0 & \\ 0 & -1 & 0 & 0
    \end{pmatrix} + 
     \begin{pmatrix}
        -1 & 3 & -1 & \\ 0 & 0 & -1 & 0
    \end{pmatrix} \equiv
     \begin{pmatrix}
        1 & 2 & -1 & \\ 0 & -1 & -1 & 0
    \end{pmatrix}.
\end{equation}
Clearly, it is not gauge invariant and must be dressed appropriately. We choose the minimal dressing \footnote{By this, we mean that we use the minimal number of fields to dress the bare monopole and construct a gauge-invariant operator. A non-minimal dressing—i.e., using additional fields—would yield an operator with a higher scaling dimension. In this work, we focus exclusively on mapping the operators with the lowest scaling dimensions.}:

\begin{equation}
\begin{pmatrix}
        \tikznode{g11}{+} & & \tikznode{g12}{2} & & \tikznode{g13}{-} & & \\ 0 & & \tikznode{g22}{-} & &\tikznode{g23}{-} & &0
    \end{pmatrix} \quad,
\end{equation}
\begin{tikzpicture}[overlay,remember picture,line width=.7pt,transform canvas={yshift=0mm}]
	\draw[fcolor,->-] (g22)--(g11);
        \draw[fcolor,->-] (g23)--(g12);
        \draw[fcolor,->-] (g13)--(g12);
\end{tikzpicture}
hence, we see that the operator: 
\begin{equation}
    \mathfrak{M}^{\left(\:\hspace{-2pt}
        \resizebox{50pt}{!}{%
        \begin{tabular}{cccc} +&+&0& \\0&0&0&0
        \end{tabular}}\right)} \gamma_1 \gamma_2 \bar{\alpha}_{1,2}
\end{equation}
is gauge invariant. 
We denote this operator by combining the GNO flux assignment and the dressing in the following way:
\begin{equation}
\mathfrak{M}^{\left(\:\hspace{-2pt}
        \resizebox{50pt}{!}{%
        \begin{tabular}{cccc} +&+&0& \\0&0&0&0
        \end{tabular}}\right)} \gamma_1 \gamma_2 \bar{\alpha}_{1,2}
\equiv
\begin{pmatrix}
        \tikznode{g11}{+} & & \tikznode{g12}{+} & & \tikznode{g13}{0} & & \\ 0 & & \tikznode{g22}{0} & &\tikznode{g23}{0} & &0
    \end{pmatrix} \quad .
\end{equation}
\begin{tikzpicture}[overlay,remember picture,line width=.7pt,transform canvas={yshift=0mm}]
	\draw[fcolor,->-] (g22)--(g11);
        \draw[fcolor,->-] (g23)--(g12);
        \draw[fcolor,->-] (g13)--(g12);
\end{tikzpicture}

As each of the three fermionic fields is quantized in a background of total magnetic flux $1$, their spin statistics changes to that of spin-0 bosons\footnote{In principle, they also have a spin-1 channel. Unless otherwise stated we dress the monopoles with the zero modes of the fermion fields, resulting in a spin-0 operator \cite{Borokhov_2002}.}. We conclude that the dressed monopole operator has Lorentz spin 0, and is the right operator to be mapped to $\varphi_{3}\bar{\varphi}_1$. 

We can readily identify the off-diagonal mesonic operators and summarize their dual descriptions---including necessary dressings---in Table~\ref{table: mesons of SU(2)_1 QCD3}.
The mapping of diagonal mesonic operators $|\elephi_i|^2$ is less straightforward because these operators are neutral under the maximal torus of $U(5)$, visible in the quiver UV Lagrangian. 
We expect that these operators are mapped to suitable linear combinations of mesonic operators of the quiver, compatible with the mass deformations discussed in Section \ref{checks}. We defer the study of the mapping of diagonal mesonic operators to future work.

\begin{center}
\renewcommand{\arraystretch}{0.6}
\begin{longtable}{ |>{\centering\arraybackslash}p{3.5cm}|>{\centering\arraybackslash}p{3.5cm}|>{\centering\arraybackslash}p{2cm}|>{\centering\arraybackslash}p{3cm}|  }
\caption{Dressed monopole operators of the planar mirror (Figure \ref{fig: SU(2)_1w5s}) and their putative mapping to mesonic operators of the electric theory. The remaining mesonic operators can be constructed by charge conjugation.}
\label{table: mesons of SU(2)_1 QCD3} \\ 

\hline \multicolumn{1}{|c|}{\textbf{GNO Flux}} & \multicolumn{1}{c|}{\textbf{Gauge Charge}} & \multicolumn{1}{c|}{\textbf{Spin}} & \multicolumn{1}{c|}{\textbf{``Electric" Meson}} \\ \hline 
\endfirsthead

\multicolumn{4}{c}%
{{ \textbf{\tablename\ \thetable{}} -- \text{continued from previous page}}} \\
\hline \multicolumn{1}{|c|}{\textbf{GNO Flux }} & \multicolumn{1}{c|}{\textbf{Gauge Charge}} & \multicolumn{1}{c|}{\textbf{Spin}} & \multicolumn{1}{c|}{\textbf{``Electric" Meson}} \\ \hline 
\endhead

\hline \multicolumn{4}{|r|}{{\textit{Continued on next page}}} \\ \hline
\endfoot

\hline \hline
\endlastfoot
\vspace{-0.5cm}
\begin{equation*}
\begin{pmatrix}
        \tikznode{g11}{+} & & \tikznode{g12}{0} & & 0 & & \\ 0 & & \tikznode{g21}{0} & &0 & &0
    \end{pmatrix}
\end{equation*}
\begin{tikzpicture}[overlay,remember picture,line width=.7pt,transform canvas={yshift=0mm}]
	\draw[fcolor,->-] (g12)--(g11);
        \draw[fcolor,->-] (g21)--(g11);
\end{tikzpicture} 
\vspace{-0.5cm}
& 
\vspace{-0.5cm}
\begin{equation*}
\begin{pmatrix}
        \tikznode{g11}{2} & & \tikznode{g12}{-} & & 0 & & \\ 0 & & \tikznode{g21}{-} & &0 & &0
    \end{pmatrix}
\end{equation*}
\begin{tikzpicture}[overlay,remember picture,line width=.7pt,transform canvas={yshift=0mm}]
	\draw[fcolor,->-] (g12)--(g11);
        \draw[fcolor,->-] (g21)--(g11);
\end{tikzpicture}
\vspace{-0.5cm}
& \vspace{0.3cm} 0 & \vspace{0.3cm} $\varphi_2\bar{\varphi}_1$ \\ \hline

\vspace{-0.5cm}\begin{equation*}
\begin{pmatrix}
        \tikznode{g11}{+} & & \tikznode{g12}{+} & & \tikznode{g13}{0} & & \\ 0 & & \tikznode{g22}{0} & &\tikznode{g23}{0} & &0
    \end{pmatrix}
\end{equation*}
\begin{tikzpicture}[overlay,remember picture,line width=.7pt,transform canvas={yshift=0mm}]
	\draw[fcolor,->-] (g22)--(g11);
        \draw[fcolor,->-] (g23)--(g12);
        \draw[fcolor,->-] (g13)--(g12);
\end{tikzpicture} 
\vspace{-0.5cm}
&
\vspace{-0.5cm}
\begin{equation*}
\begin{pmatrix}
        \tikznode{g11}{+} & & \tikznode{g12}{2} & & \tikznode{g13}{-} & & \\ 0 & & \tikznode{g22}{-} & &\tikznode{g23}{-} & &0
    \end{pmatrix}
\end{equation*}
\begin{tikzpicture}[overlay,remember picture,line width=.7pt,transform canvas={yshift=0mm}]
	\draw[fcolor,->-] (g22)--(g11);
        \draw[fcolor,->-] (g23)--(g12);
        \draw[fcolor,->-] (g13)--(g12);
\end{tikzpicture}
\vspace{-0.5cm}
& \vspace{0.3cm} 0 &  \vspace{0.3cm}  $\varphi_3\bar{\varphi}_1$ \\ \hline

\vspace{-0.5cm}
\begin{equation*}
\begin{pmatrix}
        \tikznode{g11}{+} & & \tikznode{g12}{+} & & \tikznode{g13}{+} & & \tikznode{f1}{{\color{white}+}}\\ 0 & & \tikznode{g22}{0} & &\tikznode{g23}{0} & &\tikznode{g24}{0}
    \end{pmatrix}
\end{equation*}
\begin{tikzpicture}[overlay,remember picture,line width=.7pt,transform canvas={yshift=0mm}]
	\draw[fcolor,-<-] (g11)--(g22);
        \draw[fcolor,-<-] (g12)--(g23);
        \draw[fcolor,-<-] (g13)--(f1);
        \draw[fcolor,-<-] (g13)--(g24);
\end{tikzpicture}
\vspace{-0.5cm}
& 
\vspace{-0.5cm}\begin{equation*}
\begin{pmatrix}
        \tikznode{g11}{+} & & \tikznode{g12}{+} & & \tikznode{g13}{2} & & \tikznode{f1}{{\color{white}+}}\\ 0 & & \tikznode{g22}{-} & &\tikznode{g23}{-} & & \tikznode{g24}{-}
    \end{pmatrix}
\end{equation*}
\begin{tikzpicture}[overlay,remember picture,line width=.7pt,transform canvas={yshift=0mm}]
	\draw[fcolor,-<-] (g11)--(g22);
        \draw[fcolor,-<-] (g12)--(g23);
        \draw[fcolor,-<-] (g13)--(f1);
        \draw[fcolor,-<-] (g13)--(g24);
\end{tikzpicture}
\vspace{-0.5cm}
& \vspace{0.3cm} 0 &  \vspace{0.3cm}  $\varphi_4\bar{\varphi}_1$ \\ \hline
\vspace{-0.5cm}
\begin{equation*}
\begin{pmatrix}
        \tikznode{g11}{+} & & \tikznode{g12}{+} & & \tikznode{g13}{+} & & \tikznode{f1}{{\color{white}+}}\\ 0 & & \tikznode{g22}{0} & &\tikznode{g23}{0} & &\tikznode{g24}{+}
    \end{pmatrix}
\end{equation*}
\begin{tikzpicture}[overlay,remember picture,line width=.7pt,transform canvas={yshift=0mm}]
	\draw[fcolor,-<-] (g11)--(g22);
        \draw[fcolor,-<-] (g12)--(g23);
        \draw[fcolor,-<-] (g13)--(f1);
        \draw[fcolor,-<-] (g24)--(g23);
\end{tikzpicture} 
\vspace{-0.5cm}
&
\vspace{-0.5cm}\begin{equation*}
\begin{pmatrix}
        \tikznode{g11}{+} & & \tikznode{g12}{+} & & \tikznode{g13}{+} & & \tikznode{f1}{{\color{white}+}}\\ 0 & & \tikznode{g22}{-} & &\tikznode{g23}{-2} & & \tikznode{g24}{+}
    \end{pmatrix}
\end{equation*}
\begin{tikzpicture}[overlay,remember picture,line width=.7pt,transform canvas={yshift=0mm}]
	\draw[fcolor,-<-] (g11)--(g22);
        \draw[fcolor,-<-] (g12)--(g23);
        \draw[fcolor,-<-] (g13)--(f1);
        \draw[fcolor,-<-] (g24)--(g23);
\end{tikzpicture}
\vspace{-0.5cm}
& \vspace{0.3cm} 0 &  \vspace{0.3cm}  $\varphi_5\bar{\varphi}_1$ \\ \hline
\vspace{-0.5cm}\begin{equation*}
\begin{pmatrix}
        \tikznode{g11}{0} & & \tikznode{g12}{+} & & \tikznode{g13}{0} & & \tikznode{f1}{}\\ 0 & & \tikznode{g22}{0} & &\tikznode{g23}{0} & &\tikznode{g24}{0}
    \end{pmatrix}
\end{equation*}
\begin{tikzpicture}[overlay,remember picture,line width=.7pt,transform canvas={yshift=0mm}]
	\draw[fcolor,-<-] (g12)--(g13);
        \draw[fcolor,-<-] (g12)--(g23);
        \draw[fcolor,-<-] (g12)--(g11);
\end{tikzpicture} \vspace{-0.5cm}& 
\vspace{-0.5cm}\begin{equation*}
\begin{pmatrix}
        \tikznode{g11}{-} & & \tikznode{g12}{3} & & \tikznode{g13}{-} & & \tikznode{f1}{}\\ 0 & & \tikznode{g22}{0} & &\tikznode{g23}{-} & & \tikznode{g24}{0}
    \end{pmatrix}
\end{equation*}
\begin{tikzpicture}[overlay,remember picture,line width=.7pt,transform canvas={yshift=0mm}]
	\draw[fcolor,-<-] (g12)--(g13);
        \draw[fcolor,-<-] (g12)--(g23);
        \draw[fcolor,-<-] (g12)--(g11);
\end{tikzpicture}
\vspace{-0.5cm}
& \vspace{0.3cm} 0 &  \vspace{0.3cm}  $\varphi_3\bar{\varphi}_2$ \\ \hline

\vspace{-0.5cm}\begin{equation*}
\begin{pmatrix}
        \tikznode{g11}{0} & & \tikznode{g12}{+} & & \tikznode{g13}{+} & & \tikznode{f1}{{\color{white}+}}\\ 0 & & \tikznode{g22}{0} & &\tikznode{g23}{0} & &\tikznode{g24}{0}
    \end{pmatrix}
\end{equation*}
\begin{tikzpicture}[overlay,remember picture,line width=.7pt,transform canvas={yshift=0mm}]
	\draw[fcolor,-<-] (g13)--(f1);
        \draw[fcolor,-<-] (g13)--(g24);
        \draw[fcolor,-<-] (g12)--(g23);
        \draw[fcolor,-<-] (g12)--(g11);
\end{tikzpicture} \vspace{-0.5cm}& 
\vspace{-0.5cm}\begin{equation*}
\begin{pmatrix}
        \tikznode{g11}{-} & & \tikznode{g12}{2} & & \tikznode{g13}{2} & & \tikznode{f1}{{\color{white}+}}\\ 0 & & \tikznode{g22}{0} & &\tikznode{g23}{-} & & \tikznode{g24}{-}
    \end{pmatrix}
\end{equation*}
\begin{tikzpicture}[overlay,remember picture,line width=.7pt,transform canvas={yshift=0mm}]
	\draw[fcolor,-<-] (g13)--(f1);
        \draw[fcolor,-<-] (g13)--(g24);
        \draw[fcolor,-<-] (g12)--(g23);
        \draw[fcolor,-<-] (g12)--(g11);
\end{tikzpicture}
\vspace{-0.5cm}
& \vspace{0.3cm} 0 &  \vspace{0.3cm}  $\varphi_4\bar{\varphi}_2$ \\ \hline

\vspace{-0.5cm}\begin{equation*}
\begin{pmatrix}
        \tikznode{g11}{0} & & \tikznode{g12}{+} & & \tikznode{g13}{+} & & \tikznode{f1}{{\color{white}+}}\\ 0 & & \tikznode{g22}{0} & &\tikznode{g23}{0} & &\tikznode{g24}{+}
    \end{pmatrix}
\end{equation*}
\begin{tikzpicture}[overlay,remember picture,line width=.7pt,transform canvas={yshift=0mm}]
	\draw[fcolor,-<-] (g13)--(f1);
        \draw[fcolor,-<-] (g24)--(g23);
        \draw[fcolor,-<-] (g12)--(g23);
        \draw[fcolor,-<-] (g12)--(g11);
\end{tikzpicture} \vspace{-0.5cm}& 
\vspace{-0.5cm}\begin{equation*}
\begin{pmatrix}
        \tikznode{g11}{-} & & \tikznode{g12}{2} & & \tikznode{g13}{+} & & \tikznode{f1}{{\color{white}+}}\\ 0 & & \tikznode{g22}{0} & &\tikznode{g23}{-2} & & \tikznode{g24}{+}
    \end{pmatrix}
\end{equation*}
\begin{tikzpicture}[overlay,remember picture,line width=.7pt,transform canvas={yshift=0mm}]
	\draw[fcolor,-<-] (g13)--(f1);
        \draw[fcolor,-<-] (g24)--(g23);
        \draw[fcolor,-<-] (g12)--(g23);
        \draw[fcolor,-<-] (g12)--(g11);
\end{tikzpicture}
\vspace{-0.5cm}
& \vspace{0.3cm} 0 &  \vspace{0.3cm}  $\varphi_5\bar{\varphi}_2$ \\ \hline
\vspace{-0.5cm}\begin{equation*}
\begin{pmatrix}
        \tikznode{g11}{0} & & \tikznode{g12}{0} & & \tikznode{g13}{+} & & \tikznode{f1}{{\color{white}+}}\\ 0 & & \tikznode{g22}{0} & &\tikznode{g23}{0} & &\tikznode{g24}{0}
    \end{pmatrix}
\end{equation*}
\begin{tikzpicture}[overlay,remember picture,line width=.7pt,transform canvas={yshift=0mm}]
	\draw[fcolor,-<-] (g13)--(f1);
        \draw[fcolor,-<-] (g13)--(g12);
        \draw[fcolor,-<-] (g13)--(g24);
\end{tikzpicture} \vspace{-0.5cm}& \vspace{-0.5cm}\begin{equation*}
\begin{pmatrix}
        \tikznode{g11}{0} & & \tikznode{g12}{-} & & \tikznode{g13}{3} & & \tikznode{f1}{{\color{white}+}}\\ 0 & & \tikznode{g22}{0} & &\tikznode{g23}{0} & & \tikznode{g24}{-}
    \end{pmatrix}
\end{equation*}
\begin{tikzpicture}[overlay,remember picture,line width=.7pt,transform canvas={yshift=0mm}]
	\draw[fcolor,-<-] (g13)--(f1);
        \draw[fcolor,-<-] (g13)--(g12);
        \draw[fcolor,-<-] (g13)--(g24);
\end{tikzpicture}
\vspace{-0.5cm}
& \vspace{0.3cm} 0 &  \vspace{0.3cm}  $\varphi_4\bar{\varphi}_3$ \\ \hline

\vspace{-0.5cm}\begin{equation*}
\begin{pmatrix}
        \tikznode{g11}{0} & & \tikznode{g12}{0} & & \tikznode{g13}{+} & & \tikznode{f1}{{\color{white}+}}\\ 0 & & \tikznode{g22}{0} & &\tikznode{g23}{0} & &\tikznode{g24}{+}
    \end{pmatrix}
\end{equation*}
\begin{tikzpicture}[overlay,remember picture,line width=.7pt,transform canvas={yshift=0mm}]
	\draw[fcolor,-<-] (g13)--(f1);
        \draw[fcolor,-<-] (g13)--(g12);
        \draw[fcolor,-<-] (g24)--(g23);
\end{tikzpicture} \vspace{-0.5cm}& 
\vspace{-0.5cm}\begin{equation*}
\begin{pmatrix}
        \tikznode{g11}{0} & & \tikznode{g12}{-} & & \tikznode{g13}{2} & & \tikznode{f1}{{\color{white}+}}\\ 0 & & \tikznode{g22}{0} & &\tikznode{g23}{-} & & \tikznode{g24}{+}
    \end{pmatrix}
\end{equation*}
\begin{tikzpicture}[overlay,remember picture,line width=.7pt,transform canvas={yshift=0mm}]
	\draw[fcolor,-<-] (g13)--(f1);
        \draw[fcolor,-<-] (g13)--(g12);
        \draw[fcolor,-<-] (g24)--(g23);
\end{tikzpicture}
\vspace{-0.5cm}
& 
\vspace{0.3cm} 0 &  \vspace{0.3cm}  $\varphi_5\bar{\varphi}_3$ \\ \hline
\vspace{-0.5cm}\begin{equation*}
\begin{pmatrix}
        \tikznode{g11}{0} & & \tikznode{g12}{0} & & \tikznode{g13}{0} & & \tikznode{f1}{}\\ 0 & & \tikznode{g22}{0} & &\tikznode{g23}{0} & &\tikznode{g24}{+}
    \end{pmatrix}
\end{equation*}
\begin{tikzpicture}[overlay,remember picture,line width=.7pt,transform canvas={yshift=0mm}]
	\draw[fcolor,-<-] (g24)--(g13);
        \draw[fcolor,-<-] (g24)--(g23);
\end{tikzpicture} \vspace{-0.5cm}
& 
\vspace{-0.5cm}\begin{equation*}
\begin{pmatrix}
        \tikznode{g11}{0} & & \tikznode{g12}{0} & & \tikznode{g13}{-} & & \tikznode{f1}{}\\ 0 & & \tikznode{g22}{0} & &\tikznode{g23}{-} & & \tikznode{g24}{2}
    \end{pmatrix}
\end{equation*}
\begin{tikzpicture}[overlay,remember picture,line width=.7pt,transform canvas={yshift=0mm}]
        \draw[fcolor,-<-] (g24)--(g13);
        \draw[fcolor,-<-] (g24)--(g23);
\end{tikzpicture}
\vspace{-0.5cm}
& \vspace{0.3cm} 0 &  \vspace{0.3cm}  $\varphi_5\bar{\varphi}_4$ \\ \hline

\end{longtable}
\end{center}

Let us comment on a particular feature of the monopole operators in the quiver theories considered in this paper, which is a consequence of the presence of the monopole interaction $\mathcal{V}_{monopole}$.
When looking for a condidate for the dual of an electric meson, say $\varphi_4\bar{\varphi}_3$, we may also consider the monopole:
\begin{equation}
\begin{pmatrix}
        \tikznode{g11}{0} & & \tikznode{g12}{0} & & \tikznode{g13}{0} & & \tikznode{f1}{}\\ 0 & & \tikznode{g22}{0} & &\tikznode{g23}{+} & &\tikznode{g24}{0}
    \end{pmatrix}.
\end{equation}
\begin{tikzpicture}[overlay,remember picture,line width=.7pt,transform canvas={yshift=0mm}]
	\draw[fcolor,-<-] (g23)--(g24);
        \draw[fcolor,-<-] (g23)--(g22);
        \draw[fcolor,-<-] (g23)--(g12);
\end{tikzpicture} 

Indeed, this is a spin-0 gauge invariant operator with the right charges to be mapped to $\varphi_4\bar{\varphi}_3$, similarly to the monopole listed in Table \ref{table: mesons of SU(2)_1 QCD3}.
This is possible due to the presence of the monopole potential $\mathcal{V}_{monopole}$: even though these two monopoles have different GNO flux assignments they have the same charges under the linear combination of topological symmetries preserved by $\mathcal{V}_{monopole}$.
We expect these two monopoles to mix, therefore the electric meson $\varphi_4\bar{\varphi}_3$ will be mapped to some linear combination of the two:

\begin{equation}
\alpha\begin{pmatrix}
        \tikznode{g11}{0} & & \tikznode{g12}{0} & & \tikznode{g13}{0} & & \tikznode{f1}{}\\ 0 & & \tikznode{g22}{0} & &\tikznode{g23}{+} & &\tikznode{g24}{0}
    \end{pmatrix} + \beta\begin{pmatrix} \tikznode{h11}{0} && \tikznode{h12}{0} && \tikznode {h13}{+} && \tikznode{f2}{{\color{white}+}} \\ 0 && \tikznode{h22}{0} && \tikznode{h23}{0} && \tikznode{h24}{0} \end{pmatrix} \leftrightarrow \varphi_4\bar{\varphi}_3,\;\;\; (|\alpha|^2+|\beta|^2 = 1).
\end{equation}
\begin{tikzpicture}[overlay,remember picture,line width=.7pt,transform canvas={yshift=0mm}]
	    \draw[fcolor,-<-] (g23)--(g24);
        \draw[fcolor,-<-] (g23)--(g22);
        \draw[fcolor,-<-] (g23)--(g12);
        \draw[fcolor,-<-] (h13) -- (f2);
        \draw[fcolor,-<-] (h13) -- (h24); 
        \draw[fcolor,-<-] (h13)-- (h12);
\end{tikzpicture} 
This phenomenon is ubiquitous in the quiver theories studied in this paper and the amount of monopoles that can mix between each other increases as $N$ increases due to the higher number of monopole interactions in $\mathcal{V}_{monopole}$. In Table \ref{table: mesons of SU(2)_1 QCD3} and in the rest of this paper we only report one dressed monopole that has the correct charges to be mapped to a given operator in the QCD theory\footnote{Notice that also on the QCD side there are in general many operators with the same spin and global charges. For example any operator can be multiplied by a spin-0 flavor singlet or it can be acted on by contracted derivatives.}. 
It is understood that, in most cases, there are other monopole operators with the same global charges but different GNO flux assignment and dressing, and the mapping will involve a suitable linear combination of these operators. 
It would be interesting to analyze this mixing systematically, for example by extending the techniques of \cite{Borokhov_2002,Pufu:2013vpa,Chester_2018} in the presence of monopole interactions, but we leave this to future work.

\paragraph{Conserved Currents}
We now examine how the $U(5)$ flavor symmetry currents - including the baryonic $U(1)$ subgroup - map between the electric and dual theories. These spin-1 conserved currents take the form:

\begin{equation}
    J_{jk}^\mu = i \left( \bar{\varphi}_j \partial^\mu \varphi_k - \partial^\mu \bar{\varphi}_j \varphi_k \right)
\end{equation}

where $j,k=1,...,5$, transforming in the adjoint representation of $U(5)$. Note that under charge conjugation: $(J_{jk}^\mu)^* = J_{kj}^\mu$.

The off-diagonal currents map to dressed spin-1 monopole operators in the dual theory. For example, $J_{21}^\mu$ corresponds to:

\begin{equation}
	J_{12}^{\mu}
	\dualto
    \mathfrak{M}^{
    \begin{tikzpicture}
    \matrix (A) [matrix of nodes,left delimiter=(,right delimiter={)}]
	{
		+ &[3mm] 0 &[3mm] 0 &[3mm] \\
		0 & 0 & 0 & \\
	};
	\draw[farrowstyle,->-] (A-1-2) -- (A-1-1);
	\draw[farrowstyle,->-] (A-2-2) -- (A-1-1);
    \end{tikzpicture}	
    } 
\end{equation}
where we dress with the spin-1 mode of one fermion and the spin-0 mode of the other fermion, resulting in a spin-1 operator.
Under charge conjugation, this maps to $J_{21}^\mu$:

\begin{equation}
	J_{21}^{\mu}
	\dualto
    \mathfrak{M}^{
    \begin{tikzpicture}
    \matrix (A) [matrix of nodes,left delimiter=(,right delimiter={)}]
	{
		$-$ &[3mm] 0 &[3mm] 0 &[3mm] \\
		0 & 0 & 0 & \\
	};
	\draw[farrowstyle,-<-] (A-1-2) -- (A-1-1);
	\draw[farrowstyle,-<-] (A-2-2) -- (A-1-1);
    \end{tikzpicture}	
    } 
\end{equation}
where, again, we dress with a spin-0 and a spin-1 mode.

The other off-diagonal spin 1 currents map in a similar way.
In particular, the (off-diagonal) current $J^{\mu}_{jk}$ is mapped to the same bare monopole as the meson $\varphi_j \bar{\varphi}_k$, the difference being that the former is dressed such that it has spin-1 while the latter has spin-0.

The five diagonal currents $J_{jj}^\mu$ map to topological $U(1)$ currents in the dual theory:

\begin{equation}    \label{eq:map_Jmu_SU25phi}
    J_{jj}^\mu \longleftrightarrow \frac{1}{2\pi} \epsilon^{\mu\nu\rho} \partial_\nu A^{(j)}_\rho
\end{equation}

with the gauge field combinations:

\begin{equation}
    A_\mu^{(j)} = 
    \begin{cases}       
     A^{(1,1)}_\mu & j=1 \\
     A^{(2,1)}_\mu + A^{(2,2)}_\mu & j=2 \\
     A^{(3,1)}_\mu + A^{(3,2)}_\mu & j=3 \\
     A^{(4,1)}_\mu & j=4 \\
     A^{(2,1)}_\mu & j=5 \\
    \end{cases}
\end{equation}

Here $A^{(j,k)}_\mu$ denotes the gauge field for the $U(1)$ node at row $j$ and column $k$ in quiver \eqref{fig: SU(2)_1w5s}. 
The mapping \eqref{eq:map_Jmu_SU25phi} reproduces the mapping of the maximal torus of the $U(5)$ global symmetry across the duality.

\paragraph{``Baryonic" Monopole Operators} We follow the same line of reasoning as before to study monopole operators with the right quantum numbers to be mapped to the baryons of the electric theory here. As an example, we consider the following \textit{bare} monopole operator:
\begin{equation}
    \mathfrak{M}^{\left(\:\hspace{-2pt}
        \resizebox{50pt}{!}{%
        \begin{tabular}{cccc} 0&0&0& \\+&+&0&0
        \end{tabular}}\right)}
\end{equation}
This operator has charge $+1$ under the topological symmetries of the bottom two gauge nodes; hence, it is charged under $U(1)_{X_3}+U(1)_{X_1}$, and therefore can be identified with a component of the two-index antisymmetric representation of $SU(5)$. We can calculate the gauge charge of this operator:
\begin{equation}
    \begin{pmatrix}
        0 & 0 & 0 & \\ 1 & -1 & 0 & 0
    \end{pmatrix} + 
     \begin{pmatrix}
        -1 & 0 & 0 & \\ -1 & 3 & -1 & 0
    \end{pmatrix} \equiv
     \begin{pmatrix}
        -1 & 0 & 0 & \\ 0 & 2 & -1 & 0
    \end{pmatrix}.
\end{equation}
Clearly, it is not gauge invariant and must be dressed appropriately. We again choose the minimal dressing that cancels the gauge charges of the bare monopole: 
\begin{equation}
\begin{pmatrix}
        \tikznode{g11}{-} & & \tikznode{g12}{0} & & \tikznode{g13}{0} & & \\ 0 & & \tikznode{g22}{2} & &\tikznode{g23}{-} & &0
    \end{pmatrix},
\end{equation}
\begin{tikzpicture}[overlay,remember picture,line width=.7pt,transform canvas={yshift=0mm}]
	\draw[fcolor,-<-] (g22)--(g11);
        \draw[fcolor,-<-] (g22)--(g23);
\end{tikzpicture}
Therefore the dressed monopole operator:
\begin{equation}
  \mathfrak{M}^{\left(\:\hspace{-2pt}
        \resizebox{50pt}{!}{%
        \begin{tabular}{cccc} 0&0&0& \\+&+&0&0
        \end{tabular}}\right)} \bar{\gamma}_1\bar{\alpha}_{2,2}
\equiv
\begin{pmatrix}
        \tikznode{g11}{0} & & \tikznode{g12}{0} & & \tikznode{g13}{0} & & \\ + & & \tikznode{g22}{+} & &\tikznode{g23}{0} & &0
    \end{pmatrix};
\end{equation}
\begin{tikzpicture}[overlay,remember picture,line width=.7pt,transform canvas={yshift=0mm}]
	\draw[fcolor,-<-] (g22)--(g11);
        \draw[fcolor,-<-] (g22)--(g23);
\end{tikzpicture}
is a gauge invariant Lorentz scalar and has the correct charges to be mapped to $\varphi_3\varphi_2$ (color indices have been suppressed). Similarly, we can identify the 10 baryonic operators, reported with the appropriate dressing in Table \ref{table: baryons of SU(2)_1 QCD3}.

\begin{center}
\renewcommand{\arraystretch}{0.6}
\begin{longtable}{ |>{\centering\arraybackslash}p{3.5cm}|>{\centering\arraybackslash}p{3.5cm}|>{\centering\arraybackslash}p{2cm}|>{\centering\arraybackslash}p{3cm}|  }
\caption{Dressed monopole operators of the planar mirror (Figure \ref{fig: SU(2)_1w5s}) and their putative mapping to baryonic operators of the electric theory (color indices have been suppressed). The remaining anti-baryonic operators can be constructed by charge conjugation.}
\label{table: baryons of SU(2)_1 QCD3} \\ 

\hline \multicolumn{1}{|c|}{\textbf{GNO Flux}} & \multicolumn{1}{c|}{\textbf{Gauge Charge}} & \multicolumn{1}{c|}{\textbf{Spin}} & \multicolumn{1}{c|}{\textbf{``Electric" Baryon}} \\ \hline 
\endfirsthead

\multicolumn{4}{c}%
{{ \textbf{\tablename\ \thetable{}} -- \text{continued from previous page}}} \\
\hline \multicolumn{1}{|c|}{\textbf{GNO Flux }} & \multicolumn{1}{c|}{\textbf{Gauge Charge}} & \multicolumn{1}{c|}{\textbf{Spin}} & \multicolumn{1}{c|}{\textbf{``Electric" Baryon}} \\ \hline 
\endhead

\hline \multicolumn{4}{|r|}{{\textit{Continued on next page}}} \\ \hline
\endfoot

\hline \hline
\endlastfoot
\vspace{-0.5cm}
\begin{equation*}
\begin{pmatrix}
        \tikznode{g11}{0} & & \tikznode{g12}{0} & & \tikznode{g13}{0} & & \tikznode{f1}{}\\ \tikznode{g21}{+} & & \tikznode{g22}{0} & &\tikznode{g23}{0} & &\tikznode{g24}{0}
    \end{pmatrix}
\end{equation*}
\begin{tikzpicture}[overlay,remember picture,line width=.7pt,transform canvas={yshift=0mm}]
	\draw[fcolor,-<-] (g21)--(g22);
\end{tikzpicture}
\vspace{-0.5cm} 
& \vspace{-0.5cm}\begin{equation*}
\begin{pmatrix}
        \tikznode{g11}{0} & & \tikznode{g12}{0} & & \tikznode{g13}{0} & & \tikznode{f1}{}\\ \tikznode{g21}{+} & & \tikznode{g22}{-} & &\tikznode{g23}{0} & &\tikznode{g24}{0}
    \end{pmatrix}
\end{equation*}
\begin{tikzpicture}[overlay,remember picture,line width=.7pt,transform canvas={yshift=0mm}]
	\draw[fcolor,-<-] (g21)--(g22);
\end{tikzpicture}\vspace{-0.5cm}
& \vspace{0.3cm}0 \vspace{-0.5cm}& \vspace{0.3cm}$\varphi_2\varphi_1$ \vspace{-0.5cm}\\ 
\hline

\vspace{-0.5cm}
\begin{equation*}
\begin{pmatrix}
        \tikznode{g11}{0} & & \tikznode{g12}{0} & & \tikznode{g13}{0} & & \\ + & & \tikznode{g22}{+} & &\tikznode{g23}{0} & &0
    \end{pmatrix}
\end{equation*}
\begin{tikzpicture}[overlay,remember picture,line width=.7pt,transform canvas={yshift=0mm}]
	\draw[fcolor,-<-] (g22)--(g11);
        \draw[fcolor,-<-] (g22)--(g23);
\end{tikzpicture} \vspace{-0.5cm}& 
\vspace{-0.5cm}
\begin{equation*}
\begin{pmatrix}
        \tikznode{g11}{-} & & \tikznode{g12}{0} & & \tikznode{g13}{0} & & \\ 0 & & \tikznode{g22}{2} & &\tikznode{g23}{-} & &0
    \end{pmatrix}
\end{equation*}
\begin{tikzpicture}[overlay,remember picture,line width=.7pt,transform canvas={yshift=0mm}]
	\draw[fcolor,-<-] (g22)--(g11);
        \draw[fcolor,-<-] (g22)--(g23);
\end{tikzpicture}\vspace{-0.8cm} &\vspace{0.3cm} 0 & \vspace{0.3cm}$\varphi_3\varphi_1$ \\ \hline
\vspace{-0.5cm}
\begin{equation*}
\begin{pmatrix}
        \tikznode{g11}{0} & & \tikznode{g12}{0} & & \tikznode{g13}{0} & & \tikznode{f1}{}\\ \tikznode{g21}{+} & & \tikznode{g22}{+} & &\tikznode{g23}{+} & &\tikznode{g24}{0}
    \end{pmatrix}
\end{equation*}
\begin{tikzpicture}[overlay,remember picture,line width=.7pt,transform canvas={yshift=0mm}]
	\draw[fcolor,-<-] (g22)--(g11);
        \draw[fcolor,-<-] (g23) -- (g12);
        \draw[fcolor,-<-] (g23)--(g24);
\end{tikzpicture} 
\vspace{-0.5cm}
& \vspace{-0.5cm}\begin{equation*}
\begin{pmatrix}
        \tikznode{g11}{-} & & \tikznode{g12}{-} & & \tikznode{g13}{0} & & \tikznode{f1}{}\\ \tikznode{g21}{0} & & \tikznode{g22}{+} & &\tikznode{g23}{2} & &\tikznode{g24}{-}
    \end{pmatrix}
\end{equation*}
\begin{tikzpicture}[overlay,remember picture,line width=.7pt,transform canvas={yshift=0mm}]
	\draw[fcolor,-<-] (g22)--(g11);
        \draw[fcolor,-<-] (g23) -- (g12);
        \draw[fcolor,-<-] (g23)--(g24);
\end{tikzpicture}
\vspace{-0.8cm}
& \vspace{0.3cm}0 & \vspace{0.3cm}$\varphi_4\varphi_1$ \\ 
\hline
\vspace{-0.5cm}
\begin{equation*}
\begin{pmatrix}
        \tikznode{g11}{0} & & \tikznode{g12}{0} & & \tikznode{g13}{0} & & \tikznode{f1}{}\\ \tikznode{g21}{+} & & \tikznode{g22}{+} & &\tikznode{g23}{+} & &\tikznode{g24}{+}
    \end{pmatrix}
\end{equation*}
\begin{tikzpicture}[overlay,remember picture,line width=.7pt,transform canvas={yshift=0mm}]
	\draw[fcolor,-<-] (g22)--(g11);
        \draw[fcolor,-<-] (g23) -- (g12);
        \draw[fcolor,-<-] (g24)--(g13);
\end{tikzpicture} 
\vspace{-0.5cm}
&\vspace{-0.5cm} \begin{equation*}
\begin{pmatrix}
        \tikznode{g11}{-} & & \tikznode{g12}{-} & & \tikznode{g13}{-} & & \tikznode{f1}{}\\ \tikznode{g21}{0} & & \tikznode{g22}{+} & &\tikznode{g23}{+} & &\tikznode{g24}{+}
    \end{pmatrix}
\end{equation*}
\begin{tikzpicture}[overlay,remember picture,line width=.7pt,transform canvas={yshift=0mm}]
	\draw[fcolor,-<-] (g22)--(g11);
        \draw[fcolor,-<-] (g23) -- (g12);
        \draw[fcolor,-<-] (g24)--(g13);
\end{tikzpicture}
\vspace{-0.5cm}
& \vspace{0.3cm}0 &\vspace{0.3cm} $\varphi_5\varphi_1$ \\ 
\hline
\vspace{-0.5cm}
\begin{equation*}
\begin{pmatrix}
        \tikznode{g11}{+} & & \tikznode{g12}{0} & & \tikznode{g13}{0} & & \tikznode{f1}{}\\ \tikznode{g21}{+} & & \tikznode{g22}{+} & &\tikznode{g23}{0} & &\tikznode{g24}{0}
    \end{pmatrix}
\end{equation*}
\begin{tikzpicture}[overlay,remember picture,line width=.7pt,transform canvas={yshift=0mm}]
	\draw[fcolor,-<-] (g11)--(g12);
        \draw[fcolor,-<-] (g22) -- (g23);
\end{tikzpicture} 
\vspace{-0.5cm}& \vspace{-0.5cm}\begin{equation*}
\begin{pmatrix}
        \tikznode{g11}{+} & & \tikznode{g12}{-} & & \tikznode{g13}{0} & & \tikznode{f1}{}\\ \tikznode{g21}{0} & & \tikznode{g22}{+} & &\tikznode{g23}{-} & &\tikznode{g24}{0}
    \end{pmatrix}
\end{equation*}
\begin{tikzpicture}[overlay,remember picture,line width=.7pt,transform canvas={yshift=0mm}]
	\draw[fcolor,-<-] (g11)--(g12);
        \draw[fcolor,-<-] (g22) -- (g23);
\end{tikzpicture}
\vspace{-0.5cm}
& \vspace{.3cm}0 & \vspace{.3cm}$\varphi_3\varphi_2$\\ 
\hline
\vspace{-0.5cm}
\begin{equation*}
\begin{pmatrix}
        \tikznode{g11}{+} & & \tikznode{g12}{0} & & \tikznode{g13}{0} & & \tikznode{f1}{}\\ \tikznode{g21}{+} & & \tikznode{g22}{+} & &\tikznode{g23}{+} & &\tikznode{g24}{0}
    \end{pmatrix}
\end{equation*}
\begin{tikzpicture}[overlay,remember picture,line width=.7pt,transform canvas={yshift=0mm}]
	\draw[fcolor,-<-] (g23)--(g12);
        \draw[fcolor,-<-] (g11)--(g12);
        \draw[fcolor,-<-] (g23) -- (g24);
\end{tikzpicture} 
\vspace{-0.5cm}& \vspace{-0.5cm}\begin{equation*}
\begin{pmatrix}
        \tikznode{g11}{+} & & \tikznode{g12}{-2} & & \tikznode{g13}{0} & & \tikznode{f1}{}\\ \tikznode{g21}{0} & & \tikznode{g22}{0} & &\tikznode{g23}{2} & &\tikznode{g24}{-}
    \end{pmatrix}
\end{equation*}
\begin{tikzpicture}[overlay,remember picture,line width=.7pt,transform canvas={yshift=0mm}]
	\draw[fcolor,-<-] (g23)--(g12);
        \draw[fcolor,-<-] (g11)--(g12);
        \draw[fcolor,-<-] (g23) -- (g24);
\end{tikzpicture}
\vspace{-0.5cm}
&\vspace{0.3cm} 0 &\vspace{0.3cm} $\varphi_4\varphi_2$ \\ 
\hline
\vspace{-0.5cm}\begin{equation*}
\begin{pmatrix}
        \tikznode{g11}{+} & & \tikznode{g12}{0} & & \tikznode{g13}{0} & & \tikznode{f1}{}\\ \tikznode{g21}{+} & & \tikznode{g22}{+} & &\tikznode{g23}{+} & &\tikznode{g24}{+}
    \end{pmatrix}
\end{equation*}
\begin{tikzpicture}[overlay,remember picture,line width=.7pt,transform canvas={yshift=0mm}]
	\draw[fcolor,-<-] (g23)--(g12);
        \draw[fcolor,-<-] (g11)--(g12);
        \draw[fcolor,-<-] (g24) -- (g13);
\end{tikzpicture}
\vspace{-0.5cm} 
& \vspace{-0.5cm}\begin{equation*}
\begin{pmatrix}
        \tikznode{g11}{+} & & \tikznode{g12}{-2} & & \tikznode{g13}{-} & & \tikznode{f1}{}\\ \tikznode{g21}{0} & & \tikznode{g22}{0} & &\tikznode{g23}{+} & &\tikznode{g24}{+}
    \end{pmatrix}
\end{equation*}
\begin{tikzpicture}[overlay,remember picture,line width=.7pt,transform canvas={yshift=0mm}]
	\draw[fcolor,-<-] (g23)--(g12);
        \draw[fcolor,-<-] (g11)--(g12);
        \draw[fcolor,-<-] (g24) -- (g13);
\end{tikzpicture}
\vspace{-0.5cm}
& \vspace{0.3cm}0 &\vspace{0.3cm} $\varphi_5\varphi_2$ \\ 
\hline
\vspace{-0.5cm}
\begin{equation*}
\begin{pmatrix}
        \tikznode{g11}{+} & & \tikznode{g12}{+} & & \tikznode{g13}{0} & & \tikznode{f1}{}\\ \tikznode{g21}{+} & & \tikznode{g22}{+} & &\tikznode{g23}{+} & &\tikznode{g24}{+}
    \end{pmatrix}
\end{equation*}
\begin{tikzpicture}[overlay,remember picture,line width=.7pt,transform canvas={yshift=0mm}]
	\draw[fcolor,-<-] (g24)--(g13);
        \draw[fcolor,-<-] (g12)--(g13);
\end{tikzpicture} 
\vspace{-0.5cm}
& 
\vspace{-0.5cm}\begin{equation*}
\begin{pmatrix}
        \tikznode{g11}{0} & & \tikznode{g12}{+} & & \tikznode{g13}{-2} & & \tikznode{f1}{}\\ \tikznode{g21}{0} & & \tikznode{g22}{0} & &\tikznode{g23}{0} & &\tikznode{g24}{+}
    \end{pmatrix}
\end{equation*}
\begin{tikzpicture}[overlay,remember picture,line width=.7pt,transform canvas={yshift=0mm}]
	\draw[fcolor,-<-] (g24)--(g13);
        \draw[fcolor,-<-] (g12)--(g13);
\end{tikzpicture}
\vspace{-0.5cm}
&
\vspace{0.3cm} 0 &\vspace{0.3cm} $\varphi_5\varphi_3$ \\ 
\hline
\vspace{-0.5cm}
\begin{equation*}
\begin{pmatrix}
        \tikznode{g11}{+} & & \tikznode{g12}{+} & & \tikznode{g13}{+} & & \tikznode{f1}{{\color{white}+}}\\ \tikznode{g21}{+} & & \tikznode{g22}{+} & &\tikznode{g23}{+} & &\tikznode{g24}{+}
    \end{pmatrix}
\end{equation*}
\begin{tikzpicture}[overlay,remember picture,line width=.7pt,transform canvas={yshift=0mm}]
	 \draw[fcolor,-<-] (g13)--(f1);
\end{tikzpicture} 
\vspace{-0.5cm}
& 
\vspace{-0.5cm}
\begin{equation*}
\begin{pmatrix}
        \tikznode{g11}{0} & & \tikznode{g12}{0} & & \tikznode{g13}{+} & & \tikznode{f1}{{\color{white}+}}\\ \tikznode{g21}{0} & & \tikznode{g22}{0} & &\tikznode{g23}{0} & &\tikznode{g24}{0}
    \end{pmatrix}
\end{equation*}
\begin{tikzpicture}[overlay,remember picture,line width=.7pt,transform canvas={yshift=0mm}]
        \draw[fcolor,-<-] (g13)--(f1);
\end{tikzpicture}
\vspace{-0.5cm}
& 
\vspace{0.3cm}0 
& 
\vspace{0.3cm}
$\varphi_5\varphi_4$ 
\end{longtable}
\end{center}

We reiterate that we only consider dressed minimal flux operators in the mapping. 
\subsection{Example: \texorpdfstring{$SU(2)_{-2}$ QCD$_3$ with $N_s=5$}{su22} scalars}
\begin{figure}[ht]
   \centerline{\includegraphics[width=.9\textwidth]{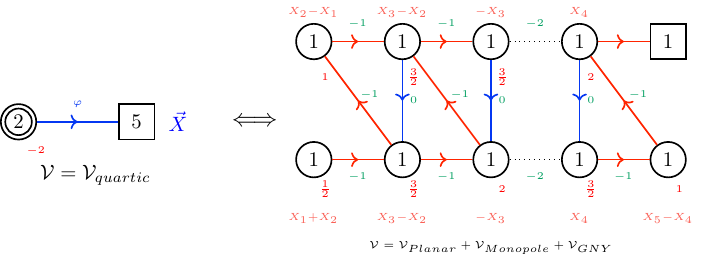}}
    \caption{The dual for $SU(2)_{-2}$ QCD$_3$ with 5 scalars is shown here. All labelling conventions follow those of Figure \ref{fig:scalar_Qcd_general}.}
    \label{fig: SU(2)_2w5s}
\end{figure}

We now consider the example of $SU(2)_{-2}$ with 5 scalars. Notice that it has a lower Chern-Simons level than that of the theory considered in Section \ref{sec: SU(2)_1w5s}; this minor difference will greatly alter the operator map. We propose the duality shown in Figure \ref{fig: SU(2)_2w5s}.

As before, we are mainly concerned with determining the monopole operators that are dual to the 24 \textbf{mesons} and 10 \textbf{baryons}. We do not show the explicit calculation and state the results in Tables \ref{table: mesons of SU(2)_2 QCD3} and \ref{table: baryons of SU(2)_2 QCD3} respectively.

\begin{center}
\renewcommand{\arraystretch}{0.6}
\begin{longtable}{ |>{\centering\arraybackslash}p{4cm}|>{\centering\arraybackslash}p{1.5cm}|>{\centering\arraybackslash}p{4cm}|>{\centering\arraybackslash}p{1.5cm}|  }
\caption{Dressed monopole operators of the planar mirror (Figure \ref{fig: SU(2)_2w5s}) and their putative mapping to mesonic operators of the electric theory. The remaining mesonic operators can be constructed by charge conjugation.}
\label{table: mesons of SU(2)_2 QCD3} \\ 

\hline \multicolumn{1}{|c|}{\textbf{GNO Flux}} & \multicolumn{1}{c|}{\textbf{Meson}} & \multicolumn{1}{c|}{\textbf{GNO Flux}} & \multicolumn{1}{c|}{\textbf{Meson}} \\ \hline 
\endfirsthead

\multicolumn{4}{c}%
{{ \textbf{\tablename\ \thetable{}} -- \text{continued from previous page}}} \\
\hline \multicolumn{1}{|c|}{\textbf{GNO Flux }} & \multicolumn{1}{c|}{\textbf{Meson}} & \multicolumn{1}{c|}{\textbf{GNO Flux}} & \multicolumn{1}{c|}{\textbf{Meson}} \\ \hline 
\endhead

\hline \multicolumn{4}{|r|}{{\textit{Continued on next page}}} \\ \hline
\endfoot

\hline \hline
\endlastfoot
\vspace{-0.5cm}
\begin{equation*}
\begin{pmatrix}
        \tikznode{g11}{+} & & \tikznode{g12}{0} & & 0 & & 0 &&\\ 0 & & \tikznode{g21}{0} & &0 & &0 && 0
    \end{pmatrix}
\end{equation*}
\begin{tikzpicture}[overlay,remember picture,line width=.7pt,transform canvas={yshift=0mm}]
	\draw[fcolor,-<-] (g11)--(g12);
        \draw[fcolor,-<-] (g11)--(g21);
\end{tikzpicture} 
\vspace{-0.5cm}
& \vspace{0.3cm}$\varphi_2\bar{\varphi}_1$ 
& 
\vspace{-0.5cm}
\begin{equation*}
\begin{pmatrix}
        \tikznode{g11}{+} & & \tikznode{g12}{+} & & \tikznode{g13}{0} & & \tikznode{g14}{0} && \tikznode{f1}{}\\ \tikznode{g21}{0} & & \tikznode{g22}{0} & & \tikznode{g23}{0} & & \tikznode{g24}{0} && \tikznode{g25}{0}
    \end{pmatrix}
\end{equation*}
\begin{tikzpicture}[overlay,remember picture,line width=.7pt,transform canvas={yshift=0mm}]
	\draw[fcolor,-<-] (g11)--(g22);
        \draw[fcolor,-<-] (g12)--(g23);
        \draw[fcolor,-<-] (g12)--(g13);
\end{tikzpicture} 
\vspace{-0.5cm}
 & \vspace{0.3cm} $\varphi_3\bar{\varphi}_1$ \\ \hline

\vspace{-0.5cm}
\begin{equation*}
\begin{pmatrix}
        \tikznode{g11}{+} & & \tikznode{g12}{+} & & \tikznode{g13}{+} & & \tikznode{g14}{+} && \tikznode{f1}{{\color{white}+}}\\ \tikznode{g21}{0} & & \tikznode{g22}{0} & & \tikznode{g23}{0} & & \tikznode{g24}{0} && \tikznode{g25}{0}
    \end{pmatrix}
\end{equation*}
\begin{tikzpicture}[overlay,remember picture,line width=.7pt,transform canvas={yshift=0mm}]
	\draw[fcolor,-<-] (g11)--(g22);
        \draw[fcolor,-<-] (g12)--(g23);
        \draw[fcolor,-<-] (g14)--(f1);
        \draw[fcolor,-<-] (g14)--(g25);
\end{tikzpicture} 
\vspace{-0.5cm}
 & \vspace{0.3cm} $\varphi_4\bar{\varphi}_1$ &
\vspace{-0.5cm}
\begin{equation*}
\begin{pmatrix}
        \tikznode{g11}{+} & & \tikznode{g12}{+} & & \tikznode{g13}{+} & & \tikznode{g14}{+} && \tikznode{f1}{{\color{white}+}}\\ \tikznode{g21}{0} & & \tikznode{g22}{0} & & \tikznode{g23}{0} & & \tikznode{g24}{0} && \tikznode{g25}{+}
    \end{pmatrix}
\end{equation*}
\begin{tikzpicture}[overlay,remember picture,line width=.7pt,transform canvas={yshift=0mm}]
	\draw[fcolor,-<-] (g11)--(g22);
        \draw[fcolor,-<-] (g12)--(g23);
        \draw[fcolor,-<-] (g14)--(f1);
        \draw[fcolor,-<-] (g25)--(g24);
\end{tikzpicture} 
\vspace{-0.5cm}
 &\vspace{0.3cm} $\varphi_5\bar{\varphi}_1$  \\ \hline
\vspace{-0.5cm}
 \begin{equation*}
\begin{pmatrix}
        \tikznode{g11}{0} & & \tikznode{g12}{+} & & \tikznode{g13}{0} & & \tikznode{g14}{0} && \tikznode{f1}{{\color{white}+}}\\ \tikznode{g21}{0} & & \tikznode{g22}{0} & & \tikznode{g23}{0} & & \tikznode{g24}{0} && \tikznode{g25}{0}
    \end{pmatrix}
\end{equation*}
\begin{tikzpicture}[overlay,remember picture,line width=.7pt,transform canvas={yshift=0mm}]
	\draw[fcolor,-<-] (g12)--(g11);
        \draw[fcolor,-<-] (g12)--(g13);
        \draw[fcolor,-<-] (g12)--(g23);
\end{tikzpicture} 
\vspace{-0.5cm}
 & \vspace{0.3cm}$\varphi_3\bar{\varphi}_2$ & \vspace{-0.5cm}
\begin{equation*}
\begin{pmatrix}
        \tikznode{g11}{0} & & \tikznode{g12}{+} & & \tikznode{g13}{+} & & \tikznode{g14}{+} && \tikznode{f1}{{\color{white}+}}\\ \tikznode{g21}{0} & & \tikznode{g22}{0} & & \tikznode{g23}{0} & & \tikznode{g24}{0} && \tikznode{g25}{0}
    \end{pmatrix}
\end{equation*}
\begin{tikzpicture}[overlay,remember picture,line width=.7pt,transform canvas={yshift=0mm}]
	\draw[fcolor,-<-] (g12)--(g11);
        \draw[fcolor,-<-] (g12)--(g23);
        \draw[fcolor,-<-] (g14)--(f1);
        \draw[fcolor,-<-] (g14)--(g25);
\end{tikzpicture} 
\vspace{-0.5cm}
 & \vspace{0.3cm}$\varphi_4\bar{\varphi}_2$ \\ \hline
\vspace{-0.5cm}
 \begin{equation*}
\begin{pmatrix}
        \tikznode{g11}{0} & & \tikznode{g12}{+} & & \tikznode{g13}{+} & & \tikznode{g14}{+} && \tikznode{f1}{{\color{white}+}}\\ \tikznode{g21}{0} & & \tikznode{g22}{0} & & \tikznode{g23}{0} & & \tikznode{g24}{0} && \tikznode{g25}{+}
    \end{pmatrix}
\end{equation*}
\begin{tikzpicture}[overlay,remember picture,line width=.7pt,transform canvas={yshift=0mm}]
	\draw[fcolor,-<-] (g12)--(g11);
        \draw[fcolor,-<-] (g12)--(g23);
        \draw[fcolor,-<-] (g14)--(f1);
        \draw[fcolor,-<-] (g25)--(g24);
\end{tikzpicture} 
\vspace{-0.5cm}
 & \vspace{0.3cm} $\varphi_5\bar{\varphi}_2$ & \vspace{-0.5cm}
 \begin{equation*}
\begin{pmatrix}
        \tikznode{g11}{0} & & \tikznode{g12}{0} & & \tikznode{g13}{+} & & \tikznode{g14}{+} && \tikznode{f1}{{\color{white}+}}\\ \tikznode{g21}{0} & & \tikznode{g22}{0} & & \tikznode{g23}{0} & & \tikznode{g24}{0} && \tikznode{g25}{0}
    \end{pmatrix}
\end{equation*}
\begin{tikzpicture}[overlay,remember picture,line width=.7pt,transform canvas={yshift=0mm}]
	\draw[fcolor,-<-] (g13)--(g12);
        \draw[fcolor,-<-] (g14)--(f1);
        \draw[fcolor,-<-] (g14)--(g25);
\end{tikzpicture} 
\vspace{-0.5cm}
 & \vspace{0.3cm}$\varphi_4\bar{\varphi}_3$ \\ \hline
\vspace{-0.5cm}
 \begin{equation*}
\begin{pmatrix}
        \tikznode{g11}{0} & & \tikznode{g12}{0} & & \tikznode{g13}{+} & & \tikznode{g14}{+} && \tikznode{f1}{{\color{white}+}}\\ \tikznode{g21}{0} & & \tikznode{g22}{0} & & \tikznode{g23}{0} & & \tikznode{g24}{0} && \tikznode{g25}{+}
    \end{pmatrix}
\end{equation*}
\begin{tikzpicture}[overlay,remember picture,line width=.7pt,transform canvas={yshift=0mm}]
	\draw[fcolor,-<-] (g13)--(g12);
        \draw[fcolor,-<-] (g14)--(f1);
        \draw[fcolor,-<-] (g25)--(g24);
\end{tikzpicture} 
\vspace{-0.5cm}
 &\vspace{0.3cm} $\varphi_5\bar{\varphi}_3$ &
\vspace{-0.5cm}
\begin{equation*}
\begin{pmatrix}
        \tikznode{g11}{0} & & \tikznode{g12}{0} & & \tikznode{g13}{0} & & \tikznode{g14}{0} && \tikznode{f1}{{\color{white}+}}\\ \tikznode{g21}{0} & & \tikznode{g22}{0} & & \tikznode{g23}{0} & & \tikznode{g24}{0} && \tikznode{g25}{+}
    \end{pmatrix}
\end{equation*}
\begin{tikzpicture}[overlay,remember picture,line width=.7pt,transform canvas={yshift=0mm}]
        \draw[fcolor,-<-] (g25)--(g14);
        \draw[fcolor,-<-] (g25)--(g24);
\end{tikzpicture} 
\vspace{-0.5cm}
 & \vspace{0.3cm}$\varphi_5\bar{\varphi}_4$ \\ \hline
\end{longtable}
\end{center}

\begin{center}
\renewcommand{\arraystretch}{0.6}
\begin{longtable}{ |>{\centering\arraybackslash}p{4cm}|>{\centering\arraybackslash}p{1.5cm}|>{\centering\arraybackslash}p{4cm}|>{\centering\arraybackslash}p{1.5cm}|  }
\caption{Dressed monopole operators of the planar mirror (Figure \ref{fig: SU(2)_2w5s}) and their putative mapping to baryonic operators of the electric theory (color indices have been suppressed). The remaining anti-baryonic operators can be constructed by charge conjugation.}
\label{table: baryons of SU(2)_2 QCD3} \\

\hline \multicolumn{1}{|c|}{\textbf{GNO Flux}} & \multicolumn{1}{c|}{\textbf{Baryon}} & \multicolumn{1}{c|}{\textbf{GNO Flux}} & \multicolumn{1}{c|}{\textbf{Baryon}} \\ \hline 
\endfirsthead

\multicolumn{4}{c}%
{{ \textbf{\tablename\ \thetable{}} -- \text{continued from previous page}}} \\
\hline \multicolumn{1}{|c|}{\textbf{GNO Flux }} & \multicolumn{1}{c|}{\textbf{Baryon}} & \multicolumn{1}{c|}{\textbf{GNO Flux}} & \multicolumn{1}{c|}{\textbf{Baryon}} \\ \hline 
\endhead

\hline \multicolumn{4}{|r|}{{\textit{Continued on next page}}} \\ \hline
\endfoot

\hline \hline
\endlastfoot
\vspace{-0.5cm}
\begin{equation*}
\begin{pmatrix}
        \tikznode{g11}{0} & & \tikznode{g12}{0} & & \tikznode{g13}{0} & & \tikznode{g14}{0} && \tikznode{f1}{{\color{white}+}}\\ \tikznode{g21}{+} & & \tikznode{g22}{0} & & \tikznode{g23}{0} & & \tikznode{g24}{0} && \tikznode{g25}{0}
    \end{pmatrix}
\end{equation*}
\begin{tikzpicture}[overlay,remember picture,line width=.7pt,transform canvas={yshift=0mm}]
        \draw[fcolor,-<-] (g21)--(g22);
\end{tikzpicture} 
\vspace{-0.5cm}
 & \vspace{0.3cm}$\varphi_2\varphi_1$ &
\vspace{-0.5cm}
\begin{equation*}
\begin{pmatrix}
        \tikznode{g11}{0} & & \tikznode{g12}{0} & & \tikznode{g13}{0} & & \tikznode{g14}{0} && \tikznode{f1}{}\\ \tikznode{g21}{+} & & \tikznode{g22}{+} & & \tikznode{g23}{0} & & \tikznode{g24}{0} && \tikznode{g25}{0}
    \end{pmatrix}
\end{equation*}
\begin{tikzpicture}[overlay,remember picture,line width=.7pt,transform canvas={yshift=0mm}]
        \draw[fcolor,-<-] (g22)--(g23); \draw[fcolor,-<-] (g22)--(g11);
\end{tikzpicture} 
\vspace{-0.5cm}
 &\vspace{0.3cm} $\varphi_3\varphi_1$ \\ 
\hline
\vspace{-0.5cm}
\begin{equation*}
\begin{pmatrix}
        \tikznode{g11}{0} & & \tikznode{g12}{0} & & \tikznode{g13}{0} & & \tikznode{g14}{0} && \tikznode{f1}{}\\ \tikznode{g21}{+} & & \tikznode{g22}{+} & & \tikznode{g23}{+} & & \tikznode{g24}{+} && \tikznode{g25}{0}
    \end{pmatrix}
\end{equation*}
\begin{tikzpicture}[overlay,remember picture,line width=.7pt,transform canvas={yshift=0mm}]
        \draw[fcolor,-<-] (g23)--(g12); \draw[fcolor,-<-] (g22)--(g11); \draw[fcolor,-<-] (g24)--(g25);
\end{tikzpicture} 
\vspace{-0.5cm}
 & \vspace{0.3cm}$\varphi_4\varphi_1$ &
\vspace{-0.5cm}
\begin{equation*}
\begin{pmatrix}
        \tikznode{g11}{0} & & \tikznode{g12}{0} & & \tikznode{g13}{0} & & \tikznode{g14}{0} && \tikznode{f1}{}\\ \tikznode{g21}{+} & & \tikznode{g22}{+} & & \tikznode{g23}{+} & & \tikznode{g24}{+} && \tikznode{g25}{+}
    \end{pmatrix}
\end{equation*}
\begin{tikzpicture}[overlay,remember picture,line width=.7pt,transform canvas={yshift=0mm}]
        \draw[fcolor,-<-] (g22)--(g11); \draw[fcolor,-<-] (g23)--(g12); \draw[fcolor,-<-] (g25)--(g14);
\end{tikzpicture} 
\vspace{-0.5cm}
 & \vspace{.3cm}$\varphi_5\varphi_1$ \\ 
\hline
\vspace{-0.5cm}
\begin{equation*}
\begin{pmatrix}
        \tikznode{g11}{+} & & \tikznode{g12}{0} & & \tikznode{g13}{0} & & \tikznode{g14}{0} && \tikznode{f1}{}\\ \tikznode{g21}{+} & & \tikznode{g22}{+} & & \tikznode{g23}{0} & & \tikznode{g24}{0} && \tikznode{g25}{0}
    \end{pmatrix}
\end{equation*}
\begin{tikzpicture}[overlay,remember picture,line width=.7pt,transform canvas={yshift=0mm}]
        \draw[fcolor,-<-] (g11)--(g12); \draw[fcolor,-<-] (g22)--(g23);
\end{tikzpicture} 
\vspace{-0.5cm}
 &\vspace{.3cm} $\varphi_3\varphi_2$ &
\vspace{-0.5cm}
\begin{equation*}
\begin{pmatrix}
        \tikznode{g11}{+} & & \tikznode{g12}{0} & & \tikznode{g13}{0} & & \tikznode{g14}{0} && \tikznode{f1}{}\\ \tikznode{g21}{+} & & \tikznode{g22}{+} & & \tikznode{g23}{+} & & \tikznode{g24}{+} && \tikznode{g25}{0}
    \end{pmatrix}
\end{equation*}
\begin{tikzpicture}[overlay,remember picture,line width=.7pt,transform canvas={yshift=0mm}]
        \draw[fcolor,-<-] (g11)--(g12); \draw[fcolor,-<-] (g23)--(g12); \draw[fcolor,-<-] (g24)--(g25);
\end{tikzpicture} 
\vspace{-0.5cm}
 & \vspace{0.3cm}$\varphi_4\varphi_2$ \\ 
\hline
\vspace{-0.5cm}
\begin{equation*}
\begin{pmatrix}
        \tikznode{g11}{+} & & \tikznode{g12}{0} & & \tikznode{g13}{0} & & \tikznode{g14}{0} && \tikznode{f1}{}\\ \tikznode{g21}{+} & & \tikznode{g22}{+} & & \tikznode{g23}{+} & & \tikznode{g24}{+} && \tikznode{g25}{+}
    \end{pmatrix}
\end{equation*}
\begin{tikzpicture}[overlay,remember picture,line width=.7pt,transform canvas={yshift=0mm}]
        \draw[fcolor,-<-] (g11)--(g12); \draw[fcolor,-<-] (g23)--(g12); \draw[fcolor,-<-] (g25)--(g14);
\end{tikzpicture} 
\vspace{-0.5cm}
 & \vspace{0.3cm}$\varphi_5\varphi_2$ &
\vspace{-0.5cm}
\begin{equation*}
\begin{pmatrix}
        \tikznode{g11}{+} & & \tikznode{g12}{+} & & \tikznode{g13}{0} & & \tikznode{g14}{0} && \tikznode{f1}{}\\ \tikznode{g21}{+} & & \tikznode{g22}{+} & & \tikznode{g23}{+} & & \tikznode{g24}{+} && \tikznode{g25}{0}
    \end{pmatrix}
\end{equation*}
\begin{tikzpicture}[overlay,remember picture,line width=.7pt,transform canvas={yshift=0mm}]
        \draw[fcolor,-<-] (g12)--(g13); \draw[fcolor,-<-] (g24)--(g25);
\end{tikzpicture} 
\vspace{-0.5cm}
 & \vspace{0.3cm}$\varphi_4\varphi_3$ \\ 
\hline
\vspace{-0.5cm}
\begin{equation*}
\begin{pmatrix}
        \tikznode{g11}{+} & & \tikznode{g12}{+} & & \tikznode{g13}{0} & & \tikznode{g14}{0} && \tikznode{f1}{}\\ \tikznode{g21}{+} & & \tikznode{g22}{+} & & \tikznode{g23}{+} & & \tikznode{g24}{+} && \tikznode{g25}{+}
    \end{pmatrix}
\end{equation*}
\begin{tikzpicture}[overlay,remember picture,line width=.7pt,transform canvas={yshift=0mm}]
        \draw[fcolor,-<-] (g12)--(g13); \draw[fcolor,-<-] (g25)--(g14);
\end{tikzpicture} 
\vspace{-0.5cm}
 &\vspace{0.3cm} $\varphi_5\varphi_3$ &
 \vspace{-0.5cm}
\begin{equation*}
\begin{pmatrix}
        \tikznode{g11}{+} & & \tikznode{g12}{+} & & \tikznode{g13}{+} & & \tikznode{g14}{+} && \tikznode{f1}{{\color{white}+}}\\ \tikznode{g21}{+} & & \tikznode{g22}{+} & & \tikznode{g23}{+} & & \tikznode{g24}{+} && \tikznode{g25}{+}
    \end{pmatrix}
\end{equation*}
\begin{tikzpicture}[overlay,remember picture,line width=.7pt,transform canvas={yshift=0mm}]
        \draw[fcolor,-<-] (g14)--(f1);
\end{tikzpicture} 
\vspace{-0.5cm} 
 &\vspace{0.3cm} $\varphi_5\varphi_4$ \\ 
\hline
\end{longtable}
\end{center}


\subsection{Example: \texorpdfstring{$SU(3)_{-1}$ QCD$_3$ with $N_s=7$}{su3} scalars}
\begin{figure}[ht]
    \centerline{\includegraphics[width=1\textwidth]{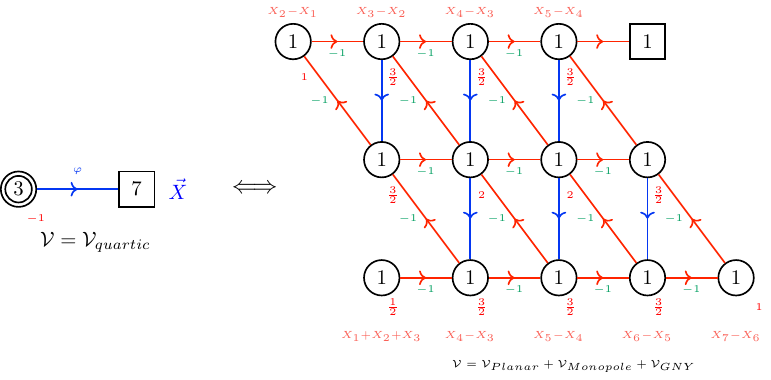}}
    \caption{The dual for $SU(3)_{-1}$ QCD$_3$ with 7 scalars is shown here. All labelling conventions follow those of Figure \ref{fig:scalar_Qcd_general}. Mixed BF couplings between gauge nodes connected by vertical scalar fields have been suppressed for brevity.}
    \label{fig: SU(3)_1w7s}
\end{figure}

We consider the case of $SU(3)_{-1}$ with $7$ scalars and its planar dual as our final example of this section. We propose the duality shown in Figure \ref{fig: SU(3)_1w7s}.

We are concerned with the mapping of 48 \textbf{mesonic operators} (identified by being charged under $U(1)_{X_i}-U(1)_{X_j}$) and $\binom{7}{3}=35$ \textbf{baryonic operators} (identified by being charged under $U(1)_{X_i}+U(1)_{X_j}+U(1)_{X_k}$), all of which are Lorentz scalars. Since the reasoning we follow is the same as before, we leave the complete operator map to the reader and only provide the GNO fluxes of the bare monopoles as a representative example of the general trends of construction and identification of these operators.

We report the monopoles mapped to mesons of the electric theory in Equation \eqref{table: mesons of SU(3)_1 QCD3}. Note that we only specify the off-diagonal elements of the meson matrix in Equation \eqref{table: mesons of SU(3)_1 QCD3} and the remaining monopoles can be constructed by charge conjugation).
\begin{equation} \label{table: mesons of SU(3)_1 QCD3}
    \begin{aligned}
        \mathbf{M} \leftrightarrow &\; \bigg\{ \mathfrak{M}^{\begin{pmatrix} + & 0 & 0 & 0 & \\ & 0 & 0 &0 &0 \\ & 0 &0 &0 &0 & 0\end{pmatrix}}, \, \mathfrak{M}^{\begin{pmatrix} + & + & 0 & 0 & \\ & 0 & 0 &0 &0 \\ & 0 &0 &0 &0 & 0\end{pmatrix}},\, \mathfrak{M}^{\begin{pmatrix} + & + & + & 0 & \\ & 0 & 0 &0 &0 \\ & 0 &0 &0 &0 & 0\end{pmatrix}},\,\mathfrak{M}^{\begin{pmatrix} + & + & + & + & \\ & 0 & 0 &0 &0 \\ & 0 &0 &0 &0 & 0\end{pmatrix}},\\ & \mathfrak{M}^{\begin{pmatrix} + & + & + & + & \\ & 0 & 0 &0 &+ \\ & 0 &0 &0 &0 & 0\end{pmatrix}},\, \mathfrak{M}^{\begin{pmatrix} + & + & + & + & \\ & 0 & 0 &0 &+ \\ & 0 &0 &0 &0 & +\end{pmatrix}},\, \mathfrak{M}^{\begin{pmatrix} 0 & + & 0 & 0 & \\ & 0 & 0 &0 &0 \\ & 0 &0 &0 &0 & 0\end{pmatrix}},\, \mathfrak{M}^{\begin{pmatrix} 0 & + & + & 0 & \\ & 0 & 0 &0 &0 \\ & 0 &0 &0 &0 & 0\end{pmatrix}},\\ &\mathfrak{M}^{\begin{pmatrix} 0 & + & + & + & \\ & 0 & 0 &0 &0 \\ & 0 &0 &0 &0 & 0\end{pmatrix}},\, \mathfrak{M}^{\begin{pmatrix} 0 & + & + & + & \\ & 0 & 0 &0 &+ \\ & 0 &0 &0 &0 & 0\end{pmatrix}},\, \mathfrak{M}^{\begin{pmatrix} 0 & + & + & + & \\ & 0 & 0 &0 &+ \\ & 0 &0 &0 &0 & +\end{pmatrix}},\, \mathfrak{M}^{\begin{pmatrix} 0 & 0 & + & 0 & \\ & 0 & 0 &0 &0 \\ & 0 &0 &0 &0 & 0\end{pmatrix}},\\ & \mathfrak{M}^{\begin{pmatrix} 0 & 0 & + & + & \\ & 0 & 0 &0 &0 \\ & 0 &0 &0 &0 & 0\end{pmatrix}},\,  \mathfrak{M}^{\begin{pmatrix} 0 & 0 & + & + & \\ & 0 & 0 &0 &+ \\ & 0 &0 &0 &0 & 0\end{pmatrix}},\, \mathfrak{M}^{\begin{pmatrix} 0 & 0 & + & + & \\ & 0 & 0 &0 &+ \\ & 0 &0 &0 &0 & +\end{pmatrix}},\, \mathfrak{M}^{\begin{pmatrix} 0 & 0 & 0 & + & \\ & 0 & 0 &0 &0 \\ & 0 &0 &0 &0 & 0\end{pmatrix}},\\ & 
        \mathfrak{M}^{\begin{pmatrix} 0 & 0 & 0 & + & \\ & 0 & 0 &0 &+ \\ & 0 &0 &0 &0 & 0\end{pmatrix}},\, \mathfrak{M}^{\begin{pmatrix} 0 & 0 & 0 & + & \\ & 0 & 0 &0 &+ \\ & 0 &0 &0 &0 &+\end{pmatrix}},\, \mathfrak{M}^{\begin{pmatrix} 0 & 0 & 0 & 0 & \\ & 0 & 0 &0 &+ \\ & 0 &0 &0 &0 & 0\end{pmatrix}},\, \mathfrak{M}^{\begin{pmatrix} 0 & 0 & 0 & 0 & \\ & 0 & 0 &0 &+ \\ & 0 &0 &0 &0 & +\end{pmatrix}},\\ & 
        \mathfrak{M}^{\begin{pmatrix} 0 & 0 & 0 & 0 & \\ & 0 & 0 &0 &0 \\ & 0 &0 &0 &0 & +\end{pmatrix}}\bigg\}
    \end{aligned}
\end{equation}
In order for these monopole to be gauge invariant they need to be appropriately dressed as explained in Appendix \ref{app:monopoles}.
The fermions used to dress the monopoles, in the chosen monopole background, have a spin-0 mode. Using this mode we dress the monopoles to obtain spin-0 operators that therefore have the correct spin to map to mesons of the QCD$_3$.

We report the monopoles mapped to baryons of the electric theory in Equation \eqref{table: baryons of SU(3)_1 QCD3}. Note that monopoles mapped to anti-baryons can be constructed by charge conjugation.
\begin{equation}\label{table: baryons of SU(3)_1 QCD3}
    \begin{aligned}
        \mathbf{B} \leftrightarrow &\; \bigg\{ \mathfrak{M}^{\begin{pmatrix} 0 & 0 & 0 & 0 & \\ & 0 & 0 &0 &0 \\ & + &0 &0 &0 & 0\end{pmatrix}},\, \mathfrak{M}^{\begin{pmatrix} 0 & 0 & 0 & 0 & \\ & 0 & 0 &0 &0 \\ & + &+ &0 &0 & 0\end{pmatrix}},\, \mathfrak{M}^{\begin{pmatrix} 0 & 0 & 0 & 0 & \\ & + & 0 &0 &0 \\ & + &+ &0 &0 & 0\end{pmatrix}},\, \mathfrak{M}^{\begin{pmatrix} + & 0 & 0 & 0 & \\ & + & 0 &0 &0 \\ & + &+ &0 &0 & 0\end{pmatrix}},\\ & 
        \mathfrak{M}^{\begin{pmatrix} 0 & 0 & 0 & 0 & \\ & 0 & 0 &0 &0 \\ & + &+ &+ &0 & 0\end{pmatrix}},\,  \mathfrak{M}^{\begin{pmatrix} 0 & 0 & 0 & 0 & \\ & + & 0 &0 &0 \\ & + &+ &+ &0 & 0\end{pmatrix}},\, \mathfrak{M}^{\begin{pmatrix} + & 0 & 0 & 0 & \\ & + & 0 &0 &0 \\ & + &+ &+ &0 & 0\end{pmatrix}},\, \mathfrak{M}^{\begin{pmatrix} 0 & 0 & 0 & 0 & \\ & + & + &0 &0 \\ & + &+ &+ &0 & 0\end{pmatrix}},\\&
        \mathfrak{M}^{\begin{pmatrix} + & 0 & 0 & 0 & \\ & + & + &0 &0 \\ & + &+ &+ &0 & 0\end{pmatrix}},\, \mathfrak{M}^{\begin{pmatrix} + & + & 0 & 0 & \\ & + & + &0 &0 \\ & + &+ &+ &0 & 0\end{pmatrix}},\, \mathfrak{M}^{\begin{pmatrix} 0 & 0 & 0 & 0 & \\ & 0 & 0 &0 &0 \\ & + &+ &+ &+ & 0\end{pmatrix}},\,
        \mathfrak{M}^{\begin{pmatrix} 0 & 0 & 0 & 0 & \\ & + & 0 &0 &0 \\ & + &+ &+ &+ & 0\end{pmatrix}},\,\\ &
        \mathfrak{M}^{\begin{pmatrix} + & 0 & 0 & 0 & \\ & + & 0 &0 &0 \\ & + &+ &+ &+ & 0\end{pmatrix}},
        \mathfrak{M}^{\begin{pmatrix} 0 & 0 & 0 & 0 & \\ & + & + &0 &0 \\ & + &+ &+ &+ & 0\end{pmatrix}},\, 
        \mathfrak{M}^{\begin{pmatrix} + & 0 & 0 & 0 & \\ & + & + &0 &0 \\ & + &+ &+ &+ & 0\end{pmatrix}},\,  \mathfrak{M}^{\begin{pmatrix} + & + & 0 & 0 & \\ & + & + &0 &0 \\ & + &+ &+ &+ & 0\end{pmatrix}},\\ &
         \mathfrak{M}^{\begin{pmatrix} 0 & 0 & 0 & 0 & \\ & + & + &+ &0 \\ & + &+ &+ &+ & 0\end{pmatrix}},\, 
         \mathfrak{M}^{\begin{pmatrix} + & 0 & 0 & 0 & \\ & + & + &+ &0 \\ & + &+ &+ &+ & 0\end{pmatrix}},\,  \mathfrak{M}^{\begin{pmatrix} + & +& 0 & 0 & \\ & + & + &+ &0 \\ & + &+ &+ &+ & 0\end{pmatrix}},\, 
         \mathfrak{M}^{\begin{pmatrix} + & + & + & 0 & \\ & + & + &+ &0 \\ & + &+ &+ &+ & 0\end{pmatrix}}, \\ & 
          \mathfrak{M}^{\begin{pmatrix} 0 & 0 & 0 & 0 & \\ & 0 & 0 &0 &0 \\ & + &+ &+ &+ & +\end{pmatrix}},\, 
           \mathfrak{M}^{\begin{pmatrix} 0 & 0 & 0 & 0 & \\ & + & 0 &0 &0 \\ & + &+ &+ &+ & +\end{pmatrix}},\, 
        \mathfrak{M}^{\begin{pmatrix} + & 0 & 0 & 0 & \\ & + & 0 &0 &0 \\ & + &+ &+ &+ & +\end{pmatrix}},\,
        \mathfrak{M}^{\begin{pmatrix} 0 & 0 & 0 & 0 & \\ & + & + &0 &0 \\ & + &+ &+ &+ & +\end{pmatrix}},\\&
         \mathfrak{M}^{\begin{pmatrix} + & 0 & 0 & 0 & \\ & + & + &0 &0 \\ & + &+ &+ &+ & +\end{pmatrix}},\,
         \mathfrak{M}^{\begin{pmatrix} + & + & 0 & 0 & \\ & + & + &0 &0 \\ & + &+ &+ &+ & +\end{pmatrix}},\,
         \mathfrak{M}^{\begin{pmatrix} 0 & 0 & 0 & 0 & \\ & + & + &+ &0 \\ & + &+ &+ &+ & +\end{pmatrix}},\,
        \mathfrak{M}^{\begin{pmatrix} + & 0 & 0 & 0 & \\ & + & + &+&0 \\ & + &+ &+ &+ & +\end{pmatrix}}, \\&
        \mathfrak{M}^{\begin{pmatrix} + & + & 0 & 0 & \\ & + & + &+ &0 \\ & + &+ &+ &+ & +\end{pmatrix}},\,\mathfrak{M}^{\begin{pmatrix} + & + & + & 0 & \\ & + & + &+ &0 \\ & + &+ &+ &+ & +\end{pmatrix}},\,
        \mathfrak{M}^{\begin{pmatrix} 0 & 0 & 0 & 0 & \\ & + & + &+ &+ \\ & + &+ &+ &+ & +\end{pmatrix}},\, 
        \mathfrak{M}^{\begin{pmatrix} + & 0 & 0 & 0 & \\ & + & + &+ &+ \\ & + &+ &+ &+ & +\end{pmatrix}},\\&
        \mathfrak{M}^{\begin{pmatrix} + & + & 0 & 0 & \\ & + & + &+ &+ \\ & + &+ &+ &+ & +\end{pmatrix}},\, \mathfrak{M}^{\begin{pmatrix} + & + & + & 0 & \\ & + & + &+ &+ \\ & + &+ &+ &+ & +\end{pmatrix}},\, 
        \mathfrak{M}^{\begin{pmatrix} + & + & + & + & \\ & + & + &+ &+ \\ & + &+ &+ &+ & +\end{pmatrix}}\,\bigg\}
    \end{aligned}
\end{equation}

These monopoles are not gauge invariant and hence they should be dressed. The dressing can be done analogously as in Table \ref{table: baryons of SU(2)_1 QCD3} (or \ref{table: baryons of SU(2)_2 QCD3}). The resulting operators have Lorentz spin 0, matching that of the baryons.

\subsection{Mapping Gauge Invariant Operators: The General Pattern}\label{sec: gen_opmap}
Before we conclude, we briefly comment on the general trend of mapping gauge invariant operators between the dual theories. We remind the reader that the proposed duality exchanges flavor and topological symmetries; hence, mesons, conserved currents and baryons of the electric theory will be mapped to suitably dressed disorder operators (monopoles) of the mirror theory. We ask the reader to keep Figure \ref{fig:scalar_Qcd_general} in mind for the rest of the discussion. We also take this opportunity to remind the reader that a specific choice of fugacities of the topological symmetry parametrizes each column.

\subsubsection*{Mesons and Conserved Currents}
Mesons and conserved currents are Lorentz scalars and vectors, respectively, that transform in the adjoint representation of the flavor symmetry group $SU(N_s)$. We now specify the general pattern of mapping these operators to monopoles of the mirror theory (Figure \ref{fig:scalar_Qcd_general}). We focus on the mapping of
the $\frac{1}{2} N_s(N_s-1)$ “off-diagonal" mesons $\varphi_I \bar{\varphi}_J$ with $J<I$. The case $J>I$ can be obtained by taking the Hermitian conjugate of these operators, which map to monopoles with negative GNO fluxes. 

A generic minimal flux mesonic operator \textbf{cannot} be charged under the topological symmetry of the first $U(1)$ gauge node in the last row of the quiver diagram. This is due to the parameterization of topological fugacities and representation-theoretic constraints. 
Modulo a subtlety arising from the presence of a monopole interaction in the dual quiver---already discussed in Section~\ref{sec: gen_proposal_bqcd}---we propose that \textit{off-diagonal} mesonic operators are mapped to monopoles carrying GNO flux under the gauge nodes in the top row and/or along the rightmost diagonal of the quiver.

To describe this mapping systematically, we find it convenient to organize the mesons $\varphi_I \bar\varphi_J$ into three distinct classes, based on the values of the flavor indices $I$ and $J$:
\begin{itemize}
    \item First, we consider the case where both $I$ and $J$ are less than $N_s - N$.
    \item Next, we study the regime where $J < N_s - N \leq I$.
    \item Finally, we address the case in which both $I$ and $J$ are greater than or equal to $N_s - N$.
\end{itemize}

\paragraph{$\varphi_I \bar\varphi_J$, where $J<I<N_s-N$:}
The simplest case is that of the meson $\varphi_{I+1}\bar{\varphi}_{I}$ with $(1\leq I \leq N_s-N-1)$.
In the mirror theory, this is mapped to a monopole with $+1$ GNO flux under the top node of the $I^{\text{th}}$ column: 

\begin{equation}\label{eq: generic_mesons0}
\varphi_{I+1}\bar{\varphi}_{I}
\dualto
\begin{tikzpicture}[baseline=(current bounding box).center]
\matrix (A) [matrix of nodes, left delimiter=(,right delimiter={)}]{
 0& \ldots & \boxed{+} & \ldots & 0 & 0 & 0 & \ldots & 0 & 0  \\
 &  &  & &  & & & & & & & {\color{black}{0}} \\
 & & & & & & & & & & & & $\color{black}{\ddots}$ \\
 & &  & & & & & & & & & & & {\color{black}{0}}\\};

\draw (A-1-1.south west) -|(A-1-10.south east);
\draw (A-1-10.south east) -| (A-2-12.south west);
\draw (A-2-12.south west) -| (A-3-13.south west);
\draw (A-3-13.south west) -| (A-4-14.south west);

\node[anchor=north east] at ([shift={(-30mm,-12mm)}]A.north east) {$\text{0 GNO Flux}$};

\draw [decorate,decoration={brace,amplitude=5pt,raise=2ex}]
  (A-1-1.north west) -- (A-1-6.north east) node[midway,xshift=0em,yshift=2em]{\tiny$N_s-N$};
\draw [decorate,decoration={brace,amplitude=5pt,raise=2ex}]
  (A-1-7.north west) -- (A-1-10.north east) node[midway,xshift=0em,yshift=2em]{\tiny$k$};
\draw [decorate,decoration={brace,amplitude=5pt,raise=1ex}]
  (A-1-10.north east) -- (A-4-14.north east) node[midway,xshift=0.8em,yshift=1em]{\tiny$N$};

\draw (A-1-3)++(0,-1) node (l1) {$I$};
\draw (l1) edge[->] (A-1-3);
\end{tikzpicture}
\end{equation}

Thereafter, $+1$ charges must propagate contiguously from left to right.
A meson of the form $\varphi_{I+1+A}\bar{\varphi}_I$ $(A\neq0)$ is dual to a monopole with the following charge assignment: 
\begin{equation}\label{eq: generic_mesons1}
\varphi_{I+1+A}\bar{\varphi}_I \dualto
\begin{tikzpicture} [baseline=(current bounding box).center]
\matrix (A) [matrix of nodes, left delimiter=(,right delimiter={)}]{
 0& \ldots & + & \ldots & + & 0 & 0 & \ldots & 0 & 0  \\
 &  &  & &  & & & & & & & {\color{black}{0}} \\
 & & & & & & & & & & & & $\color{black}{\ddots}$ \\
 & &  & & & & & & & & & & & {\color{black}{0}}\\};

\draw (A-1-1.south west) -|(A-1-10.south east);
\draw (A-1-10.south east) -| (A-2-12.south west);
\draw (A-2-12.south west) -| (A-3-13.south west);
\draw (A-3-13.south west) -| (A-4-14.south west);

\node[anchor=north east] at ([shift={(-30mm,-18mm)}]A.north east) {$\text{0 GNO Flux}$};

\draw [decorate,decoration={brace,amplitude=5pt,raise=2ex}]
  (A-1-1.north west) -- (A-1-6.north east) node[midway,xshift=0em,yshift=2em]{\tiny$N_s-N$};
\draw [decorate,decoration={brace,amplitude=5pt,raise=1ex}]
  (A-1-5.south east) -- (A-1-3.south west) node[midway,xshift=0em,yshift=-2em]{\tiny$A$};
\draw [decorate,decoration={brace,amplitude=5pt,raise=2ex}]
  (A-1-7.north west) -- (A-1-10.north east) node[midway,xshift=0em,yshift=2em]{\tiny$k$};
\draw [decorate,decoration={brace,amplitude=5pt,raise=1ex}]
  (A-1-10.north east) -- (A-4-14.north east) node[midway,xshift=0.8em,yshift=1em]{\tiny$N$};
\end{tikzpicture}
\end{equation}

\paragraph{$\varphi_I \bar\varphi_J$, where $J<N_s-N\leq I$:}
In this case, the dual monopole has a contiguous string of $+1$ GNO fluxes starting from the $J^{\text{th}}$ column and ending on the $(I+k)^{\text{th}}$ column. In particular, each of the $k$ nodes inside the region of the quiver connected by BF coupling has GNO flux $+1$.
The first case of this type is the meson $\varphi_{N_s-N+1}\bar{\varphi}_{I}$, which is mapped to a monopole of the form:

\begin{equation}\label{eq: generic_mesons2}
\varphi_{N_s-N+1}\bar{\varphi}_{I} \dualto
\begin{tikzpicture}[baseline=(current bounding box).center]
\matrix (A) [matrix of nodes, left delimiter=(,right delimiter={)}]{
 0& \ldots & + & \ldots & + & + & + & \ldots & + & +  \\
 &  &  & &  & & & & & & & {\color{black}{0}} \\
 & & & & & & & & & & & & $\color{black}{\ddots}$ \\
 & &  & & & & & & & & & & & {\color{black}{0}}\\};

\draw (A-1-1.south west) -|(A-1-10.south east);
\draw (A-1-10.south east) -| (A-2-12.south west);
\draw (A-2-12.south west) -| (A-3-13.south west);
\draw (A-3-13.south west) -| (A-4-14.south west);

\node[anchor=north east] at ([shift={(-30mm,-12mm)}]A.north east) {$\text{0 GNO Flux}$};

\draw [decorate,decoration={brace,amplitude=5pt,raise=2ex}]
  (A-1-1.north west) -- (A-1-6.north east) node[midway,xshift=0em,yshift=2em]{\tiny$N_s-N$};
\draw [decorate,decoration={brace,amplitude=5pt,raise=2ex}]
  (A-1-7.north west) -- (A-1-10.north east) node[midway,xshift=0em,yshift=2em]{\tiny$k$};
\draw [decorate,decoration={brace,amplitude=5pt,raise=1ex}]
  (A-1-10.north east) -- (A-4-14.north east) node[midway,xshift=0.8em,yshift=1em]{\tiny$N$};
\end{tikzpicture}
\end{equation}

Once a charge has been assigned to a column, no additional charges can be assigned to the same column in lower rows. The propagation of fluxes for lower rows follows a strict ``staircase'' pattern, i.e.: charges in the lower rows can only be assigned to the rightmost column. Thus, a generic mesonic monopole mapped to $\varphi_{N_s-N+1+J}\bar{\varphi}_I$ ($1\leq J \leq N$) will have the following flux assignment:

\begin{equation}\label{eq: generic_mesons}
\varphi_{N_s-N+1+J}\bar{\varphi}_I \dualto
\begin{tikzpicture}[baseline=(current bounding box).center]
\matrix (A) [matrix of nodes, left delimiter=(,right delimiter={)}]{
 0& \ldots & + &+ & + & \ldots & + & +  \\
 &  &  & &  & & & & &  {\color{black}{+}} \\
 & & & & & & & & & & $\color{black}{\ddots}$ \\
 & &  & & & & & & & & &  {\color{black}{+}}\\
& & & & & & & & & & & & {\color{black}{0}}\\};

\draw (A-1-1.south west) -|(A-1-8.south east);
\draw (A-1-8.south east) -| (A-2-10.south west);
\draw (A-2-10.south west) -| (A-3-11.south west);
\draw (A-3-11.south west) -| (A-4-12.south west);
\draw (A-4-12.south west) -| (A-5-13.south west);

\node[anchor=north east] at ([shift={(-30mm,-12mm)}]A.north east) {$\text{0 GNO Flux}$};

\draw [decorate,decoration={brace,amplitude=5pt,raise=2ex}]
  (A-1-1.north west) -- (A-1-4.north east) node[midway,xshift=0em,yshift=2em]{\tiny$N_s-N$};
\draw [decorate,decoration={brace,amplitude=5pt,raise=2ex}]
  (A-1-5.north west) -- (A-1-8.north east) node[midway,xshift=0em,yshift=2em]{\tiny$k$};
\draw [decorate,decoration={brace,amplitude=5pt,raise=1ex}]
  (A-2-10.north east) -- (A-4-12.north east) node[midway,xshift=0.8em,yshift=1em]{\tiny$J$};
\end{tikzpicture}
\end{equation} 

\paragraph{$\varphi_I \bar\varphi_J$, where $N_s-N\leq J < I $:}
Mesons corresponding to values of $I \geq N_s - N + 1$ will instead have their $+1$ flux assigned to the rightmost column of the lower rows, and subsequent charge assignment will follow the usual staircase assignment. Thus, the meson $\varphi_{I}\bar{\varphi}_{J}$ ($N_s-N+1\leq J < I \leq N_s$) will be mapped to a monopole of the form:

\begin{equation}\label{eq: generic_mesons3}
\varphi_{I}\bar{\varphi}_{J} \dualto
\begin{tikzpicture}[baseline=(current bounding box).center]
\matrix (A) [matrix of nodes, left delimiter=(,right delimiter={)}]{
 0& \ldots & 0 &0 & 0 & \ldots & 0 & 0  \\
 &  &  & &  & & & & &  {\color{black}{+}} \\
 & & & & & & & & & & $\color{black}{\ddots}$ \\
 & &  & & & & & & & & &  {\color{black}{+}}\\
& & & & & & & & & & & & {\color{black}{0}}\\};

\draw (A-1-1.south west) -|(A-1-8.south east);
\draw (A-1-8.south east) -| (A-2-10.south west);
\draw (A-2-10.south west) -| (A-3-11.south west);
\draw (A-3-11.south west) -| (A-4-12.south west);
\draw (A-4-12.south west) -| (A-5-13.south west);

\node[anchor=north east] at ([shift={(-30mm,-12mm)}]A.north east) {$\text{0 GNO Flux}$};

\draw [decorate,decoration={brace,amplitude=5pt,raise=2ex}]
  (A-1-1.north west) -- (A-1-4.north east) node[midway,xshift=0em,yshift=2em]{\tiny$N_s-N$};
\draw [decorate,decoration={brace,amplitude=5pt,raise=2ex}]
  (A-1-5.north west) -- (A-1-8.north east) node[midway,xshift=0em,yshift=2em]{\tiny$k$};
\draw [decorate,decoration={brace,amplitude=5pt,raise=1ex}]
  (A-2-10.north east) -- (A-4-12.north east) node[midway,xshift=1.4em,yshift=1em]{\tiny$I-J$};
\end{tikzpicture}
\end{equation}

Generically, all the  monopoles considered above will have spin-0 and spin-1 channels after they have been dressed - the spin-0 operators correspond to the off-diagonal elements of the meson matrix, while the spin-1 operators correspond to conserved currents in the original electric theory. We remind the reader that the diagonal elements $|\varphi_i|^2$ are mapped to mass deformations (refer Section \ref{checks} for a more detailed discussion on mass deformations) in the dual theory, while the diagonal currents map to topological $U(1)$ currents (refer Section \ref{sec: SU(2)_1w5s} for an example of the dressing) in the mirror.

\subsubsection*{Baryons}
Baryons are Lorentz scalars that transform in the rank-$N_s$ antisymmetric representation of the flavor symmetry group $SU(N_s)$. We describe how these operators map to minimal flux operators in the planar mirror theory. We focus on the mapping of $\binom{N_s}{N}$ baryons to monopoles in the dual theory. Anti-baryons can be obtained by taking the Hermitian conjugate of these operators, which would map to monopoles with negative GNO fluxes.

\paragraph{$(\prod_{j=1}^{N-1}\varphi_j)\varphi_{N+I}$, where $0\leq I\leq N_s+k-N-1$:}
A generic minimal flux baryonic monopole operator \textbf{must} carry charge $+1$ under the topological symmetry of the first $U(1)$ gauge node in the last row of the quiver diagram\footnote{The topological symmetry associated to this node is parametrized by $\sum_{k=1}^N X_k$, whereas all other nodes have topological symmetries parametrized by $X_i - X_{i+1}$. Thus, the group theoretic properties of a monopole acquiring charge $+1$ under this specific node follow from representation theory.}.

Accordingly, the baryon $\prod_{j=1}^{N}\varphi_j=\varphi_{1}\ldots\varphi_{N}$ is dual to the minimal flux configuration:
\begin{equation}\label{eq: generic_baryons0}
\prod_{j=1}^{N}\varphi_j\dualto
\begin{tikzpicture}[baseline=(current bounding box).center]
\matrix (A) [matrix of nodes, left delimiter=(,right delimiter={)}]{
 &  &  &  &  &  &  &  &  &   \\
 &  &  & &  & & & & & & & {\color{white}{+}} \\
 & & & & & & & & & & & & $\color{white}{\ddots}$ \\
 & & & & + &  & & & & & & & & & & {\color{white}{+}}\\};

\draw (A-4-5.north west) -| (A-4-5.south east);
\node[anchor=north east] at ([shift={(-2mm,-2mm)}]A.north east) {$\text{0 GNO Flux}$};

\end{tikzpicture}
\end{equation}

Thereafter, the charge assignment must propagate contiguously from left to right in a row. A baryon of the form $\bigg(\prod_{j=1}^{N-1}\varphi_j\bigg)\varphi_{N+I}=\varphi_{1}\ldots\varphi_{N-1}\varphi_{N+I}$ is dual to the flux configuration:
\begin{equation}\label{eq: generic_baryons1}
\bigg(\prod_{j=1}^{N-1}\varphi_j\bigg)\varphi_{N+I}\dualto
\begin{tikzpicture}[baseline=(current bounding box).center]
\matrix (A) [matrix of nodes, left delimiter=(,right delimiter={)}]{
 &  &  &  &  &  &  &  &  &   \\
 &  &  & &  & & & & & & & {\color{white}{+}} \\
 & & & & & & & & & & & & $\color{white}{\ddots}$ \\
 & & & & + & + &\ldots &+ & & & & & & & & {\color{white}{+}}\\};

\draw (A-4-5.north west) -| (A-4-8.south east);
\node[anchor=north east] at ([shift={(-2mm,-2mm)}]A.north east) {$\text{0 GNO Flux}$};

\draw [decorate,decoration={brace,amplitude=5pt,raise=1ex}]
  (A-4-8.south east) -- (A-4-5.south east) node[midway,xshift=0em,yshift=-2em]{\tiny$I$};

\end{tikzpicture}
\end{equation} 
As before, care must be taken if there is a BF-coupled region (of length $k$), and each of the $k$ nodes inside the region of the quiver connected only by BF couplings must have GNO flux $+1$. For instance, the baryon $\varphi_{1}\ldots\varphi_{N-1}\varphi_{N_s-N+1}$ is dual to the flux configuration:
\begin{equation}\label{eq: generic_baryons2}
\bigg(\prod_{j=1}^{N-1}\varphi_j\bigg)\varphi_{N_s-N+1}\dualto
\begin{tikzpicture}[baseline=(current bounding box).center]
\matrix (A) [matrix of nodes, left delimiter=(,right delimiter={)}]{
 &  &  &  &  &  &  &  &  &   \\
 &  &  & &  & & & & & & & {\color{white}{+}} \\
 & & & & & & & & & & & & $\color{white}{\ddots}$ \\
 & & & & + & + &\ldots &+ & +&\ldots &+ &+ & & & & {\color{white}{+}}\\};

\draw (A-4-5.north west) -| (A-4-12.south east);
\node[anchor=north east] at ([shift={(-2mm,-2mm)}]A.north east) {$\text{0 GNO Flux}$};

\draw [decorate,decoration={brace,amplitude=5pt,raise=1ex}]
  (A-4-8.south east) -- (A-4-5.south east) node[midway,xshift=0em,yshift=-2em]{\tiny$N_s-N+1$};
\draw [decorate,decoration={brace,amplitude=5pt,raise=1ex}]
  (A-4-11.south east) -- (A-4-8.south east) node[midway,xshift=0em,yshift=-2em]{\tiny$k$};

\end{tikzpicture}
\end{equation} 

\paragraph{$(\prod_{k=1}^{N-J}\varphi_j)(\prod_{k=N-J+2}^{N+1}\varphi_{k})$, where $1\leq J\leq N-1$:}
The flux assignment in higher rows follows a ``staircase" pattern. For instance, the baryon $\bigg(\prod_{j=1}^{N-2}\varphi_j\bigg)\varphi_{N}\varphi_{N+1}=\varphi_{1}\ldots\varphi_{N-2}\varphi_{N}\varphi_{N+1}$ is dual to the flux configuration:

\begin{equation}\label{eq: generic_baryons3}
\bigg(\prod_{j=1}^{N-2}\varphi_j\bigg)\varphi_{N}\varphi_{N+1}\dualto
\begin{tikzpicture}[baseline=(current bounding box).center]
\matrix (A) [matrix of nodes, left delimiter=(,right delimiter={)}]{
 &  &  &  &  &  &  &  &  &   \\
 &  &  & {\color{white}+}&  & & & & & & & {\color{white}{+}} \\
 & & & &+ & & & & & & & & $\color{white}{\ddots}$ \\
 & & & & + & + &\ldots &0 & 0&\ldots &0 &0 &\ldots &0 & & {\color{white}{+}}\\};

\draw (A-2-4.south west) -| (A-3-5.south east);
\draw (A-4-6.north west) -| (A-4-6.south east);
\node[anchor=north east] at ([shift={(-2mm,-2mm)}]A.north east) {$\text{0 GNO Flux}$};

\draw [decorate,decoration={brace,amplitude=5pt,raise=1ex}]
  (A-4-8.south east) -- (A-4-5.south east) node[midway,xshift=0em,yshift=-2em]{\tiny$N_s-N+1$};
\draw [decorate,decoration={brace,amplitude=5pt,raise=1ex}]
  (A-4-11.south east) -- (A-4-8.south east) node[midway,xshift=0em,yshift=-2em]{\tiny$k$};
\draw [decorate,decoration={brace,amplitude=5pt,raise=1ex}]
  (A-4-14.south east) -- (A-4-11.south east) node[midway,xshift=0em,yshift=-2em]{\tiny$N$};

\end{tikzpicture}
\end{equation} 

Thus, the baryon $\bigg(\prod_{k=1}^{N-J}\varphi_k\bigg)\bigg(\prod_{k=N-J+2}^{N+1}\varphi_k\bigg)=\varphi_{1}\ldots\varphi_{N-J}\varphi_{N-J+2}\ldots\varphi_{N+1}$ is dual to the flux configuration:

\begin{equation}\label{eq: generic_baryons4}
\bigg(\prod_{k=1}^{N-J}\varphi_k\bigg)\bigg(\prod_{k=N-J+2}^{N+1}\varphi_k\bigg)\dualto
\begin{tikzpicture}[baseline=(current bounding box).center]
\matrix (A) [matrix of nodes, left delimiter=(,right delimiter={)}]{
 0 &  &  &  &  &  &  &  &  &   \\
 &  $\ddots$&  &  &  &  &  &  &  &   \\
 &  & + &  &  &  &  &  &  &   \\
 &  &  & $\ddots$&  & & & & & & & {\color{white}{+}} \\
 & & & &+ & & & & & & & & $\color{white}{\ddots}$ \\
 & & & & + & + &\ldots &0 & 0&\ldots &0 &0 &\ldots &0 & & {\color{white}{+}}\\};

\draw (A-2-2.south west) -|(A-3-3.south east);
\draw (A-3-3.south east) -| (A-4-4.south east);
\draw (A-4-4.south east) -| (A-5-5.south east);
\draw (A-6-6.north west) -| (A-6-6.south east);
\node[anchor=north east] at ([shift={(-2mm,-2mm)}]A.north east) {$\text{0 GNO Flux}$};

\draw [decorate,decoration={brace,amplitude=5pt,raise=1ex}]
 (A-6-8.south east) -- (A-6-5.south east) node[midway,xshift=0em,yshift=-2em]{\tiny$N_s-N+1$};
\draw [decorate,decoration={brace,amplitude=5pt,raise=1ex}]
 (A-6-11.south east) -- (A-6-8.south east) node[midway,xshift=0em,yshift=-2em]{\tiny$k$};
\draw [decorate,decoration={brace,amplitude=5pt,raise=1ex}]
  (A-6-14.south east) -- (A-6-11.south east) node[midway,xshift=0em,yshift=-2em]{\tiny$N$};
\draw [decorate,decoration={brace,amplitude=5pt,raise=4ex}]
 (A-5-5.north east) -- (A-3-3.north east) node[midway,xshift=-1.8em,yshift=-2em]{\tiny$J$};

\end{tikzpicture}
\end{equation} 

\paragraph{Mapping a generic baryon:}
Unlike mesons, baryons allow column reuse: a single column may host fluxes in multiple rows, provided the staircase rule is respected. Thus, a generic baryonic monopole  will have the following flux assignment:

\begin{equation}
\bigg(\prod_{m=1}^{N-J-1}\varphi_m\bigg)\bigg(\prod_{l=1}^J\varphi_{N+b_l-l}\bigg)\varphi_{N+I}\dualto
\begin{tikzpicture}[baseline=(current bounding box).center]
\matrix (A) [matrix of nodes, left delimiter=(,right delimiter={)}]{
 0 &  &  &  &  &  &    \\
 &  $\ddots$&  &  &  &     \\
 &  & + & \ldots & + & 0 &\ldots  &    \\
 &  &  & $\ddots$&  & & & & &  \\
 & & & &+ &+ & \ldots& + &0 &\ldots & &  \\
 & & & & + & + &\ldots & \ldots & +&\ldots & + & 0 &\ldots & \\};


\draw [decorate,decoration={brace,amplitude=5pt,raise=1ex}]
(A-6-11.south east) -- (A-6-5.south east) node[midway,xshift=0em,yshift=-2em]{\tiny$I$};

\draw [decorate,decoration={brace,amplitude=5pt,raise=.5ex}]
 (A-5-5.north west) -- (A-5-8.north east) node[midway,xshift=0em,yshift=1.2em]{\tiny$b_1$};

\draw [decorate,decoration={brace,amplitude=5pt,raise=.5ex}]
 (A-3-3.north west) -- (A-3-5.north east) node[midway,xshift=0em,yshift=1.2em]{\tiny$b_J$};
 
\draw [decorate,decoration={brace,amplitude=5pt,raise=4ex}]
 (A-5-5.north east) -- (A-3-3.north east) node[midway,xshift=-1.8em,yshift=-2em]{\tiny$J$};

\end{tikzpicture}
\end{equation}


We note that $k\geq0$ does not change the counting of the mesonic and baryonic operators.
\section{Planar Abelian Duals of CS-QCD\texorpdfstring{$_3$}{3} with Bosons and Fermions}\label{sec: qcd_b&f}
In this section, we generalize the proposal of the previous section to non-Abelian theories with fundamental scalars and fermions. We anticipate that adding a fermionic flavor in the electric theory corresponds to ``bosonizing" a column in the planar theory. We provide several examples of this here. 

\subsection{General Proposal}
\begin{figure}[ht]
     \centerline{\includegraphics[width=1.3\textwidth]{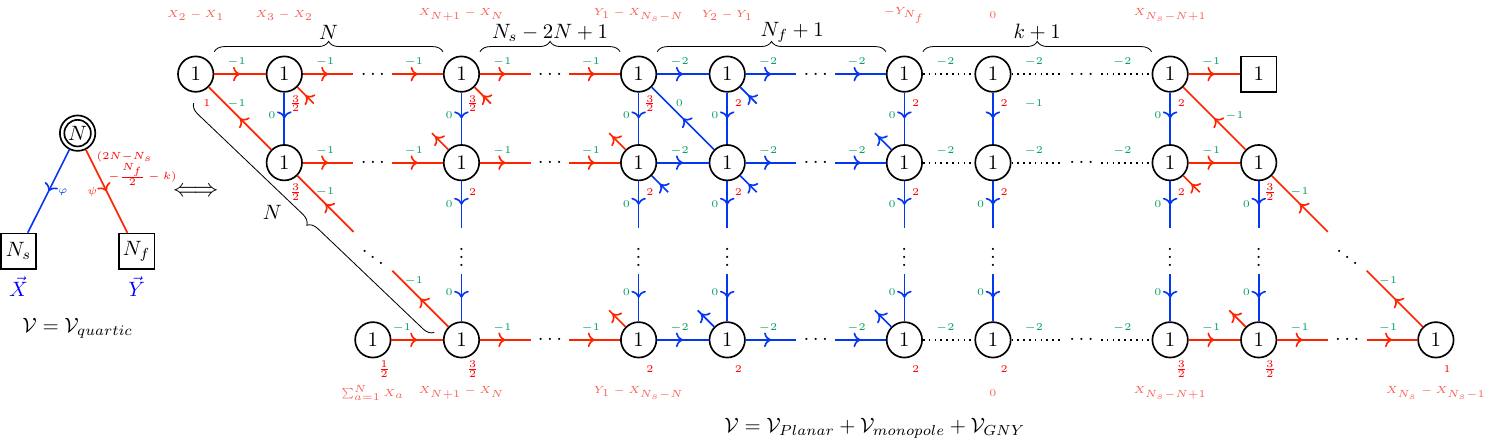}}
    \caption{The proposed dual for $SU(N)$ QCD$_3$ with $N_s$ scalars and $N_f$ fermions at Chern-Simons level $2N-N_s-\frac{N_f}{2}-k$ is shown here. As before, single/double circles (squares) correspond to $U/SU$ gauge (flavor) symmetries, scalar (fermion) bifundamental fields are indicated in blue (red), Chern-Simons levels of gauge nodes are indicated in maroon, while mixed BF couplings between gauge nodes are indicated in green. Dotted lines indicate the absence of any bifundamental matter fields and only the presence of BF interactions. Note that the Chern-Simons level of each gauge node is equal to minus half the sum of all BF couplings with the surrounding nodes.
    We also indicate the fugacities of the topological symmetries of each column in orange.
    }
    \label{fig:scalar_Qcd_spicy}
\end{figure}

We propose the following planar, Abelian dual for $SU(N)$ QCD$_3$ with $N_s$ scalars and $N_f$ fermions at Chern-Simons level $2N-N_s-\frac{N_f}{2}-k$ in Figure \ref{fig:scalar_Qcd_spicy}. The proposed duality is believed to hold in the following range of parameters\footnote{We consider some examples with $N_s < 2N$ in Sections \ref{sec: SU(3)_0 w N_s=5} and \ref{sec: SU(2)_(-1/2) QCD}.}:
\begin{equation}
    N_s + N_f  \geq 2N, \qquad N_s \geq 2N, \qquad k\geq 0
\end{equation}
 This proposal should be seen as a generalization of the one considered for scalar QCD$_3$ obtained by replacing $N_f$ scalars with $N_f$ fermions and replacing the corresponding fermionic columns in the planar mirror with purely bosonic columns.

We briefly comment on the potential $\mathcal{V}$ shown in Figure \ref{fig:scalar_Qcd_spicy}. In addition to quartic scalar interactions (Equation \ref{eq:quarticV}), mixed scalar-fermion interactions are also generated radiatively in the electric QCD$_3$ theory and are captured by the electric potential $\mathcal{V}_{quartic}:$
\begin{equation}
    \begin{aligned}
    \mathcal{V}_{quartic} &= \sum_{\alpha,\beta=1}^N\sum_{I,J=1}^{N_s}\sum_{A,B=1}^{N_f}\big[\varphi^{\alpha I}\bar{\varphi}_{\alpha J}\varphi^{\beta J}\bar{\varphi}_{\beta I}+(\varphi^{\alpha I}\bar{\varphi}_{\alpha I})(\varphi^{\beta J}\bar{\varphi}_{\beta J}) + \varphi^{\alpha I}\bar{\varphi}_{\beta I}\bar{\psi}_{\alpha A}\psi^{\beta A}\\ & + (\varphi^{\alpha I}\bar{\varphi}_{\alpha I})(\psi^{\beta A}\bar{\psi}_{\beta A})\big].
    \end{aligned}
\end{equation}
On the other hand, in addition to the usual Yukawa interactions, the monopole potential, and the GNY potential, an extra contribution is captured by $\mathcal{V}_{Planar}$ in the mirror theory. $\mathcal{V}_{Planar}$ contains a cubic scalar interaction ($\phi_i\phi_j\phi_k$+$c.c.$) for every closed planar triangle. There is a term with a coefficient of $+1$ ($-1$) for each triangle closed clockwise (anticlockwise). 
$\mathcal{V}_{monopole}$ involves every monopole with GNO fluxes $+1/-1$ under two nodes connected by a vertical scalar, analogous to the case discussed in Section \ref{sec: non-Abelian}, and are dressed as explained in Appendix \ref{app:monopoles}.
As before, we claim $\mathcal{V}_{monopole}$ and $\mathcal{V}_{Planar}$ are essential to match the rank of the global symmetry on the electric side. In the presence of these terms, the UV symmetry of the quiver theory is $U(1)_{top}^{N_s}\times U(1)_{top}^{N_f}$, where each $U(1)$ factor corresponds to the topological symmetry of the $N_s+N_f-1$ columns of the quiver, and the additional $U(1)$ factor is the topological symmetry of the bottom-most gauge node. We also claim the symmetry enhances in the IR:
\begin{equation}
    U(1)^{N_s+N_f}_{top}\to U(N_s)\times U(N_f) 
\end{equation}

which is mapped precisely to the flavor symmetries of the electric theory. Furthermore, we expect the faithful symmetry to be \cite{Benini:2017aed}:
\begin{equation}
    \frac{U(N_s)\times U(N_f)}{\mathbb{Z}_N}\rtimes \mathbb{Z}_2^{\mathcal{C}}
\end{equation}
where $\mathbb{Z}_N$ is the center of $SU(N)$ and $\mathbb{Z}_2^{\mathcal{C}}$ is a charge conjugation symmetry.

In the rest of this section, we make these ideas more precise by considering some examples.
\paragraph{A Note on Higher Spin Baryons and Dual Monopoles}
We take a moment to comment on higher spin baryonic operators and their dual flux configurations here. 
Consider a generic baryon of the electric theory:
\begin{equation}
    (\chi_1)^{i_1}(\chi_2)^{i_2} \cdots (\chi_F)^{i_F}
\end{equation}
where each $\chi_i$ is either a boson or fermion, and the powers satisfy $\sum_{k=1}^F i_k = N$ \footnote{N.B.: If multiple copies of the same scalar appear, derivatives must be inserted; otherwise, the antisymmetric contraction vanishes.}. This baryon maps to the following monopole operator in the mirror theory:
\begin{equation}
\begin{tikzpicture}
\matrix (A) [matrix of nodes, left delimiter=(,right delimiter={)}]{
 +&  +&  \dots &  &+  & + &  &  &  &   \\
 &  +&  \dots& & + & +& +& & & & & {\color{white}{+}} \\
 & & & & + & +& \dots&  & & & & & \dots &0 &0 &  $\color{white}{\ddots}$ \\
 & & & & + & + & \dots& & & & & & & \dots & 0& 0 {\color{white}{+}}\\};

\draw (A-4-5.north west) -| (A-4-5.south east);
\draw [decorate,decoration={brace,amplitude=4pt,raise=1ex}]
  (A-1-1.north west) -- (A-1-1.north east) node[midway,xshift=0em,yshift=1.5em]{\tiny$j_1$};
  \draw [decorate,decoration={brace,amplitude=4pt,raise=1ex}]
  (A-1-2.north west) -- (A-1-2.north east) node[midway,xshift=0em,yshift=1.5em]{\tiny$j_2$};
  \node[midway,xshift=-5em,yshift=4em] at (A-1-1.north west) {\tiny$\dots$};
  \draw [decorate,decoration={brace,amplitude=4pt,raise=1ex}]
  (-1.1,1.2) -- (-.6,1.2) node[midway,xshift=0em,yshift=1.5em]{\tiny$j_N$};
  \node at (0,1.5) {\tiny$\dots$};
  \draw [decorate,decoration={brace,amplitude=4pt,raise=1ex}]
  (3.1,-1.1) -- (2.8,-1.1) node[midway,xshift=0em,yshift=-1.5em]{\tiny$j_{F-1}$};
\end{tikzpicture}
\end{equation} 
Here, $j_k$ denotes the number of $+$ in column $k$ (excluding the base corner $+$ which is always present). The $j_k$ are related to the baryonic charges $i_k$ via:
\begin{equation}
    \begin{cases}
    j_1 &= 1 - i_1 \\
    j_2 &= 2 - i_1 - i_2 \\
    \vdots \\
    j_N &= N - \sum_{k=1}^N i_k \\
    j_{N+1} &= N - \sum_{k=1}^{N+1} i_k \\
    \vdots \\
    j_{F-1} &= N - \sum_{k=1}^{F-1} i_k
    \end{cases}
\end{equation}

some $j_k$ may be negative—for instance, if $i_1 > 1$, then $j_1 < 0$. In such cases, negative fluxes (denoted by “$-$” symbols) must appear in the corresponding column. The fluxes within each column should be distributed across rows in a way that ensures the resulting monopole is \emph{minimally} dressed. We will see some explicit examples of this now. 

Finally, we note that the number of spin-$J$ baryons that can be built without derivatives in $SU(N)$ QCD$_3$ with $N_s$ scalars and $N_f$ fermions is $\binom{N_f+2J}{2J}\times\binom{N_s}{N-2J}$, constructed by (anti-)symmetrizing the (bosonic) fermionic fields. 
Consider in particular the spin-$J$ baryons:
\begin{equation}
\psi_{\{i_1} \psi_{i_2} \dots \psi_{i_{2J} \} } \varphi_{ [j_i} \varphi_{ j_2} \dots \varphi_{j_{N-2J}]}
\end{equation}
where the flavor indices of scalars $j_n$ are antisymmetrized and the flavor indices of fermions $i_n$ are symmetrized. The fermions are contracted to give a spin-$J$ operator. This operator transform in the $2J$-index symmetric representation of $SU(N_f)$ and in the $(N-2J)$-index antisymmetric representation of $SU(N_s)$. The operator in the highest weight:
\begin{equation}
(\psi_{1})^{2J}
\varphi_{1} \varphi_{2} \dots \varphi_{N-2J}
\end{equation}
is mapped to the following monopole on the mirror side:
\begin{equation}
\begin{tikzpicture}
\matrix (A) [matrix of nodes, left delimiter=(,right delimiter={)}]{
 0&  0&  \dots &  &0 & 0 &  & {\color{white}{+}} &  &   \\
 &  0&  \dots& &  & & & & & & & {\color{white}{+}} \\
  &  &  +& & + & +& \dots& + & & & & {\color{white}{+}} \\
 & & & & + & +& \dots& + & & & & & \dots &0 &0 &  $\color{white}{\dots}$ \\
 & & & & + & + & \dots& + & & & & & & \dots & 0& 0 {\color{white}{+}}\\};

\draw [decorate,decoration={brace,amplitude=4pt,raise=1ex}]
 (A-3-8.east)--  (A-5-8.east) node[midway,xshift=1em,yshift=0em]{\tiny$J$};

  \draw [decorate,decoration={brace,amplitude=4pt,raise=1ex}]
 (A-5-8.south east) -- (A-5-8.south west) node[midway,xshift=0em,yshift=-1.5em]{\tiny$N_s - N$};
 
  
\end{tikzpicture}
\end{equation}

\subsection{Example: \texorpdfstring{$SU(3)_{-\frac{1}{2}}$ QCD$_3$ with $N_s=6$ scalars and $N_f=1$}{su32} fermion}
\begin{figure}[ht]
    \centerline{\includegraphics[width=1.1\textwidth]{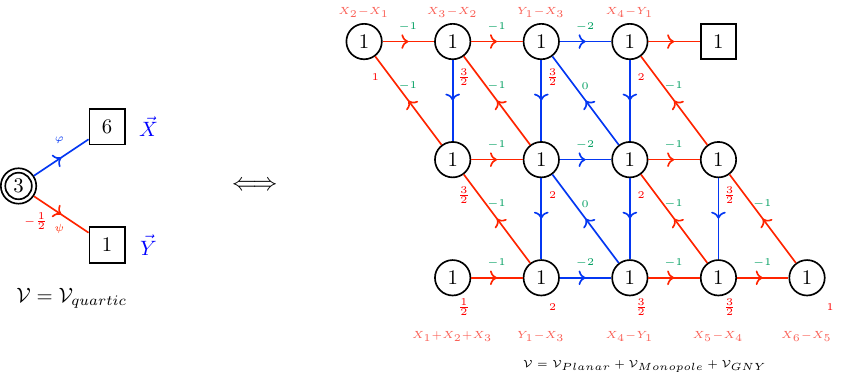}}
    \caption{The dual for $SU(3)_{-\frac{1}{2}}$ QCD$_3$ with 6 scalars and 1 fermion is shown here. All labelling conventions are the same as Figure \ref{fig:scalar_Qcd_spicy}, and BF couplings between nodes connected by a vertical scalar field have been suppressed for brevity.}
    \label{fig: SU(3)_1/2 6s1fs}
\end{figure}

Let us begin with the example of $SU(3)_{-\frac{1}{2}}$ with 6 scalars and 1 fermion, and its planar dual description. We propose the duality shown in Figure \ref{fig: SU(3)_1/2 6s1fs}.

We are mainly concerned with mapping the following gauge invariant operators of the electric theory:

\begin{enumerate}
    \item There are 48 \textbf{mesonic operators} of which:
    \begin{enumerate}
        \item 35 spin-0 mesons of the form $\varphi_i\bar{\varphi}_j$, identified by  $X_i-X_j$ (including their charge conjugates). 
        \item 12 spin-$\frac{1}{2}$ mesons of the form $\varphi_i\bar{\psi}$, identified by  $X_i - Y_1$ (including their charge conjugates).
        \item 1 meson with a spin-0 and spin-1 channel $\psi\bar{\psi}$.
    \end{enumerate}
    As already discussed in the previous cases the mapping of diagonal mesons, which are neutral under the maximal torus of the global symmetry, is less straightforward, and we leave it to future work. The mapping of these off-diagonal mesons is given in Table \ref{table: mesons of SU(3)_1/2 QCD3}.
    \item There are many \textbf{baryonic operators}:
    \begin{enumerate}
        \item $\binom{6}{3}=20$ spin-0 baryons of the form $\varphi_i\varphi_j\varphi_k$, identified by $X_i+X_j+X_k$. 
        \item $6$ baryons with a spin-0 and spin-1 channel of the form $\psi\psi\phi_i$, identified by $2Y_1 + X_i$.
        \item $\binom{6}{2}=15$ spin-$\frac{1}{2}$ baryons of the form $\psi\varphi_i\varphi_j$, identified by $Y_1+X_i+X_j$.
        \item $1$ baryon with spin-$\frac{3}{2}$ channels, identified by $3Y_1$, constructed by symmetrizing $\psi$ in the flavor and spinor indices.
    \end{enumerate}

    The mapping of baryons is given in Table \ref{table: baryons of SU(3)_1/2 QCD3}.
\end{enumerate}

\begin{center}
\renewcommand{\arraystretch}{0.6}
\begin{longtable}{ |>{\centering\arraybackslash}p{4cm}|>{\centering\arraybackslash}p{1.5cm}|>{\centering\arraybackslash}p{4cm}|>{\centering\arraybackslash}p{1.5cm}|  }
\caption{Dressed monopole operators of the planar mirror (Figure \ref{fig: SU(3)_1/2 6s1fs}) and their putative mapping to mesonic operators of the electric theory. The remaining mesonic operators can be constructed by charge conjugation.}
\label{table: mesons of SU(3)_1/2 QCD3} \\ 

\hline \multicolumn{1}{|c|}{\textbf{GNO Flux}} & \multicolumn{1}{c|}{\textbf{Meson}} & \multicolumn{1}{c|}{\textbf{GNO Flux}} & \multicolumn{1}{c|}{\textbf{Meson}} \\ \hline 
\endfirsthead

\multicolumn{4}{c}%
{{ \textbf{\tablename\ \thetable{}} -- \text{continued from previous page}}} \\
\hline \multicolumn{1}{|c|}{\textbf{GNO Flux }} & \multicolumn{1}{c|}{\textbf{Meson}} & \multicolumn{1}{c|}{\textbf{GNO Flux}} & \multicolumn{1}{c|}{\textbf{Meson}} \\ \hline 
\endhead

\hline \multicolumn{4}{|r|}{{\textit{Continued on next page}}} \\ \hline
\endfoot

\hline \hline
\endlastfoot
\vspace{-0.5cm}
\begin{equation*}
\begin{pmatrix}

        \tikznode{g11}{+} & & \tikznode{g12}{0} & & \tikznode{g13}{0} & & \tikznode{g14}{0} & & \tikznode{f1}{{\color{white}+}} & \tikznode{f}{}\\ & & \tikznode{g21}{0} & &  \tikznode{g22}{0} & & \tikznode{g23}{0} & & \tikznode{g24}{0} & & \tikznode{h}{}\\ & & \tikznode{g31}{0} & &\tikznode{g32}{0} & & \tikznode{g33}{0} & & \tikznode{g34}{0} & & \tikznode{g35}{0}
    \end{pmatrix}
\end{equation*}
\begin{tikzpicture}[overlay,remember picture,line width=.7pt,transform canvas={yshift=0mm}]
        \draw[fcolor,->-] (g12)--(g11);\draw[fcolor,->-] (g21)--(g11);
\end{tikzpicture} \vspace{-0.5cm}
 & \vspace{0.3cm}$\varphi_2\bar{\varphi}_1$ & \vspace{-0.5cm}
 \begin{equation*}
\begin{pmatrix}

        \tikznode{g11}{+} & & \tikznode{g12}{+} & & \tikznode{g13}{0} & & \tikznode{g14}{0} & & \tikznode{f1}{{\color{white}+}} & \tikznode{f}{}\\ & & \tikznode{g21}{0} & &  \tikznode{g22}{0} & & \tikznode{g23}{0} & & \tikznode{g24}{0} & & \tikznode{h}{}\\ & & \tikznode{g31}{0} & &\tikznode{g32}{0} & & \tikznode{g33}{0} & & \tikznode{g34}{0} & & \tikznode{g35}{0}
    \end{pmatrix}
\end{equation*}
\begin{tikzpicture}[overlay,remember picture,line width=.7pt,transform canvas={yshift=0mm}]
       \draw[fcolor,->-] (g13)--(g12);\draw[fcolor,->-] (g22)--(g12);\draw[fcolor,->-] (g21)--(g11);
       \end{tikzpicture} \vspace{-0.5cm}
 & \vspace{.3cm}$\varphi_3\bar{\varphi}_1$ \\
\hline
\end{longtable}
\end{center}

\begin{center}
\renewcommand{\arraystretch}{0.6}
\begin{longtable}{ |>{\centering\arraybackslash}p{4cm}|>{\centering\arraybackslash}p{1.5cm}|>{\centering\arraybackslash}p{4cm}|>{\centering\arraybackslash}p{1.5cm}|  }
\caption{Dressed monopole operators of the planar mirror (Figure \ref{fig: SU(3)_1/2 6s1fs}) and their putative mapping to baryonic operators of the electric theory (all indices have been suppressed). The remaining anti-baryonic operators can be constructed by charge conjugation.}
\label{table: baryons of SU(3)_1/2 QCD3} \\ 

\hline \multicolumn{1}{|c|}{\textbf{GNO Flux}} & \multicolumn{1}{c|}{\textbf{Baryon}} & \multicolumn{1}{c|}{\textbf{GNO Flux}} & \multicolumn{1}{c|}{\textbf{Baryon}} \\ \hline 
\endfirsthead

\multicolumn{4}{c}%
{{ \textbf{\tablename\ \thetable{}} -- \text{continued from previous page}}} \\
\hline \multicolumn{1}{|c|}{\textbf{GNO Flux }} & \multicolumn{1}{c|}{\textbf{Baryon}} & \multicolumn{1}{c|}{\textbf{GNO Flux}} & \multicolumn{1}{c|}{\textbf{Baryon}} \\ \hline 
\endhead

\hline \multicolumn{4}{|r|}{{\textit{Continued on next page}}} \\ \hline
\endfoot

\hline \hline
\endlastfoot
\vspace{-0.5cm}
\begin{equation*}
\begin{pmatrix}

        \tikznode{g11}{0} & & \tikznode{g12}{0} & & \tikznode{g13}{0} & & \tikznode{g14}{0} & & \tikznode{f1}{{\color{white}+}} & \tikznode{f}{}\\ & & \tikznode{g21}{0} & &  \tikznode{g22}{0} & & \tikznode{g23}{0} & & \tikznode{g24}{0} & & \tikznode{h}{}\\ & & \tikznode{g31}{+} & &\tikznode{g32}{0} & & \tikznode{g33}{0} & & \tikznode{g34}{0} & & \tikznode{g35}{0}
    \end{pmatrix}
\end{equation*}
\begin{tikzpicture}[overlay,remember picture,line width=.7pt,transform canvas={yshift=0mm}]
        \draw[fcolor,->-] (g32)--(g31);
\end{tikzpicture} \vspace{-0.5cm}
 & \vspace{0.3cm}$\varphi_1\varphi_2\varphi_3$ & \vspace{-0.5cm}
\begin{equation*}
\begin{pmatrix}

        \tikznode{g11}{0} & & \tikznode{g12}{0} & & \tikznode{g13}{0} & & \tikznode{g14}{0} & & \tikznode{f1}{{\color{white}+}} & \tikznode{f}{}\\ & & \tikznode{g21}{0} & &  \tikznode{g22}{0} & & \tikznode{g23}{0} & & \tikznode{g24}{0} & & \tikznode{h}{}\\ & & \tikznode{g31}{+} & &\tikznode{g32}{+} & & \tikznode{g33}{0} & & \tikznode{g34}{0} & & \tikznode{g35}{0}
    \end{pmatrix}
\end{equation*}
\begin{tikzpicture}[overlay,remember picture,line width=.7pt,transform canvas={yshift=0mm}]
       \draw[fcolor,->-] (g21)--(g32);\draw[bcolor,->-] (g33)--(g32);
       \end{tikzpicture} \vspace{-0.5cm}
 &\vspace{0.3cm} $\varphi_1\varphi_2\psi$ \\
\hline
 
\end{longtable}
\end{center}

\subsection{Example: \texorpdfstring{$SU(2)_{-\frac{5}{2}}$ with $N_s=5$ scalars and $N_f=1$}{su24} fermion}
\begin{figure}[ht]
    \centerline{\includegraphics[width=1.1\textwidth]{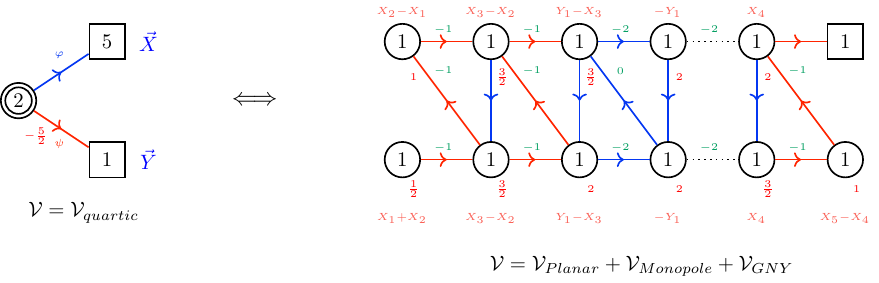}}
    \caption{The dual for $SU(2)_{-\frac{5}{2}}$ QCD$_3$ with 5 scalars and 1 fermion is shown here. All labelling conventions are the same as Figure \ref{fig:scalar_Qcd_spicy}, and BF couplings between nodes connected by a vertical scalar field have been suppressed for brevity.}
    \label{fig: SU(2) 5s1f}
\end{figure}

We now consider the example of $SU(2)_{-\frac{5}{2}}$ with $5$ scalars and $1$ fermion. Notice that this theory has $k=1$. We propose the duality shown in Figure \ref{fig: SU(2) 5s1f}.

We are mainly concerned with the mapping of $\binom{5}{2}=10$ spin-0 mesons, $5$ spin-$\frac{1}{2}$ mesons, and $\binom{5}{2}=10$ spin-0 baryons, $5$ spin-$\frac{1}{2}$ baryons and $1$ baryon with spin-1 channel. Since the details of the dressings are identical to the discussions preceding this section, we leave the complete operator map to the reader and provide the GNO fluxes of the bare monopoles are a representative example of the general trends of identifying these operators.

The mapping of the off-diagonal spin-0 mesons is provided in Equation \ref{eq: spin-0 mesons of SU(2) w5s1f} (the remaining mesons can be identified by charge conjugation):

\begin{equation}\label{eq: spin-0 mesons of SU(2) w5s1f}
    \begin{aligned}
        \mathbf{M}|_{\text{spin-0}} \leftrightarrow &\; \bigg\{ \mathfrak{M}^{\begin{pmatrix} + & 0 & 0 & 0 & 0 & \\ 0 & 0 & 0 & 0 & 0 & 0\end{pmatrix}},\, \mathfrak{M}^{\begin{pmatrix} + & + & 0 & 0 & 0 & \\ 0 & 0 & 0 & 0 & 0 & 0\end{pmatrix}},\, \mathfrak{M}^{\begin{pmatrix} + & + & + & + & + & \\ 0 & 0 & 0 & 0 & 0 & 0\end{pmatrix}},\,\mathfrak{M}^{\begin{pmatrix} + & + & + & + & + & \\ 0 & 0 & 0 & 0 & 0 & +\end{pmatrix}},\\ & \mathfrak{M}^{\begin{pmatrix} 0 & + & 0 & 0 & 0 & \\ 0 & 0 & 0 & 0 & 0 & 0\end{pmatrix}},\, \mathfrak{M}^{\begin{pmatrix} 0 & + & + & + & + & \\ 0 & 0 & 0 & 0 & 0 & 0\end{pmatrix}},\, \mathfrak{M}^{\begin{pmatrix} 0 & + & + & + & + & \\ 0 & 0 & 0 & 0 & 0 & +\end{pmatrix}},\, \mathfrak{M}^{\begin{pmatrix} 0 & 0 & + & + & + & \\ 0 & 0 & 0 & 0 & 0 & 0\end{pmatrix}},\\ & \mathfrak{M}^{\begin{pmatrix} 0 & 0 & + & + & + & \\ 0 & 0 & 0 & 0 & 0 & +\end{pmatrix}},\, \mathfrak{M}^{\begin{pmatrix} 0 & 0 & 0 & 0 & 0 & \\ 0 & 0 & 0 & 0 & 0 & +\end{pmatrix}}\bigg\}.
    \end{aligned}
\end{equation}
A generic spin-0 mesonic monopole operator will be dressed as follows\footnote{We indicate the presence of only BF couplings in the second example by keeping a dotted line explicit between the two nodes.}:
\begin{equation}
\mathfrak{M}^{\begin{pmatrix}

        \tikznode{g11}{0} & & \tikznode{g12}{+} & & \tikznode{g13}{0} & & \tikznode{g14}{0} & & \tikznode{g15}{0} && \tikznode{f1}{{\color{white}+}} \\  \tikznode{g21}{0} & &  \tikznode{g22}{0} & & \tikznode{g23}{0} & & \tikznode{g24}{0} & & \tikznode{g25}{0} && \tikznode{g26}{0}
    \end{pmatrix}},\,\, \mathfrak{M}^{\begin{pmatrix}

        \tikznode{gg11}{0} & & \tikznode{gg12}{0} & & \tikznode{gg13}{+} & & \tikznode{gg14}{+} & & \tikznode{gg15}{+} && \tikznode{ff1}{{\color{white}+}} \\  \tikznode{gg21}{0} & &  \tikznode{gg22}{0} & & \tikznode{gg23}{0} & & \tikznode{gg24}{0} & & \tikznode{gg25}{0} && \tikznode{gg26}{0}
    \end{pmatrix}}
\end{equation}
\begin{tikzpicture}[overlay,remember picture,line width=.7pt,transform canvas={yshift=0mm}]
        \draw[fcolor,->-] (g11)--(g12);
        \draw[fcolor,->-] (g13)--(g12);
        \draw[fcolor,->-] (g23)--(g12);

        \draw[dotted] (gg14)--(gg15);
        \draw[fcolor,->-] (gg12)--(gg13);
        \draw[fcolor,->-] (ff1)--(gg15);
        \draw[fcolor,->-] (gg26)--(gg15);
        
\end{tikzpicture}

The mapping of the off-diagonal spin-$\frac{1}{2}$ mesons is provided in Equation \ref{eq: spin-1/2 mesons of SU(2) w5s1f} (the remaining mesons can be identified by charge conjugation):

\begin{equation}\label{eq: spin-1/2 mesons of SU(2) w5s1f}
    \begin{aligned}
        \mathbf{M}|_{\text{spin-}\frac{1}{2}} \leftrightarrow &\; \bigg\{ \mathfrak{M}^{\begin{pmatrix} + & + & + & 0 & 0 & \\ 0 & 0 & 0 & 0 & 0 & 0\end{pmatrix}},\, \mathfrak{M}^{\begin{pmatrix} 0 & + & + & 0 & 0 & \\ 0 & 0 & 0 & 0 & 0 & 0\end{pmatrix}},\, \mathfrak{M}^{\begin{pmatrix} 0 & 0 & + & 0 & 0 & \\ 0 & 0 & 0 & 0 & 0 & 0\end{pmatrix}},\,\mathfrak{M}^{\begin{pmatrix} 0 & 0 & 0 & + & + & \\ 0 & 0 & 0 & 0 & 0 & 0\end{pmatrix}},\\ & \mathfrak{M}^{\begin{pmatrix} 0 & 0 & 0 & + & + & \\ 0 & 0 & 0 & 0 & 0 & +\end{pmatrix}}\bigg\}.
    \end{aligned}
\end{equation}
A generic spin-$\frac{1}{2}$ meson will be dressed as follows:
\begin{equation}
\mathfrak{M}^{\begin{pmatrix}

        \tikznode{g11}{0} & & \tikznode{g12}{0} & & \tikznode{g13}{+} & & \tikznode{g14}{0} & & \tikznode{g15}{0} && \tikznode{f1}{{\color{white}+}} \\  \tikznode{g21}{0} & &  \tikznode{g22}{0} & & \tikznode{g23}{0} & & \tikznode{g24}{0} & & \tikznode{g25}{0} && \tikznode{g26}{0}
    \end{pmatrix}},\,\, \mathfrak{M}^{\begin{pmatrix}

        \tikznode{gg11}{0} & & \tikznode{gg12}{0} & & \tikznode{gg13}{0} & & \tikznode{gg14}{+} & & \tikznode{gg15}{+} && \tikznode{ff1}{{\color{white}+}} \\  \tikznode{gg21}{0} & &  \tikznode{gg22}{0} & & \tikznode{gg23}{0} & & \tikznode{gg24}{0} & & \tikznode{gg25}{0} && \tikznode{gg26}{0}
    \end{pmatrix}}
\end{equation}
\begin{tikzpicture}[overlay,remember picture,line width=.7pt,transform canvas={yshift=0mm}]
        \draw[fcolor,->-] (g12)--(g13);
        \draw[bcolor,->-] (g14)--(g13);

        \draw[dotted] (gg14)--(gg15);
        \draw[bcolor,->-] (gg13)--(gg14);
        \draw[fcolor,->-] (ff1)--(gg15);
        \draw[fcolor,->-] (gg26)--(gg15);
        
\end{tikzpicture} 

The mapping of the spin-0 baryons is provided in Equation \ref{eq: spin-0 baryons of SU(2) w5s1f} (the remaining antibaryons can be identified by charge conjugation):

\begin{equation}\label{eq: spin-0 baryons of SU(2) w5s1f}
    \begin{aligned}
        \mathbf{B}|_{\text{spin-0}} \leftrightarrow &\; \bigg\{ \mathfrak{M}^{\begin{pmatrix} 0 & 0 & 0 & 0 & 0 & \\ + & 0 & 0 & 0 & 0 & 0\end{pmatrix}},\, \mathfrak{M}^{\begin{pmatrix} 0 & 0 & 0 & 0 & 0 & \\ + & + & 0 & 0 & 0 & 0\end{pmatrix}},\, \mathfrak{M}^{\begin{pmatrix} 0 & 0 & 0 & 0 & 0 & \\ + & + & + & + & + & 0\end{pmatrix}},\,\mathfrak{M}^{\begin{pmatrix} 0 & 0 & 0 & 0 & 0 & \\ + & + & + & + & + & +\end{pmatrix}},\\ & \mathfrak{M}^{\begin{pmatrix} + & 0 & 0 & 0 & 0 & \\ + & + & 0 & 0 & 0 & 0\end{pmatrix}},\,\mathfrak{M}^{\begin{pmatrix} + & 0 & 0 & 0 & 0 & \\ + & + & + & + & + & 0\end{pmatrix}},\,\mathfrak{M}^{\begin{pmatrix} + & + & 0 & 0 & 0 & \\ + & + & + & + & + & 0\end{pmatrix}},\,\mathfrak{M}^{\begin{pmatrix} + & + & + & + & + & \\ + & + & + & + & + & +\end{pmatrix}},\\ & \mathfrak{M}^{\begin{pmatrix} + &0 & 0 & 0 & 0 & \\ + & + & + & + & + & +\end{pmatrix}},\,\mathfrak{M}^{\begin{pmatrix} + &+ & 0 & 0 & 0 & \\ + & + & + & + & + & +\end{pmatrix}}\,\bigg\}.
    \end{aligned}
\end{equation}

A generic spin-0 baryon will be dressed as follows:
\begin{equation}
\mathfrak{M}^{\begin{pmatrix}

        \tikznode{g11}{+} & & \tikznode{g12}{0} & & \tikznode{g13}{0} & & \tikznode{g14}{0} & & \tikznode{g15}{0} && \tikznode{f1}{{\color{white}+}} \\  \tikznode{g21}{+} & &  \tikznode{g22}{+} & & \tikznode{g23}{0} & & \tikznode{g24}{0} & & \tikznode{g25}{0} && \tikznode{g26}{0}
    \end{pmatrix}},\,\, \mathfrak{M}^{\begin{pmatrix}

        \tikznode{gg11}{0} & & \tikznode{gg12}{0} & & \tikznode{gg13}{0} & & \tikznode{gg14}{0} & & \tikznode{gg15}{0} && \tikznode{ff1}{{\color{white}+}} \\  \tikznode{gg21}{+} & &  \tikznode{gg22}{+} & & \tikznode{gg23}{+} & & \tikznode{gg24}{+} & & \tikznode{gg25}{+} && \tikznode{gg26}{0}
    \end{pmatrix}}
\end{equation}
\begin{tikzpicture}[overlay,remember picture,line width=.7pt,transform canvas={yshift=0mm}]
        \draw[fcolor,->-] (g12)--(g11);
        \draw[fcolor,->-] (g23)--(g22);

        \draw[dotted] (gg24)--(gg25);
        \draw[fcolor,->-] (gg11)--(gg22);
        \draw[fcolor,->-] (gg12)--(gg23);
        \draw[fcolor,->-] (gg26)--(gg25);
        
\end{tikzpicture} 

The mapping of the spin-$\frac{1}{2}$ baryons is provided in Equation \ref{eq: spin-1/2 baryons of SU(2) w5s1f} (the remaining antibaryons can be identified by charge conjugation):

\begin{equation}\label{eq: spin-1/2 baryons of SU(2) w5s1f}
    \begin{aligned}
        \mathbf{B}|_{\text{spin-}\frac{1}{2}} \leftrightarrow &\; \bigg\{ \mathfrak{M}^{\begin{pmatrix} 0 & 0 & 0 & 0 & 0 & \\ + & + & + & 0 & 0 & 0\end{pmatrix}},\, \mathfrak{M}^{\begin{pmatrix} + & 0 & 0 & 0 & 0 & \\ + & + & + & 0 & 0 & 0\end{pmatrix}},\, \mathfrak{M}^{\begin{pmatrix} + & + & 0 & 0 & 0 & \\ + & + & + & 0 & 0 & 0\end{pmatrix}},\,\mathfrak{M}^{\begin{pmatrix} + & + & + & 0 & 0 & \\ + & + & + & + & + & 0\end{pmatrix}},\\ & \mathfrak{M}^{\begin{pmatrix} + & + & + & 0 & 0 & \\ + & + & + & + & + & +\end{pmatrix}}\,\bigg\}.
    \end{aligned}
\end{equation}
A generic spin-$\frac{1}{2}$ baryon will be dressed as follows:
\begin{equation}
\mathfrak{M}^{\begin{pmatrix}

        \tikznode{g11}{+} & & \tikznode{g12}{0} & & \tikznode{g13}{0} & & \tikznode{g14}{0} & & \tikznode{g15}{0} && \tikznode{f1}{{\color{white}+}} \\  \tikznode{g21}{+} & &  \tikznode{g22}{+} & & \tikznode{g23}{+} & & \tikznode{g24}{0} & & \tikznode{g25}{0} && \tikznode{g26}{0}
    \end{pmatrix}},\,\, \mathfrak{M}^{\begin{pmatrix}

        \tikznode{gg11}{+} & & \tikznode{gg12}{+} & & \tikznode{gg13}{+} & & \tikznode{gg14}{0} & & \tikznode{gg15}{0} && \tikznode{ff1}{{\color{white}+}} \\  \tikznode{gg21}{+} & &  \tikznode{gg22}{+} & & \tikznode{gg23}{+} & & \tikznode{gg24}{+} & & \tikznode{gg25}{+} && \tikznode{gg26}{0}
    \end{pmatrix}}
\end{equation}
\begin{tikzpicture}[overlay,remember picture,line width=.7pt,transform canvas={yshift=0mm}]
        \draw[fcolor,->-] (g12)--(g11);
        \draw[bcolor,->-] (g24)--(g23);

        \draw[dotted] (gg24)--(gg25);
        \draw[bcolor,->-] (gg14)--(gg13);
        \draw[fcolor,->-] (gg26)--(gg25);
        
\end{tikzpicture} 
Finally, the baryon $\psi^2$ obtained by symmetrizing the flavor and spin indices is mapped to the dressed monopole:
\begin{equation}
\mathfrak{M}^{\begin{pmatrix}

        \tikznode{g11}{+} & & \tikznode{g12}{+} & & \tikznode{g13}{+} & & \tikznode{g14}{0} & & \tikznode{g15}{0} && \tikznode{f1}{{\color{white}+}} \\  \tikznode{g21}{+} & &  \tikznode{g22}{+} & & \tikznode{g23}{+} & & \tikznode{g24}{0} & & \tikznode{g25}{0} && \tikznode{g26}{0}
    \end{pmatrix}}
\end{equation}
\begin{tikzpicture}[overlay,remember picture,line width=.7pt,transform canvas={yshift=0mm}]
        \draw[bcolor,->-] (g14)--(g13);
        \draw[bcolor,->-] (g24)--(g23);
        
\end{tikzpicture}

\section{Planar Abelian Duals of Unitary CS-QCD\texorpdfstring{$_3$}{3} with Bosons and Fermions}\label{sec: U(n)_b&f}
\begin{figure}[ht]
    \centerline{\includegraphics[width=1.3\textwidth]{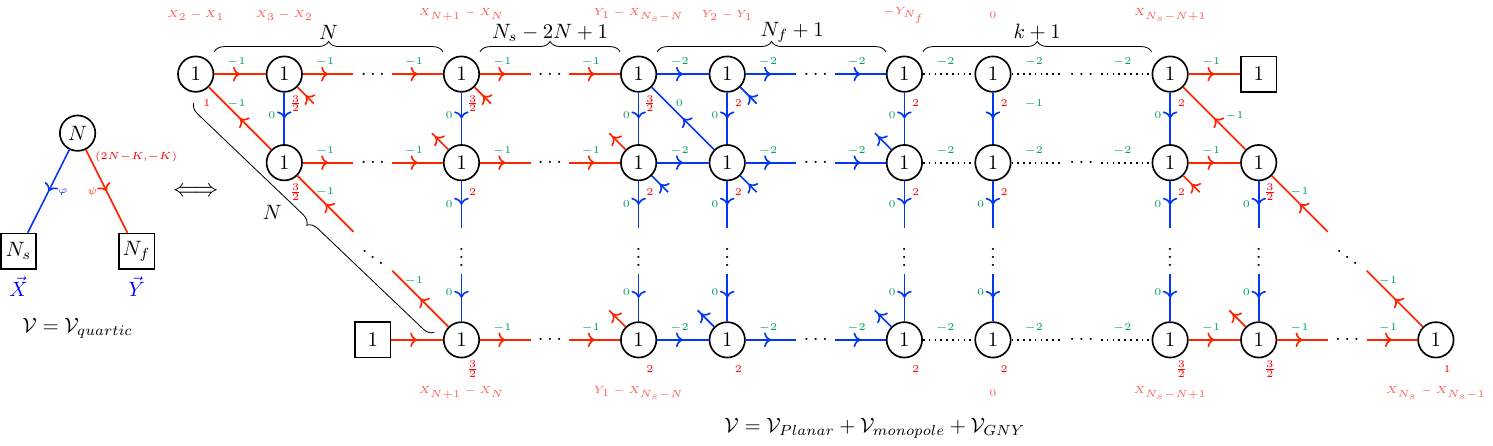}}
    \caption{The dual of $U(N)$ QCD$_3$ with $N_s$ scalars and $N_f$ fermions at Chern-Simons level $(2N-K,\ -K)$, where for clarity we define $K = N_s + \frac{N_f}{2} + k$, is shown here. All labelling conventions are the same as Figure \ref{fig:scalar_Qcd_spicy}.}
    \label{fig:U(N)_(Ns+Nf+k-2n)_l=2}
\end{figure}

We have so far focused on constructing duals of $SU(N)$ QCD$_3$ with mixed matter content. These results, however, extend naturally to the case of \textit{unitary} QCD$_3$ by gauging the baryonic symmetry. 
From the discussion in Section \ref{sec: non-Abelian}, we know that the baryonic symmetry of the $SU(N)$ QCD$_3$ maps to the topological symmetry of the bottom-left gauge node of the planar-abelian quiver (Figure \ref{fig:scalar_Qcd_general}). 
Gauging of this topological symmetry with zero CS level introduces a $U(1)_0$ sector with no charged matter whose path integral results in a delta-function, which effectively ungauges the $U(1)$ bottom-left gauge symmetry. 
While we did not keep track of background CS levels when presenting the duality in Figure \ref{fig:scalar_Qcd_general}, we expect the duality to involve a non-zero CS level for the baryonic symmetry.
In analogy with the SUSY result of \cite{Benvenuti:2024seb,Benvenuti:2025a},
we expect a background CS at level $-2$ for the baryonic symmetry on the QCD side of \ref{fig:scalar_Qcd_general}. 
Therefore
we propose that the dual description for $U(N)$ QCD$_3$ with $N_s$ critical scalars and $N_f$ fermions, where the theory is defined at Chern-Simons level $(2N-N_s-\frac{N_f}{2}-k, -N_s-\frac{N_f}{2}-k)$
\footnote{We adopt the standard notation:
\[
U(N)_{(k,\,k+\ell N)} \cong \frac{SU(N)_k \times U(1)_{N(k+\ell N)}}{\mathbb{Z}_N},
\]
which corresponds to the Chern-Simons action
\[
- \frac{i k}{4\pi} \int \text{tr}(A \wedge dA) - \frac{i \ell}{4\pi} \int \text{tr}(A) \wedge \text{tr}(dA).
\]} 
is the planar-abelian quiver depicted in Figure~\ref{fig:U(N)_(Ns+Nf+k-2n)_l=2}.

\begin{figure}[ht]
    \centering
    \centerline{\includegraphics[width=1.3\textwidth]{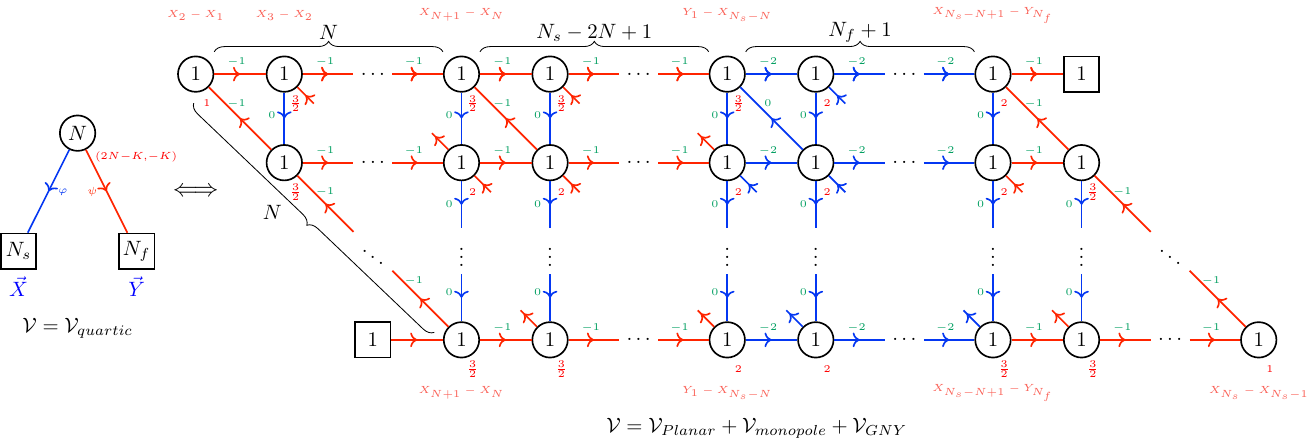}}
    \caption{The dual of $U(N)$ QCD$_3$ with $N_s$ scalars and $N_f$ fermions at Chern-Simons level $(2N-K,\ -K)$, with $K = N_s + \frac{N_f}{2}$, is shown here. All labeling conventions follow those of Figure~\ref{fig:scalar_Qcd_spicy}.}
    \label{fig:U(N)_(Ns+Nf-2n)_l=2}
\end{figure}

A particularly interesting special case arises when $k = 0$, shown in Figure~\ref{fig:U(N)_(Ns+Nf-2n)_l=2}. In this setting, the planar mirror theory contains a gauge-invariant mesonic operator built from bifundamental fields that connect the two $U(1)$ flavor nodes. This meson must map to a monopole operator in the electric $U(N)$ QCD$_3$ theory that is neutral under the $S(U(N_s)\times U(N_f))$ global symmetry.
Consider the bare monopole $\mon$ with GNO fluxes $(+,0,\dots,0)$ in the $U(N)_{2N-N_s-\tfrac{N_f}{2},\;-N_s-\tfrac{N_f}{2}}$ QCD$_3$ that transforms as the highest weight of the $(N_s-2N)$-index symmetric conjugate representation of $SU(N)$ and has charge $N_s$ under the diagonal $U(1)$. This monopole can be made gauge invariant by dressing with $N_s$ modes of the scalars $\elephi$, where two sets of $N$ scalars are antisymmetrized in gauge indices and the other $N_s-2N$ are symmetrized in gauge indices. 
By further symmetrizing all the scalars in $SU(N_s)$ indices we obtain operators that are neutral under $S(U(N_s)\times U(N_f))$ and have charge $-1$ under the topological symmetry, which are mapped to the long mesons described above. 
Therefore, we have the mapping:
\begin{equation}
\begin{array}{r}
\mon^{+}
\overbrace{
\partial_{\bullet}\elephi_{1}^{[1}
\partial_{\bullet}\elephi_{1}^{2}
\dots
\partial_{\bullet}\elephi_{1}^{N_s-2N}
}^{N_s-2N}
\;%
\overbrace{
\elephi_{1} ^{N_s-2N+1}
\elephi_{2} ^{N_s-2N+2}
\dots
\elephi_{N} ^{N_s-N}
}^{N}
\;%
\overbrace{
\elephi_{1} ^{N_s-N+1}
\elephi_{2} ^{N_s-N+2}
\dots
\elephi_{N} ^{N_s]}
}^{N}
\\
\dualto \text{long mesons}
\end{array}
\end{equation}
where lower indices are $SU(N)$ gauge indices and upper indices are $SU(N_s)$ flavor indices, which are completely antisymmetrized.
The modes with gauge index $1$ ``see" the GNO flux and have half-integer spin, while the other modes have spin-0. Then the spin of this operator mod $1$ is given by $\tfrac{1}{2}(N_s-2N +2) \text{ mod }1$. This matches the integrality of the spin of the long mesons, because each meson contains an even (odd) number of fermions if $N_s$ is even (odd).

Prior to gauging the baryonic $U(1)$ symmetry, one can also introduce a Chern-Simons term for it at level $\Delta_{\ell}$, following the discussion in~\cite{witten2003sl2zactionthreedimensionalconformal}. This leads to a $U(N)$ gauge theory with CS level $(2N-N_s-\frac{N_f}{2}-k,\,\Delta_{\ell}N-N_s-\frac{N_f}{2}-k)$. In the mirror dual, this corresponds to the addition of a new $U(1)_{\Delta_{\ell}}$ gauge node, coupled to the rest of the quiver via BF interactions. 


\newpage
\section{Mass Deformations and RG Flow Across Dualities}\label{sec:massdef}

In this section, we focus on studying the mapping of mass deformations across dualities, which serves as an important consistency check for our proposed dualities.
We consider mass deformations for the fundamental fields in the QCD theory and propose a corresponding deformation in the Abelian dual that is partly inspired by the structure of the SUSY dualities of \cite{Benvenuti:2025a,Benvenuti:2025b} and is a natural generalization of the Abelian case.

We also investigate the topological phases obtained by giving mass to all the charged fields in the QCD$_3$ theory and provide a consistent mapping of these deformations for some values of $N$, $k$ and the number of charged fields.

\subsection{Mass for a Single Field}
\label{checks}
A simple deformation that can be studied in the QCD$_3$ theory on the LHS of Figure \ref{fig:scalar_Qcd_spicy}, consists of giving a mass to a boson or to a fermion that can be either positive or negative. One could think of it as a linear potential term for a mesonic operator of the form $|\elephi|^2$ or $\bar{\psi}\psi$. In principle, knowing the map of these operators across the duality \ref{fig:scalar_Qcd_spicy} gives the knowledge of the deformation that is turned on in the dual theory. However, as we partially comment in Subsection \ref{sec: gen_opmap}, the map for these operators is not trivial\footnote{We recall that the reason behind the difficulty of this map is that these operators are not charged under the Cartan subgroup of the global symmetry. Indeed, in the dual theory, there are a large number of operators, carrying the correct spin, that are not charged under the global symmetry, and in principle any linear combination of them is a valid candidate as the target of the map.}. Partially inspired by the SUSY ancestor of the duality, for which the deformations will be analyzed in \cite{Benvenuti:2025b}, we consider examples to discuss massive deformations of the duality \ref{fig:scalar_Qcd_spicy}.

\subsubsection*{An Electric Scalar with Positive Mass}
\begin{figure}[ht]
   \centerline{\includegraphics[width=.9\textwidth]{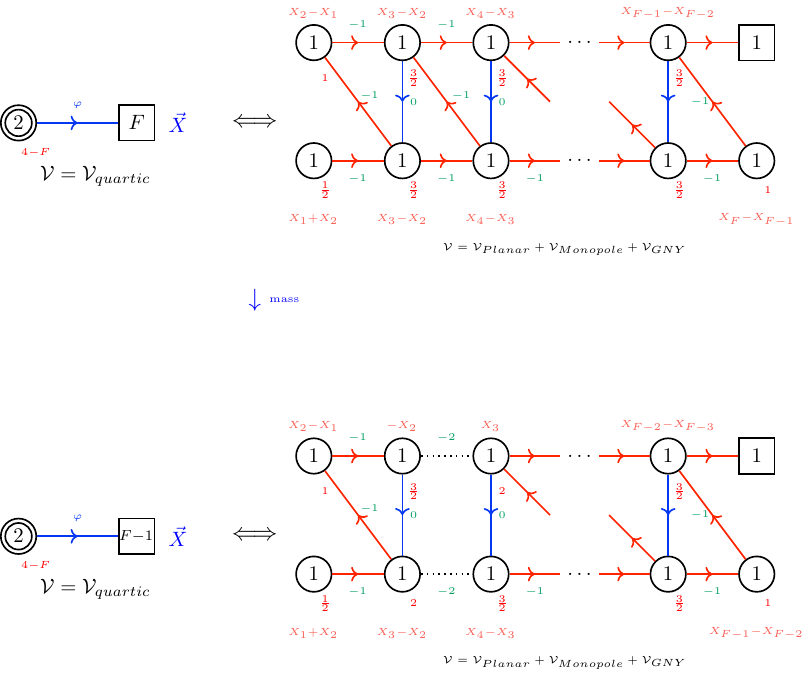}}
    \caption{Schematic depiction of the duality (bottom row) resulting from the RG flow induced by integrating out a single scalar field in $SU(2)_{4-F}$ QCD$_3$ with $F$ scalar fields and the corresponding fermionic column in its planar dual (top row). Notice that CS and BF interactions have been generated according to the sign of the corresponding fermion masses. The resulting duality is exactly what we propose for $SU(2)$ with $F-1$ scalars at CS level $4-F$.}
    \label{fig: SU(2)wFs_RGCheck}
\end{figure}

Concretely, consider \( SU(2) \) at CS level \( 4-F \) with \( F \) scalars and its planar mirror dual (top row of Figure \ref{fig: SU(2)wFs_RGCheck}). Suppose we turn on a positive mass term \( m^2 |\phi_3|^2 \) in the electric \( \text{QCD}_3 \) theory for one of the scalar fields ($\phi_3$). We claim that this deformation maps to real mass terms for the fermions in one of the columns of the dual theory as:
\begin{equation}\label{eq: mass_boson}
    m^2 |\phi_3|^2 \quad \leftrightarrow \quad m( \bar{\psi}_1 \psi_1 - \bar{\psi}_2 \psi_2 + \bar{\psi}_3 \psi_3) \,,
    \qquad 
    m>0
\end{equation}
where $\magpsi_1$ and $\magpsi_3$ are the horizontal fermions in the second column and $\magpsi_2$ is the diagonal fermion in the second column.
Upon integrating out the massive scalar field, the electric theory flows to \( SU(2) \) \( \text{QCD}_3 \) with \( F - 1 \) scalars at CS level \( 4 - F \). On the mirror side, we integrate out the fermions in the second column, leaving behind only BF couplings between the horizontal gauge nodes, as shown in the bottom row of Figure \ref{fig: SU(2)wFs_RGCheck}. The resulting duality is exactly the one we propose for $SU(N)$ with $F-1$ scalars at CS level $2N-k-(F-1)$, where $N=2$ and $k=1$ in this case.
Notice that this is a natural generalization of the Abelian case in Figure \ref{fig:Abelian_dual_bosons_fermions}, where giving a positive mass to a boson on the QED side corresponds to giving a positive mass to a fermion in the quiver side. 
The choice of signs for the fermionic masses in \eqref{eq: mass_boson} reproduces the BF couplings expected from the general proposal \ref{fig:scalar_Qcd_general}.

Immediate generalizations to electric gauge groups with higher rank follow from this observation. We reiterate that integrating out an electric scalar with a positive mass results in the integration of a column of fermions in the bulk of its planar mirror, leaving behind only horizontal BF couplings between gauge nodes.

One can immediately realize that the number of columns composed of fermions in the bulk of the the dual quiver is $N_s - 2N$, where $N_s$ is the number of scalars. Therefore we can use this strategy to integrate out only $N_s-2N$ bosons out of $N_s$. Therefore we can not use this strategy to flow to QCD$_3$ with $N_s < 2N$, which indeed is beyond the range of validity of the duality \ref{fig:scalar_Qcd_spicy}. 
In principle one could study the effect of further massive deformations to go beyond the range $N_s \geq 2N$, this is a quite non-trivial exercise and we leave it for a future work. Although, we propose examples of dualities with $N_s < 2N$ in Section \ref{sec: ns<2n}.

One could also consider a negative mass for an electric scalar, which would condense and Higgs the gauge group $SU(N)_{2N-N_s-k}\to SU(N-1)_{2(N-1)-(N_s-1)-k'}$, where $k' = k-1$. We stress that the map of the deformation in \eqref{eq: mass_boson} works only for $m > 0$.
It would be interesting to see how this may be seen in the planar mirror theory, but it is beyond the scope of this paper.

\subsubsection*{An Electric Fermion with Positive Mass}
\begin{figure}[ht]
   \centerline{\includegraphics[width=1.2\textwidth]{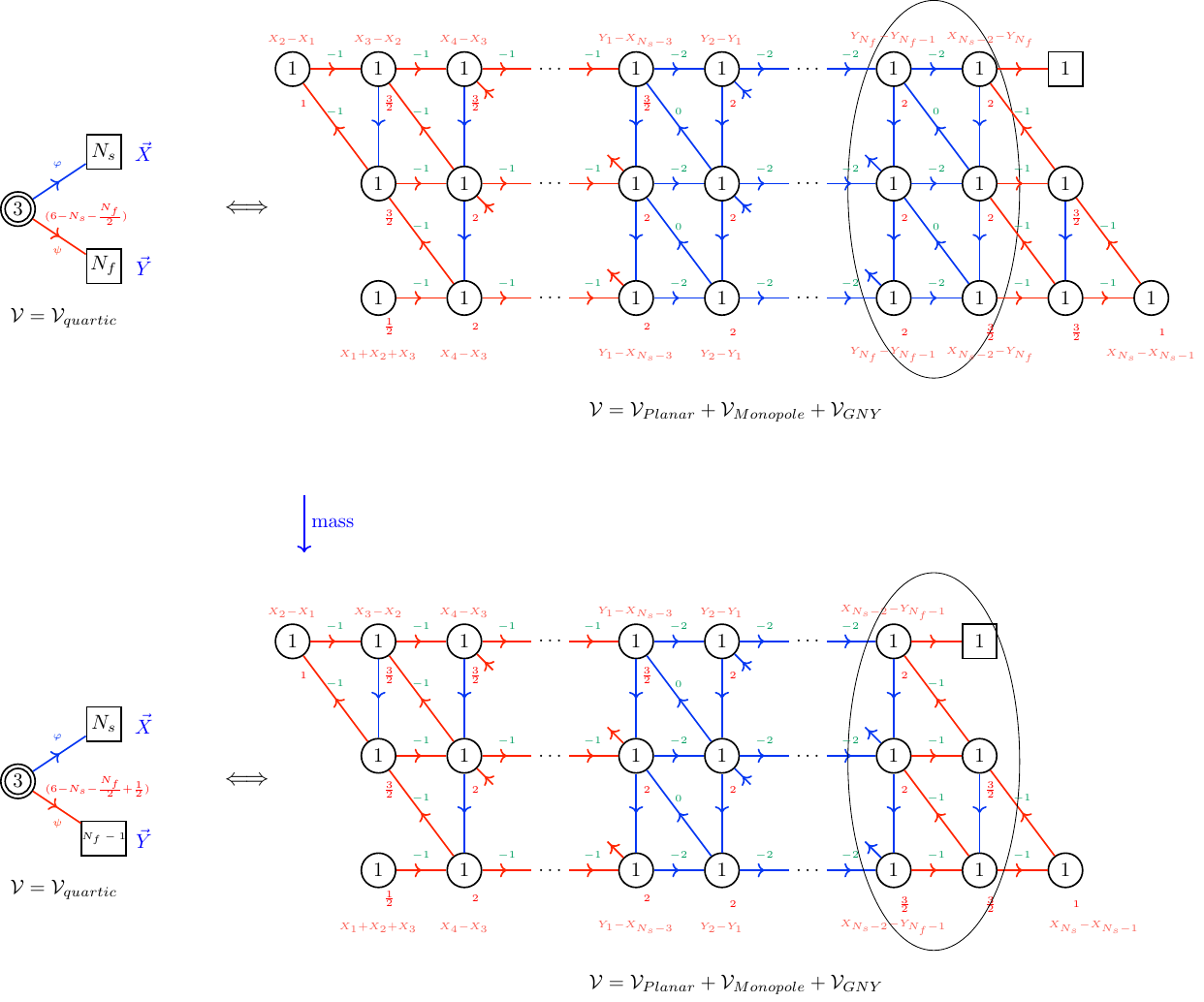}}
    \caption{Schematic depiction of the duality (bottom row) resulting from the RG flow induced by integrating out a single fermion with a positive mass in $SU(3)_{6-N_s-\frac{N_f}{2}}$ QCD$_3$ with $N_s$ scalar fields and $N_f$ fermionic fields and the corresponding bosonic column in its planar dual (top row).  The resulting duality is exactly what we propose for $SU(3)$ with $N_s$ scalars and $N_{F}-1$ fermions at CS level $6-N_s-\frac{N_f}{2}+\frac{1}{2}$.}
    \label{fig: SU(3)wFs_RGCheck2}
\end{figure}

 Concretely, consider $SU(3)$ at CS level $6-N_s-\frac{N_f}{2}$ with $N_s$ scalars and $N_f$ fermions, and its planar mirror dual (top row of Figure \ref{fig: SU(3)wFs_RGCheck2}). Suppose we turn on a positive mass term for a fermionic field $m\bar{\psi}_{N_f}\psi_{N_f}$. We claim this deformation maps to a negative mass term for the horizontal bosonic fields in a column of the mirror theory:
\begin{equation}
    m\bar{\psi}_{N_f}\psi_{N_F} \quad \leftrightarrow \quad - m^2 (|\phi_H^{(1)}|^2 +|\phi_H^{(2)}|^2 + |\phi_H^{(3)}|^2) \,.
\end{equation}

Upon integrating out the massive fermion, the electric theory flows to $SU(3)$ QCD$_3$ with $N_s$ scalars and $N_f-1$ fermions at CS level $6-N_s-\frac{N_f}{2}+\frac{1}{2}$. On the mirror side, the deformation cause the horizontal scalars to take a negative mass and thus a VEV, Higgsing pairs of $U(1)$ gauge groups connected by an horizontal scalar to the diagonal $U(1)$.
This effectively identifies the two columns of $U(1)$ gauge groups to the left and right of the column of scalars. Furthermore, some of the cubic interactions for the scalars turn into effective masses, which allow one to integrate out the $2$ diagonal scalars and $2$ of the vertical scalars. The resulting quiver is shown in the bottom row of Figure \ref{fig: SU(3)wFs_RGCheck2}. The resulting duality coincides with our proposed dual for $SU(N)$ with $N_s$ scalars and $N_f-1$ fermions at CS level $2N-N_s-\frac{(N_f-1)}{2}$, where $N=3$ in this case.

Immediate generalizations to electric gauge groups with higher rank follow from this observation. We reiterate that integrating out an electric fermion with a positive mass results in Higgsing the horizontal bosons in a column in the bulk of its planar mirror, thereby shortening the quiver. 
One can check that CS and BF coupling of the resulting theory agree with our proposal (Figure \ref{fig:scalar_Qcd_spicy}), providing a consistency check of our results.

\subsubsection*{An Electric Fermion with Negative Mass}
\begin{figure}[ht]
   \centerline{\includegraphics[width=1.2\textwidth]{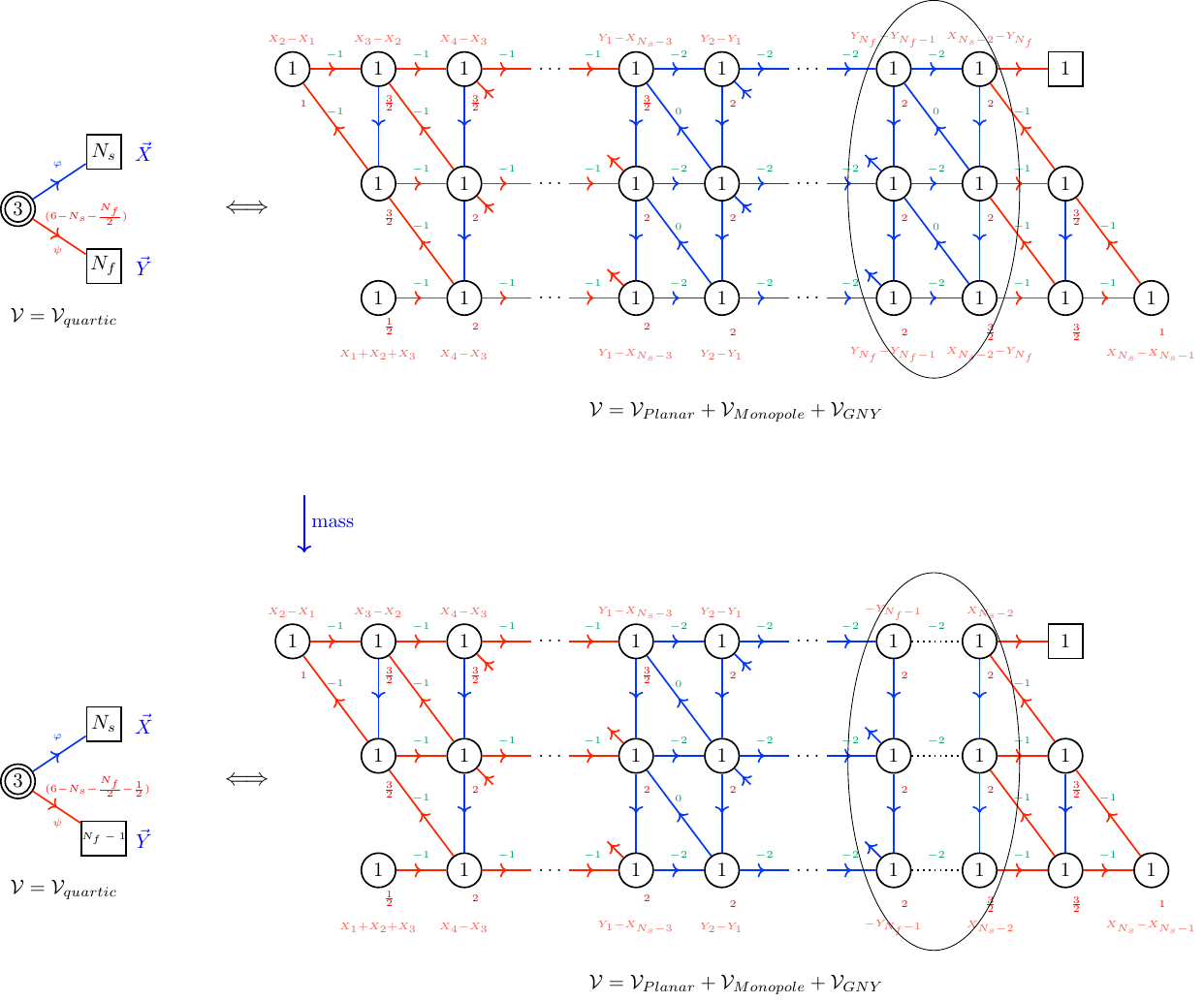}}
    \caption{Schematic depiction of the duality (bottom row) resulting from the RG flow induced by integrating out a single fermion with negative mass in $SU(3)_{6-N_s-\frac{N_f}{2}}$ QCD$_3$ with $N_s$ scalar fields and $N_f$ fermionic fields and the corresponding bosonic column in its planar dual (top row).  The resulting duality is exactly what we propose for $SU(3)$ with $N_s$ scalars and $N_{F}-1$ fermions at CS level $6-N_s-\frac{N_f}{2}-\frac{1}{2}$.}
    \label{fig: SU(3)wFs_RGCheck1}
\end{figure}

Concretely, consider $SU(3)$ at CS level $6-N_s-\frac{N_f}{2}$ with $N_s$ scalars and $N_f$ fermions, and its planar mirror dual (top row of Figure \ref{fig: SU(3)wFs_RGCheck1}). Suppose we turn on a negative mass term for a fermionic field $m\bar{\psi}_{N_f}\psi_{N_f}$. We claim this deformation maps to a positive mass term for the bosonic fields in a column of the mirror theory:
\begin{equation}
    -m\bar{\psi}_{N_f}\psi_{N_F} \quad \leftrightarrow \quad m^2(|\phi_{diag}^{(1)}|^2+|\phi_{diag}^{(2)}|^2+|\phi_H^{(1)}|^2 + |\phi_H^{(2)}|^2 + |\phi_H^{(3)}|^2) \,.
\end{equation}

Upon integrating out the massive fermion, the electric theory flows to $SU(3)$ QCD$_3$ with $N_s$ scalars and $N_f-1$ fermions at CS level $6-N_s-\frac{N_f}{2}-\frac{1}{2}$. On the mirror side, the same deformations leads to the integration of a column of bosons, leaving behind only BF couplings between horizontal gauge nodes, as shown in the bottom row of Figure \ref{fig: SU(3)wFs_RGCheck1}. The resulting duality is exactly the we propose for $SU(N)$ with $N_s$ scalars and $N_f-1$ fermions at CS level $2N-N_s-\frac{(N_f-1)}{2}-k$, where $N=3$ and $k=1$ in this case.

Immediate generalizations to electric gauge groups with higher rank follow from this observation. We reiterate that integrating out an electric fermion with a negative mass results in the integration of a column of bosons in the bulk of its planar mirror, leaving behind only BF couplings between horizontal gauge nodes.

We conclude by mentioning that in the duality \ref{fig:scalar_Qcd_spicy}, each fermion in the QCD$_3$ corresponds to a column of bosons in the dual theory. Thereforem the results summarized in Figure \ref{fig: SU(3)wFs_RGCheck2} and \ref{fig: SU(3)wFs_RGCheck1} can be used to study the effect on the dual theory of a mass for any number of fermions in the QCD$_3$.

\subsection{A consistency check via particle-vortex} 
\begin{figure}[ht]
\centerline{\includegraphics[width=.6\textwidth]{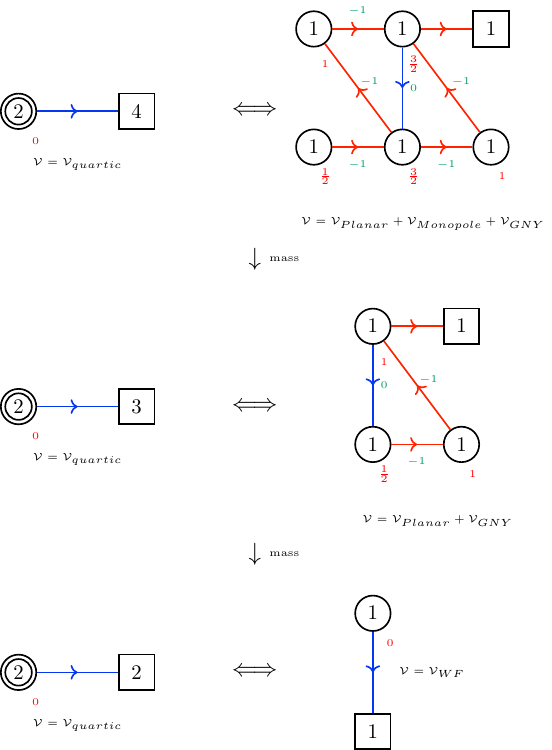}}
    \caption{Schematic depiction of the duality conjectured for $SU(2)_0$ with 2 scalars. We omit writing topological fugacities for brevity. As per our conventions, fermionic (bosonic) fields are shown in red (blue), and (mixed) Chern-Simons interactions in maroon (green).}
    \label{fig: Tetrality}
\end{figure}

Consider \( SU(2) \) at CS level \( 0 \) with \( 4 \) scalars and its planar mirror dual (top row of Figure \ref{fig: Tetrality}). Suppose we turn on a positive mass term \( m^2 |\phi_1|^2 \) in the electric \( \text{QCD}_3 \) theory for the scalar field \( \phi_1 \). We claim that this deformation maps to the mirror theory, resulting in the duality shown in the middle row of Figure \ref{fig: Tetrality}.

We can further deform the resulting QCD$_3$ by giving a mass also the $\phi_3$ scalar field, resulting in a $SU(2)_0$ QCD$_3$ with only two scalars. We claim that this deformation corresponds to giving mass to the remaining fermions, leading to the duality in the last row of Figure \ref{fig: Tetrality}. Notice that in this last step we actually remain with one scalar and two $U(1)$ gauge nodes, the diagonal combination decouples as a $U(1)_1$ which is an almost trivial theory.

At this point the $U(1)_0$ theory with one scalar can be also dualized using particle-vortex \cite{Einhorn:1977qv,Peskin:1977kp,Fisher:1989dnp, Savit:1979ny}, resulting in a \( O(2) \) Wilson-Fisher scalar. This in turn is also related, via bosonization, to a $U(1)_{\frac{1}{2}}$ theory with a fundamental fermion \eqref{eq:bosonization_compact}. The duality between $SU(2)_0$ with two scalars and $U(1)_{\frac{1}{2}}$ with one fermion, is consistent with a duality proposed in \cite{Benini:2017aed}.

\newpage

We can generalize this result further to $SU(N)_0$ with $N$ scalars by starting from $SU(N)_0$ with $2N$ scalars and integrating out $N$ of them, and find that $SU(N)_0$ with $N$ scalars is also dual to an $O(2)$ Wilson-Fisher scalar.

\subsection{TQFT Dualities from Mass Deformations} 
\label{tqfts}

Instead of considering mass deformations for a single boson or fermion, we can consider a mass deformation for all of them. In particular we could consider a positive mass for all the scalars and a homogeneous mass for the fermions in the QCD$_3$, which always leads to a TQFT, and track the effect in the dual theory that should flow to a dual TQFT to match the IR phase. A general analysis shows that the resulting TQFT duality obtained is quite complicated. 
A possible strategy is to consider gapped phases of the $\mathcal{N}=2$ theories involved in the ancestor duality \cite{Benvenuti:2025a}. 
In particular one can study the flow from the ancestor $\mathcal{N}=2$ duality to a TQFT duality which we believe is equivalent to the TQFT duality we would reach starting from our proposed dual pair.
The problem of studying RG flows in the SUSY case will be addressed in detail in \cite{Benvenuti:2025b}, here we anticipate the result in the following two cases, consisting in the SQCD theories whose Abelian planar dual contains one or zero “bulk” columns.

We start by discussing  the TQFT duality obtained starting from the duality for $SU(N)_{-1}$ with $N_s=2N$ scalars and zero fermions  \ref{fig:scalar_Qcd_general}. 
Turning on positive masses for all the scalars results in $SU(N)_{-1}$ CS theory.
The planar Abelian dual theory has a single empty bulk column, see Figure \ref{fig:scalar_Qcd_general}\footnote{The analysis discussed in this Section extends naturally to other QCD theories whose Abelian dual has a single “bulk" column.}.

As explained above we can consider a chiral-planar SUSY duality \cite{Benvenuti:2025a} where the electric theory flows to the same $SU(N)_{-1}$ TQFT. There the mass deformation can be followed on the planar Abelian side \cite{Benvenuti:2025b}, resulting in a $U(1)^2$ CS thery with CS matrix:

\begin{equation}    \label{eq:K_matrix}
    \mathbf{K} = \begin{pmatrix} N & -N \\ -N & N+1 \end{pmatrix};
\end{equation}

\begin{figure}[ht]
   \centerline{\includegraphics[width=.8\textwidth]{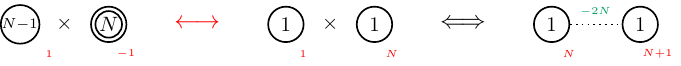}}
    \caption{Duality between \( SU(N)_{-1} \times U(N-1)_{1} \) CS theory and the Abelian CS theory associated to the CS matrix \eqref{eq:K_matrix}. (Mixed) Chern-Simons interactions are indicated in (green) red. 
    The red arrow indicates level/rank duality for $SU(N)_{-1}$, while the double arrow indicates a duality between Abelian TQFTs described in the main body.
    }
    \label{fig: TQFT1}
\end{figure}

This is the TQFT duality reported in quiver notation in Figure \ref{fig: TQFT1}.
We expect that this is the same TQFT duality one would obtain by deforming the non-SUSY duality for $SU(N)_{-1}$ with $N_s=2N$ scalars and zero fermions.
In the remainder of this Section we show that this duality can be proved via known TQFT dualities, providing a check for the flows discussed above.

The 1-form symmetry group of the Abelian CS theory associated to $\mathbf{K}$ is \( \mathbb{Z}^{(1)}_N \), as can be deduced from the Smith normal form of \( \mathbf{K} \) \cite{Delmastro:2019vnj}, matching the $SU(N)_{-1}$ on the electric side. 
The duality between the Abelian CS theory associated to \( \mathbf{K} \) and $SU(N)_{-1}$ can be proved by first applying level/rank duality (red arrow in \ref{fig: TQFT1}), and then relating the resulting Abelian theories (double arrow in \ref{fig: TQFT1}).
The last step requires us to find two \( 2 \times 2 \) matrices \( \mathbf{Q} \) and \( \mathbf{P} \) with integer coefficients, as described in \cite{Delmastro:2019vnj}, satisfying the condition
\begin{equation}
    \mathbf{Q}^{t} \mathbf{K}^{-1}_L \mathbf{Q} - \mathbf{K}^{-1} = \mathbf{P},
\end{equation}
where \( \mathbf{K}^{-1}_L = \text{diag}(1, \frac{1}{N}) \) and \( \mathbf{K}^{-1} \) are the inverse Chern-Simons matrices of the two dual theories, respectively. The following matrices satisfy these requirements:
\begin{equation}
    \mathbf{Q} = \begin{pmatrix} -1 & 1 \\ 1 & 0 \end{pmatrix}, \quad \mathbf{P} = \begin{pmatrix} 0 & -2 \\ -2 & 0 \end{pmatrix},
\end{equation}
where the diagonal elements of \( \mathbf{P} \) are even, as required for spin preservation. 
The matrix \( \mathbf{Q} \) is the (left-)invertible matrix specifying the bijective mapping of the \( 2|\text{det}(\mathbf{K})| \) independent Wilson lines, from the dual side to the electric side, i.e., \( \vec{\alpha}_L = \mathbf{Q} \vec{\alpha} \), where \( \vec{\alpha}_L \) and \( \vec{\alpha} \) are the charge vectors of the Wilson line on the electric and dual side, respectively.
Notice that the sequence of dualities depicted in \ref{fig: TQFT1} requires an additional SPT $U(N-1)_1$ on the LHS. 
It would be interesting to keep track of the invertible backgrounds generated by the deformation of the quiver theory considered here in order to determine the precise STP factors involved in the duality for $SU(N)_{-1}$ with $2N+1$ scalars, but we leave this analsis to future work.
\\

Similarly one can consider the case of $SU(N)_{0}$ QCD with $N_s=2N$ scalars. By giving positive masses to all the scalars the theory flows to $SU(N)_{0}$ YM.
The analysis of the flow to the TQFT from the corresponding SUSY theory, together with its planar Abelian dual, predicts that the dual flows to $U(1)_1$, which is an almost trivial TQFT matching the electric side.

\newpage

\section{Towards the \texorpdfstring{$N_s<2N$ }{nsn} Regime}\label{sec: ns<2n}
In this section, we attempt to study theories where the number of scalars $N_s$ is less than $2N$. We find that we are able to consistently \textit{bosonize} two columns of height $N-1$ outside of the bulk of the quiver gauge theory, which means that in the duality \ref{fig:scalar_Qcd_spicy} we replace the tallest column in the two triangular sections with a column made only by bosons. Each replacement of columns using this strategy consists of decreasing/increasing the number of bosons/fermions by 1 in duality \ref{fig:scalar_Qcd_spicy}. We stress that this trick can be used to reduce at most the number of bosons by 2.

This strategy allows us to propose a quiver which is dual to a QCD$_3$ theory with $N_s=2N-1$ or $N_s = 2N-2$ scalars and $N_f$ fermions such that $N_s + N_f \geq 2N$, generalizing the dualities discussed in Section \ref{sec: qcd_b&f}. 
Extending this procedure to even lower $N_s$ by considering quivers with more bosonic columns in the outer triangle would require a generalization of our prescription to construct monopole interactions. 
We leave this to future work.

We illustrate this proposal using two examples, one with $N_s = 2N-1$ and one with $N_s = 2N-2$. In particular, we notice that the $SU(2)$ QCD$_3$ when $N_s=2$ and $N_f \geq 2$ is dual to a completely bosonic planar theory.


\subsection{Example: \texorpdfstring{$SU(3)_{0}$ QCD$_3$ with $N_s=5$ scalars and $N_f=2$}{su33} fermions}\label{sec: SU(3)_0 w N_s=5}

\begin{figure}[ht]
    \centerline{\includegraphics[width=.9\textwidth]{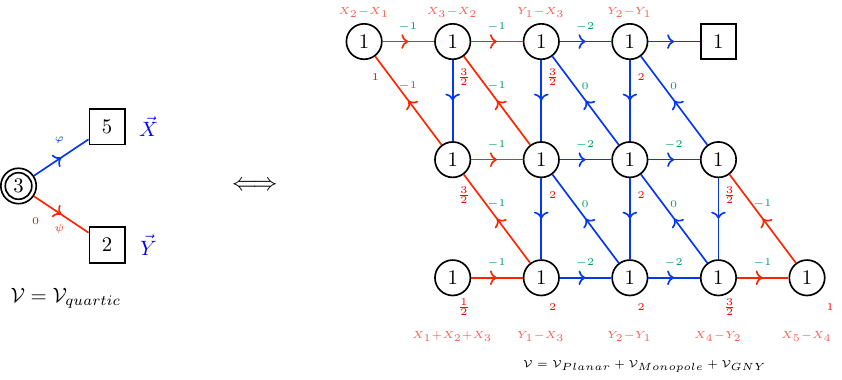}}
    \caption{The dual for $SU(3)_0$ QCD$_3$ with 5 scalars and 2 fermions is shown here. All labelling conventions are the same as Figure \ref{fig:scalar_Qcd_spicy}, and BF couplings between nodes connected by a vertical scalar field have been suppressed for brevity.}
    \label{fig: SU(3)_0 5s2fs}
\end{figure}

We consider the example of $SU(3)_{0}$ with 5 scalars and 2 fermions, and its planar dual description. We propose the duality shown in Figure \ref{fig: SU(3)_0 5s2fs}. Notice that the second-to-last column has been ``bosonized" in the sense that it is made only of scalar fields, similar to the column in the bulk.

Following the spirit of Sections \ref{sec: non-Abelian} and \ref{sec: qcd_b&f} we propose a map between mesons and baryons with monopole operators, which is reported in Tables \ref{table: mesons of SU(3)_0 QCD3} and \ref{table: baryons of SU(3)_0 QCD3} respectively. 
We reiterate that the monopoles reported there are representatives of all monopole operators with given global charge and spin.

\begin{center}
\renewcommand{\arraystretch}{0.6}
\begin{longtable}{ |>{\centering\arraybackslash}p{4cm}|>{\centering\arraybackslash}p{1.5cm}|>{\centering\arraybackslash}p{4cm}|>{\centering\arraybackslash}p{1.5cm}|  }
\caption{Dressed monopole operators of the planar mirror (Figure \ref{fig: SU(3)_0 5s2fs}) and their putative mapping to mesonic operators of the electric theory. The remaining mesonic operators can be constructed by charge conjugation.}
\label{table: mesons of SU(3)_0 QCD3} \\ 

\hline \multicolumn{1}{|c|}{\textbf{GNO Flux}} & \multicolumn{1}{c|}{\textbf{Meson}} & \multicolumn{1}{c|}{\textbf{GNO Flux}} & \multicolumn{1}{c|}{\textbf{Meson}} \\ \hline 
\endfirsthead

\multicolumn{4}{c}%
{{ \textbf{\tablename\ \thetable{}} -- \text{continued from previous page}}} \\
\hline \multicolumn{1}{|c|}{\textbf{GNO Flux }} & \multicolumn{1}{c|}{\textbf{Meson}} & \multicolumn{1}{c|}{\textbf{GNO Flux}} & \multicolumn{1}{c|}{\textbf{Meson}} \\ \hline 
\endhead

\hline \multicolumn{4}{|r|}{{\textit{Continued on next page}}} \\ \hline
\endfoot

\hline \hline
\endlastfoot
\vspace{-0.5cm}
\begin{equation*}
\begin{pmatrix}

        \tikznode{g11}{+} & & \tikznode{g12}{0} & & \tikznode{g13}{0} & & \tikznode{g14}{0} & & \tikznode{f1}{{\color{white}+}} & \tikznode{f}{}\\ & & \tikznode{g21}{0} & &  \tikznode{g22}{0} & & \tikznode{g23}{0} & & \tikznode{g24}{0} & & \tikznode{h}{}\\ & & \tikznode{g31}{0} & &\tikznode{g32}{0} & & \tikznode{g33}{0} & & \tikznode{g34}{0} & & \tikznode{g35}{0}
    \end{pmatrix}
\end{equation*}
\begin{tikzpicture}[overlay,remember picture,line width=.7pt,transform canvas={yshift=0mm}]
        \draw[fcolor,->-] (g12)--(g11);\draw[fcolor,->-] (g21)--(g11);
\end{tikzpicture} \vspace{-0.5cm}
 & \vspace{.3cm}$\varphi_2\bar{\varphi}_1$ & \vspace{-0.5cm}
 \begin{equation*}
\begin{pmatrix}

        \tikznode{g11}{+} & & \tikznode{g12}{+} & & \tikznode{g13}{0} & & \tikznode{g14}{0} & & \tikznode{f1}{{\color{white}+}} & \tikznode{f}{}\\ & & \tikznode{g21}{0} & &  \tikznode{g22}{0} & & \tikznode{g23}{0} & & \tikznode{g24}{0} & & \tikznode{h}{}\\ & & \tikznode{g31}{0} & &\tikznode{g32}{0} & & \tikznode{g33}{0} & & \tikznode{g34}{0} & & \tikznode{g35}{0}
    \end{pmatrix}
\end{equation*}
\begin{tikzpicture}[overlay,remember picture,line width=.7pt,transform canvas={yshift=0mm}]
       \draw[fcolor,->-] (g13)--(g12);\draw[fcolor,->-] (g22)--(g12);\draw[fcolor,->-] (g21)--(g11);
       \end{tikzpicture} \vspace{-0.5cm}
 & \vspace{0.3cm}$\varphi_3\bar{\varphi}_1$ \\
\hline
\end{longtable}
\end{center}

\begin{center}
\renewcommand{\arraystretch}{0.6}
\begin{longtable}{ |>{\centering\arraybackslash}p{4cm}|>{\centering\arraybackslash}p{1.5cm}|>{\centering\arraybackslash}p{4cm}|>{\centering\arraybackslash}p{1.5cm}|  }
\caption{Dressed monopole operators of the planar mirror (Figure \ref{fig: SU(3)_0 5s2fs}) and their putative mapping to baryonic operators of the electric theory (all indices have been suppressed). The remaining anti-baryonic operators can be constructed by charge conjugation.}
\label{table: baryons of SU(3)_0 QCD3} \\ 

\hline \multicolumn{1}{|c|}{\textbf{GNO Flux}} & \multicolumn{1}{c|}{\textbf{Baryon}} & \multicolumn{1}{c|}{\textbf{GNO Flux}} & \multicolumn{1}{c|}{\textbf{Baryon}} \\ \hline 
\endfirsthead

\multicolumn{4}{c}%
{{ \textbf{\tablename\ \thetable{}} -- \text{continued from previous page}}} \\
\hline \multicolumn{1}{|c|}{\textbf{GNO Flux }} & \multicolumn{1}{c|}{\textbf{Baryon}} & \multicolumn{1}{c|}{\textbf{GNO Flux}} & \multicolumn{1}{c|}{\textbf{Baryon}} \\ \hline 
\endhead

\hline \multicolumn{4}{|r|}{{\textit{Continued on next page}}} \\ \hline
\endfoot

\hline \hline
\endlastfoot
\vspace{-0.5cm}
\begin{equation*}
\begin{pmatrix}

        \tikznode{g11}{0} & & \tikznode{g12}{0} & & \tikznode{g13}{0} & & \tikznode{g14}{0} & & \tikznode{f1}{{\color{white}+}} & \tikznode{f}{}\\ & & \tikznode{g21}{0} & &  \tikznode{g22}{0} & & \tikznode{g23}{0} & & \tikznode{g24}{0} & & \tikznode{h}{}\\ & & \tikznode{g31}{+} & &\tikznode{g32}{0} & & \tikznode{g33}{0} & & \tikznode{g34}{0} & & \tikznode{g35}{0}
    \end{pmatrix}
\end{equation*}
\begin{tikzpicture}[overlay,remember picture,line width=.7pt,transform canvas={yshift=0mm}]
        \draw[fcolor,->-] (g32)--(g31);
\end{tikzpicture} \vspace{-0.5cm}
 & \vspace{0.3cm}$\varphi_1\varphi_2\varphi_3$ & \vspace{-0.5cm}
\begin{equation*}
\begin{pmatrix}

        \tikznode{g11}{0} & & \tikznode{g12}{0} & & \tikznode{g13}{0} & & \tikznode{g14}{0} & & \tikznode{f1}{{\color{white}+}} & \tikznode{f}{}\\ & & \tikznode{g21}{0} & &  \tikznode{g22}{0} & & \tikznode{g23}{0} & & \tikznode{g24}{0} & & \tikznode{h}{}\\ & & \tikznode{g31}{+} & &\tikznode{g32}{+} & & \tikznode{g33}{0} & & \tikznode{g34}{0} & & \tikznode{g35}{0}
    \end{pmatrix}
\end{equation*}
\begin{tikzpicture}[overlay,remember picture,line width=.7pt,transform canvas={yshift=0mm}]
       \draw[fcolor,->-] (g21)--(g32);\draw[bcolor,->-] (g33)--(g32);
       \end{tikzpicture} \vspace{-0.5cm}
 & \vspace{0.3cm}$\varphi_1\varphi_2\psi_1$ \\
\hline
 
\end{longtable}
\end{center}

\subsection{Example: \texorpdfstring{$SU(2)_{\frac{1}{2}}$ QCD$_3$ with $N_s=2$ scalars and $N_f = 3$}{su25} fermions} 
\label{sec: SU(2)_(-1/2) QCD}

\begin{figure}[ht]
    \centerline{\includegraphics[width=.9\textwidth]{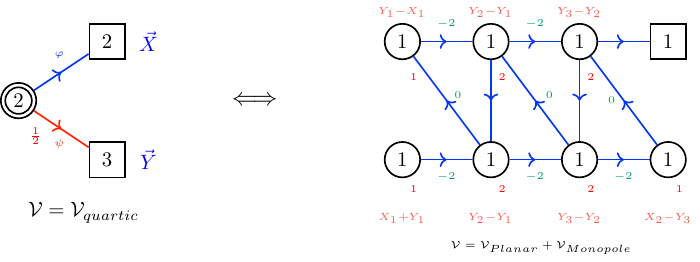}}
    \caption{The dual for $SU(2)_{\frac{1}{2}}$ QCD$_3$ with 2 scalars and 3 fermions is shown here. All labelling conventions are the same as Figure \ref{fig:scalar_Qcd_spicy}, and BF couplings between nodes connected by a vertical scalar field have been suppressed for brevity.}
    \label{fig: SU(2)_w2s3f}
\end{figure}

We finally consider the example of $SU(2)_{\frac{1}{2}}$ with $2$ scalars and $3$ fermions, and its planar dual description. We propose the duality shown in Figure \ref{fig: SU(2)_w2s3f}. Notice that the planar dual is completely bosonic. 

As before, we are concerned with determining the monopole operators dual to the mesons and baryons, and we report the results in Tables \ref{table: mesons of SU(2)_1/2 QCD3} and \ref{table: baryons of SU(2)_1/2 QCD3} respectively.

\begin{center}
\renewcommand{\arraystretch}{0.6}
\begin{longtable}{ |>{\centering\arraybackslash}p{4cm}|>{\centering\arraybackslash}p{1.5cm}|>{\centering\arraybackslash}p{4cm}|>{\centering\arraybackslash}p{1.5cm}|  }
\caption{Dressed monopole operators of the planar mirror (Figure \ref{fig: SU(2)_w2s3f}) and their putative mapping to mesonic operators of the electric theory. The remaining mesonic operators can be constructed by charge conjugation.}
\label{table: mesons of SU(2)_1/2 QCD3} \\ 

\hline \multicolumn{1}{|c|}{\textbf{GNO Flux}} & \multicolumn{1}{c|}{\textbf{Meson}} & \multicolumn{1}{c|}{\textbf{GNO Flux}} & \multicolumn{1}{c|}{\textbf{Meson}} \\ \hline 
\endfirsthead

\multicolumn{4}{c}%
{{ \textbf{\tablename\ \thetable{}} -- \text{continued from previous page}}} \\
\hline \multicolumn{1}{|c|}{\textbf{GNO Flux }} & \multicolumn{1}{c|}{\textbf{Meson}} & \multicolumn{1}{c|}{\textbf{GNO Flux}} & \multicolumn{1}{c|}{\textbf{Meson}} \\ \hline 
\endhead

\hline \multicolumn{4}{|r|}{{\textit{Continued on next page}}} \\ \hline
\endfoot

\hline \hline
\endlastfoot
\vspace{-.5cm}
\begin{equation*}
\begin{pmatrix}
        \tikznode{g11}{+} & & \tikznode{g12}{0} & & \tikznode{g13}{0} & & \tikznode{f1}{{\color{white}0}} \\ 0 & & \tikznode{g21}{0} & & \tikznode{g22}{0} & &\tikznode{g23}{0} 
    \end{pmatrix}
\end{equation*}
\begin{tikzpicture}[overlay,remember picture,line width=.7pt,transform canvas={yshift=0mm}]
	\draw[bcolor,-<-] (g11)--(g12);
\end{tikzpicture} 
\vspace{-.5cm}
& \vspace{.3cm}$\psi_1\bar{\varphi}_1$ & 
\vspace{-.5cm}
\begin{equation*}
\begin{pmatrix}
        \tikznode{g11}{+} & & \tikznode{g12}{+} & & \tikznode{g13}{0} & & \tikznode{f1}{{\color{white}0}} \\ 0 & & \tikznode{g21}{0} & & \tikznode{g22}{0} & &\tikznode{g23}{0} 
    \end{pmatrix}
\end{equation*}
\begin{tikzpicture}[overlay,remember picture,line width=.7pt,transform canvas={yshift=0mm}]
        \draw[bcolor,-<-] (g12)--(g13);
\end{tikzpicture} 
\vspace{-.5cm}
 & \vspace{.3cm}$\psi_2\bar{\varphi}_1$ \\ \hline
\vspace{-.5cm}
\begin{equation*}
\begin{pmatrix}
       \tikznode{g11}{+} & & \tikznode{g12}{+} & & \tikznode{g13}{+} & & \tikznode{f1}{{\color{white}0}} \\ 0 & & \tikznode{g21}{0} & & \tikznode{g22}{0} & &\tikznode{g23}{0} 
    \end{pmatrix}
\end{equation*}
\begin{tikzpicture}[overlay,remember picture,line width=.7pt,transform canvas={yshift=0mm}]
        \draw[bcolor,-<-] (g13)--(f1);
\end{tikzpicture} \vspace{-.5cm}
 & \vspace{.3cm}$\psi_3\bar{\varphi}_1$ &
\vspace{-.5cm}
\begin{equation*}
\begin{pmatrix}
       \tikznode{g11}{+} & & \tikznode{g12}{+} & & \tikznode{g13}{+} & & \tikznode{f1}{{\color{white}0}} \\ 0 & & \tikznode{g21}{0} & & \tikznode{g22}{0} & &\tikznode{g23}{+} 
    \end{pmatrix}
\end{equation*}
\begin{tikzpicture}[overlay,remember picture,line width=.7pt,transform canvas={yshift=0mm}]
        \draw[bcolor,-<-] (g13)--(f1);
        \draw[bcolor,-<-] (g23)--(g22);
\end{tikzpicture} \vspace{-.5cm}
 & \vspace{.3cm}$\varphi_2\bar{\varphi}_1$  \\ \hline
\vspace{-.5cm}
 \begin{equation*}
\begin{pmatrix}
       \tikznode{g11}{0} & & \tikznode{g12}{+} & & \tikznode{g13}{0} & & \tikznode{f1}{{\color{white}0}} \\ 0 & & \tikznode{g21}{0} & & \tikznode{g22}{0} & &\tikznode{g23}{0} 
    \end{pmatrix}
\end{equation*}
\begin{tikzpicture}[overlay,remember picture,line width=.7pt,transform canvas={yshift=0mm}]
	\draw[bcolor,-<-] (g12)--(g11);
        \draw[bcolor,-<-] (g12)--(g13);
\end{tikzpicture} \vspace{-.5cm}
 & \vspace{.3cm}$\psi_2\bar{\psi}_1$ &\vspace{-.5cm}
\begin{equation*}
\begin{pmatrix}
       \tikznode{g11}{0} & & \tikznode{g12}{+} & & \tikznode{g13}{+} & & \tikznode{f1}{{\color{white}0}} \\ 0 & & \tikznode{g21}{0} & & \tikznode{g22}{0} & &\tikznode{g23}{0} 
    \end{pmatrix}
\end{equation*}
\begin{tikzpicture}[overlay,remember picture,line width=.7pt,transform canvas={yshift=0mm}]
	\draw[bcolor,-<-] (g12)--(g11);
    
        \draw[bcolor,-<-] (g13)--(f1);
\end{tikzpicture} \vspace{-.5cm}
 & \vspace{.3cm}$\psi_3\bar{\psi}_1$ \\ \hline
\vspace{-.5cm}
 \begin{equation*}
\begin{pmatrix}
       \tikznode{g11}{0} & & \tikznode{g12}{0} & & \tikznode{g13}{0} & & \tikznode{f1}{{\color{white}0}} \\ \tikznode{gg}{0} & & \tikznode{g21}{+} & & \tikznode{g22}{+} & &\tikznode{g23}{+} 
    \end{pmatrix}
\end{equation*}
\begin{tikzpicture}[overlay,remember picture,line width=.7pt,transform canvas={yshift=0mm}]
	\draw[bcolor,-<-] (g21)--(gg);
\end{tikzpicture} \vspace{-.5cm}
 & \vspace{.3cm}$\varphi_2\bar{\psi}_1$ &
 \begin{equation*}
\begin{pmatrix}
       \tikznode{g11}{0} & & \tikznode{g12}{0} & & \tikznode{g13}{+} & & \tikznode{f1}{{\color{white}0}} \\ 0 & & \tikznode{g21}{0} & & \tikznode{g22}{0} & &\tikznode{g23}{0} 
    \end{pmatrix}
\end{equation*}
\begin{tikzpicture}[overlay,remember picture,line width=.7pt,transform canvas={yshift=0mm}]
	\draw[bcolor,-<-] (g13)--(g12);
        \draw[bcolor,-<-] (g13)--(f1);
\end{tikzpicture} \vspace{-.5cm}
 & \vspace{.3cm}$\psi_3\bar{\psi}_2$ \\ \hline
\vspace{-.5cm}
 \begin{equation*}
\begin{pmatrix}
      \tikznode{g11}{0} & & \tikznode{g12}{0} & & \tikznode{g13}{0} & & \tikznode{f1}{{\color{white}0}} \\ 0 & & \tikznode{g21}{0} & & \tikznode{g22}{+} & &\tikznode{g23}{+} 
    \end{pmatrix}
\end{equation*}
\begin{tikzpicture}[overlay,remember picture,line width=.7pt,transform canvas={yshift=0mm}]
        \draw[bcolor,->-] (g21)--(g22);
\end{tikzpicture} \vspace{-.5cm}
 & \vspace{.3cm}$\varphi_2\bar{\psi}_2$ & 
\vspace{-.5cm}
\begin{equation*}
\begin{pmatrix}
       \tikznode{g11}{0} & & \tikznode{g12}{0} & & \tikznode{g13}{0} & & \tikznode{f1}{{\color{white}0}} \\ 0 & & \tikznode{g21}{0} & & \tikznode{g22}{0} & &\tikznode{g23}{+} 
    \end{pmatrix}
\end{equation*}
\begin{tikzpicture}[overlay,remember picture,line width=.7pt,transform canvas={yshift=0mm}]
        \draw[bcolor,-<-] (g23)--(g22);
\end{tikzpicture} \vspace{-.5cm}
 & \vspace{.3cm}$\varphi_2\bar{\psi}_3$ \\ \hline
\end{longtable}
\end{center}

\begin{center}
\renewcommand{\arraystretch}{0.6}
\begin{longtable}{ |>{\centering\arraybackslash}p{4cm}|>{\centering\arraybackslash}p{1.5cm}|>{\centering\arraybackslash}p{4cm}|>{\centering\arraybackslash}p{1.5cm}|  }
\caption{Dressed monopole operators of the planar mirror (Figure \ref{fig: SU(2)_w2s3f}) and their putative mapping to baryonic operators of the electric theory. The remaining anti-baryonic operators can be constructed by charge conjugation.}
\label{table: baryons of SU(2)_1/2 QCD3} \\ 

\hline \multicolumn{1}{|c|}{\textbf{GNO Flux}} & \multicolumn{1}{c|}{\textbf{Baryon}} & \multicolumn{1}{c|}{\textbf{GNO Flux}} & \multicolumn{1}{c|}{\textbf{Baryon}} \\ \hline 
\endfirsthead

\multicolumn{4}{c}%
{{ \textbf{\tablename\ \thetable{}} -- \text{continued from previous page}}} \\
\hline \multicolumn{1}{|c|}{\textbf{GNO Flux }} & \multicolumn{1}{c|}{\textbf{Baryon}} & \multicolumn{1}{c|}{\textbf{GNO Flux}} & \multicolumn{1}{c|}{\textbf{Baryon}} \\ \hline 
\endhead

\hline \multicolumn{4}{|r|}{{\textit{Continued on next page}}} \\ \hline
\endfoot

\hline \hline
\endlastfoot
\vspace{-.5cm}
\begin{equation*}
\begin{pmatrix}
        \tikznode{g11}{0} & & \tikznode{g12}{0} & & \tikznode{g13}{0} & & \tikznode{f1}{{\color{white}0}} \\ \tikznode{g21}{+} & & \tikznode{g22}{0} & & \tikznode{g23}{0} & &\tikznode{g24}{0} 
    \end{pmatrix}
\end{equation*}
\begin{tikzpicture}[overlay,remember picture,line width=.7pt,transform canvas={yshift=0mm}]
	\draw[bcolor,->-] (g22)--(g21);
\end{tikzpicture} 
\vspace{-.5cm}
& 
\vspace{.3cm}$\varphi_1\psi_1$ & 
\vspace{-.5cm}
\begin{equation*}
\begin{pmatrix}
        \tikznode{g11}{0} & & \tikznode{g12}{0} & & \tikznode{g13}{0} & & \tikznode{f1}{{\color{white}0}} \\ \tikznode{g21}{+} & & \tikznode{g22}{+} & & \tikznode{g23}{0} & &\tikznode{g24}{0} 
    \end{pmatrix}
\end{equation*}
\begin{tikzpicture}[overlay,remember picture,line width=.7pt,transform canvas={yshift=0mm}]
	\draw[bcolor,->-] (g23)--(g22);
\end{tikzpicture} 
\vspace{-.5cm}
& 
\vspace{.3cm}$\varphi_1\psi_2$ \\ \hline
\vspace{-.5cm}
\begin{equation*}
\begin{pmatrix}
        \tikznode{g11}{0} & & \tikznode{g12}{0} & & \tikznode{g13}{0} & & \tikznode{f1}{{\color{white}0}} \\ \tikznode{g21}{+} & & \tikznode{g22}{+} & & \tikznode{g23}{+} & &\tikznode{g24}{0} 
    \end{pmatrix}
\end{equation*}
\begin{tikzpicture}[overlay,remember picture,line width=.7pt,transform canvas={yshift=0mm}]
	\draw[bcolor,->-] (g24)--(g23);
\end{tikzpicture} 
\vspace{-.5cm}
& \vspace{.3cm}$\varphi_1\psi_3$ & 
\vspace{-.5cm}
\begin{equation*}
\begin{pmatrix}
        \tikznode{g11}{0} & & \tikznode{g12}{0} & & \tikznode{g13}{0} & & \tikznode{f1}{{\color{white}0}} \\ \tikznode{g21}{+} & & \tikznode{g22}{+} & & \tikznode{g23}{+} & &\tikznode{g24}{+} 
    \end{pmatrix}
\end{equation*}
\vspace{-.5cm}
& \vspace{.3cm}$\varphi_1\varphi_2$ \\ \hline
\vspace{-.5cm}
\begin{equation*}
\begin{pmatrix}
        \tikznode{g11}{+} & & \tikznode{g12}{0} & & \tikznode{g13}{0} & & \tikznode{f1}{{\color{white}0}} \\ \tikznode{g21}{+} & & \tikznode{g22}{+} & & \tikznode{g23}{0} & &\tikznode{g24}{0} 
    \end{pmatrix}
\end{equation*}
\begin{tikzpicture}[overlay,remember picture,line width=.7pt,transform canvas={yshift=0mm}]
	\draw[bcolor,->-] (g23)--(g22);
        \draw[bcolor,->-] (g12)--(g11);
\end{tikzpicture} 
\vspace{-.5cm}
& \vspace{.3cm}$\psi_1\psi_2$ & 
\begin{equation*}
\begin{pmatrix}
        \tikznode{g11}{+} & & \tikznode{g12}{0} & & \tikznode{g13}{0} & & \tikznode{f1}{{\color{white}0}} \\ \tikznode{g21}{+} & & \tikznode{g22}{+} & & \tikznode{g23}{+} & &\tikznode{g24}{0} 
    \end{pmatrix}
\end{equation*}
\begin{tikzpicture}[overlay,remember picture,line width=.7pt,transform canvas={yshift=0mm}]
	\draw[bcolor,->-] (g24)--(g23);
    \draw[bcolor,->-] (g12)--(g11);
\end{tikzpicture} 
\vspace{-.5cm}
& \vspace{.3cm}$\psi_1\psi_3$ \\ \hline
\vspace{-.5cm}
\begin{equation*}
\begin{pmatrix}
        \tikznode{g11}{+} & & \tikznode{g12}{0} & & \tikznode{g13}{0} & & \tikznode{f1}{{\color{white}0}} \\ \tikznode{g21}{+} & & \tikznode{g22}{+} & & \tikznode{g23}{+} & &\tikznode{g24}{+} 
    \end{pmatrix}
\end{equation*}
\begin{tikzpicture}[overlay,remember picture,line width=.7pt,transform canvas={yshift=0mm}]
	\draw[bcolor,->-] (g12)--(g11);
\end{tikzpicture} 
\vspace{-.5cm}
&\vspace{.3cm} $\varphi_2\psi_1$ & 
\vspace{-.5cm}
\begin{equation*}
\begin{pmatrix}
        \tikznode{g11}{+} & & \tikznode{g12}{+} & & \tikznode{g13}{0} & & \tikznode{f1}{{\color{white}0}} \\ \tikznode{g21}{+} & & \tikznode{g22}{+} & & \tikznode{g23}{+} & &\tikznode{g24}{0} 
    \end{pmatrix}
\end{equation*}
\begin{tikzpicture}[overlay,remember picture,line width=.7pt,transform canvas={yshift=0mm}]
	\draw[bcolor,->-] (g24)--(g23);
    \draw[bcolor,->-] (g13)--(g12);
\end{tikzpicture} 
\vspace{-.5cm}
& \vspace{.3cm}$\psi_2\psi_3$ \\ \hline
\vspace{-.5cm}
\begin{equation*}
\begin{pmatrix}
        \tikznode{g11}{+} & & \tikznode{g12}{+} & & \tikznode{g13}{0} & & \tikznode{f1}{{\color{white}0}} \\ \tikznode{g21}{+} & & \tikznode{g22}{+} & & \tikznode{g23}{+} & &\tikznode{g24}{+} 
    \end{pmatrix}
\end{equation*}
\begin{tikzpicture}[overlay,remember picture,line width=.7pt,transform canvas={yshift=0mm}]
	\draw[bcolor,->-] (g13)--(g12);
\end{tikzpicture} 
\vspace{-.5cm}
&\vspace{.3cm} $\varphi_2\psi_2$ & 
\vspace{-.5cm}
\begin{equation*}
\begin{pmatrix}
        \tikznode{g11}{+} & & \tikznode{g12}{+} & & \tikznode{g13}{+} & & \tikznode{f1}{{\color{white}0}} \\ \tikznode{g21}{+} & & \tikznode{g22}{+} & & \tikznode{g23}{+} & &\tikznode{g24}{+} 
    \end{pmatrix}
\end{equation*}
\begin{tikzpicture}[overlay,remember picture,line width=.7pt,transform canvas={yshift=0mm}]
	\draw[bcolor,->-] (f1)--(g13);
\end{tikzpicture} 
\vspace{-.5cm}
& \vspace{.3cm}$\varphi_2\psi_3$ \\ \hline
\vspace{-.5cm}
\begin{equation*}
\begin{pmatrix}
        \tikznode{g11}{+} & & \tikznode{g12}{0} & & \tikznode{g13}{0} & & \tikznode{f1}{{\color{white}0}} \\ \tikznode{g21}{+} & & \tikznode{g22}{0} & & \tikznode{g23}{0} & &\tikznode{g24}{0} 
    \end{pmatrix}
\end{equation*}
\begin{tikzpicture}[overlay,remember picture,line width=.7pt,transform canvas={yshift=0mm}]
	\draw[bcolor,->-] (g22)--(g21);
    \draw[bcolor,->-] (g12)--(g11);
\end{tikzpicture} 
\vspace{-.5cm}
& \vspace{.3cm}$\psi_1^2$ & \vspace{-.5cm}
\begin{equation*}
\begin{pmatrix}
        \tikznode{g11}{+} & & \tikznode{g12}{+} & & \tikznode{g13}{0} & & \tikznode{f1}{{\color{white}0}} \\ \tikznode{g21}{+} & & \tikznode{g22}{+} & & \tikznode{g23}{0} & &\tikznode{g24}{0} 
    \end{pmatrix}
\end{equation*}
\begin{tikzpicture}[overlay,remember picture,line width=.7pt,transform canvas={yshift=0mm}]
	\draw[bcolor,->-] (g23)--(g22);
    \draw[bcolor,->-] (g13)--(g12);
\end{tikzpicture} 
\vspace{-.5cm}
& 
\vspace{.3cm}$\psi_2^2$ \\ \hline
\vspace{-.5cm}
\begin{equation*}
\begin{pmatrix}
        \tikznode{g11}{+} & & \tikznode{g12}{+} & & \tikznode{g13}{+} & & \tikznode{f1}{{\color{white}0}} \\ \tikznode{g21}{+} & & \tikznode{g22}{+} & & \tikznode{g23}{+} & &\tikznode{g24}{0} 
    \end{pmatrix}
\end{equation*}
\begin{tikzpicture}[overlay,remember picture,line width=.7pt,transform canvas={yshift=0mm}]
	\draw[bcolor,->-] (g24)--(g23);
    \draw[bcolor,->-] (f1)--(g13);
\end{tikzpicture} 
\vspace{-.5cm}
&\vspace{.3cm}$\psi_3^2$ & & \\ \hline

\end{longtable}
\end{center}
Notice that some of the monopole operators in Table \ref{table: baryons of SU(2)_1/2 QCD3} seem to factorize, in the sense that they can be thought of as the product of two independent gauge invariant monopoles. 
We recall, however, that there are additional monopoles with the same global charges and spin which will, in general, mix. The monopoles reported in Table \ref{table: baryons of SU(2)_1/2 QCD3} are representatives of such linear combinations.
\\

In line with our observation in the last section, we can consistently ``bosonize" the entire quiver. This is a special feature of $SU(2)$, and we anticipate that the planar dual of $SU(2)_{2-\frac{N_f}{2}-k}$ ($k\geq 0$) with $2$ scalars and $N_f$ fermions will be a fully bosonic Abelian quiver gauge theory.

\acknowledgments

We are grateful to Antonio Amariti and Pierluigi Niro for useful conversations. AS thanks Davide Bason and Leonardo Goller for helpful discussions and Davide Morgante for help with making the diagrams (created with TikZ \cite{Ellis_2017}). 
SB, RC, SP and SR are Partially supported by the MUR-PRIN grant No. 2022NY2MXY (Finanziato dall’Unione europea- Next Generation EU, Missione 4 Componente 1 CUP H53D23001080006).

\appendix
\section{Properties of Monopole Operators}
\label{app:monopoles}
In this appendix we review various properties of monopole operators in $3d$ gauge theories, which are crucial for the operator mapping across the dualities of this paper.

\subsection{Chern-Simons Fields Coupled to Fermions}
\label{sec: conventions}
Consider a gauge theory with \( N_f \) massless fermions \( \psi \), described by the Lagrangian
\begin{equation}
    \sum_{j=1}^{N_f} i \Bar{\psi}_j \slashed{D}_A \psi_j + \frac{k_0}{4\pi} A dA,
    \label{lagr:1}
\end{equation}
where \( A \) is a \( U(1) \) gauge field, and \( k_0 \) is the \textit{bare} Chern-Simons (CS) level, quantized as an integer. The presence of fermions and the CS term implies an anomaly in the time-reversal symmetry \( \mathcal{T} \) \cite{Witten_2016, seiberg2016gappedboundaryphasestopological}. When quantizing the theory, a regulator for the fermion determinant that explicitly breaks time-reversal invariance (e.g., Pauli-Villars regularization) introduces a phase \( \exp\left(-\frac{i \pi}{2} \eta(A)\right) \) in the partition function, where \( \eta \) is the APS \( \eta \)-invariant \cite{Witten_2016}. For our purposes, this phase can be interpreted as a shift in the bare CS level of \( U(1)_A \) by \( -\frac{N_f}{2} \), accompanied by the addition of a mixed gravitational CS term \( -N_f \, CS_g \), which avoids Dirac string singularities \cite{seiberg2016gappedboundaryphasestopological}. We denote this theory as \( U(1)_{k_0-N_f/2} \), where the subscript \(k= k_0-N_f/2 \) represents the shifted CS level, which can be a half-integer.

Introducing a real mass \( m \) for all fermions and integrating them out generates a CS interaction. At low energies, the effective CS level becomes
\begin{equation}
    k_{IR} = k_0-\frac{N_f}{2} + \frac{N_f}{2} \text{sign}(m),
\end{equation}
which is always an integer. Additionally, a gravitational CS term is induced:
\begin{equation}
    N_f \, CS_g \, \text{sign}(m).
\end{equation}

Of particular interest is the case where \( k = 0 \). When a positive mass \( (m > 0) \) is turned on for all fermions, the flow leads to a trivial theory with a vanishing Lagrangian. Conversely, a negative mass \( (m < 0) \) results in:
\begin{equation}
    -\frac{N_f}{4\pi} A dA - 2N_f \, CS_g.
\end{equation}
The gravitational term \( -2N_f \, CS_g \) is sometimes denoted as \( U(N_f)_1 \) in the literature.
Time reversal acts on a fermion as:
\begin{equation}
\mathcal{T}:
i \bar\psi \cancel{D}_A \psi 
\quad \to \quad
i \bar\psi \cancel{D}_A \psi  + \frac{1}{4\pi} AdA + 2 CS_g
\end{equation}

In the present paper a major role is played by quiver-like theories, where fermions can be charged under multiple gauge groups.
Consider then a $U(1)^{(a)} \times U(1)^{(b)}$ gauge theory with a fermion $\psi$ of charge $+1$ and $-1$ under the two gauge groups, bare CS levels $k_{0,a}$ and $k_{0,b}$ and mixed CS level $k_{0,ab}$:
\begin{equation}    \label{app:two_node_generic_lag}
    \mathcal{L} = i \bar{\psi} \slashed{D}_{a-b} \psi + \frac{k_{0,a}}{4\pi} a da + \frac{k_{0,b}}{4\pi} b db + \frac{k_{0,ab}}{4\pi} a db,
\end{equation}
the regularization of the fermion determinant produces a $-\tfrac{1}{2}$ shift for the gauge field $a-b$:
\begin{equation}
-\frac{1}{8\pi}(a-b) d (a-b) = -\frac{1}{8\pi}ada -\frac{1}{8\pi} bdb + \frac{1}{4\pi} adb
\end{equation}
resulting in the shifted levels:
\begin{equation}
k_a = k_{0,a} - \frac{1}{2} \,,
\quad
k_b = k_{0,b} - \frac{1}{2} \,,
\quad
k_{ab} = k_{0,ab} + 1 \,.
\end{equation}
Furthermore a mass deformation for the fermion produces an additional $\pm \tfrac{1}{2}$ shift in the CS levels $k_a$, $k_b$ and a $\mp 1$ shift in the mixed CS level $k_{ab}$.
In the quiver notation used throughout this paper the theory \eqref{app:two_node_generic_lag} and its mass deformations are denoted as follows\footnote{In the quiver notation used in this paper we do not keep track of background gravitational CS levels $\CSg$.}:
\begin{equation}    \label{app:two_node_generic_quiv}
\begin{tikzpicture}[baseline=(current bounding box).center]
	\nodeCS(0,0)(g1,1, $k_a$)
	\nodeCS(2,0)(g2,1, $k_b$)
	\path[fcolor,draw,->-] (g1) -- node[midway,above,BFcolor]  {$_{k_{ab}}$} (g2);
\end{tikzpicture}
\quad \xrightarrow[]{m\bar\psi \psi} \quad
\begin{cases}
\begin{tikzpicture}[baseline=(current bounding box).center]
	\nodeCS(0,0)(g1,1, $\scriptstyle k_a+\tfrac{1}{2}$)
	\nodeCS(2,0)(g2,1, $\scriptstyle k_b+\tfrac{1}{2}$)
	\path[draw,BFline] (g1) -- node[midway,above,BFcolor]  {$\scriptstyle _{k_{ab}-1}$} (g2);
\end{tikzpicture},
\quad m>0
\\
\begin{tikzpicture}[baseline=(current bounding box).center]
	\nodeCS(0,0)(g1,1, $\scriptstyle k_a-\tfrac{1}{2}$)
	\nodeCS(2,0)(g2,1, $\scriptstyle k_b-\tfrac{1}{2}$)
	\path[black,draw,dashed] (g1) -- node[midway,above,BFcolor]  {$\scriptstyle _{k_{ab}+1}$} (g2);
\end{tikzpicture},
\quad m<0
\end{cases}
\end{equation}

\subsection{Quantum Numbers of Monopole Operators}
\label{sec: quantum numbers of monopoles}
In three dimensions, local disorder operators can be defined in the presence of a \( U(1) \) gauge group \cite{Borokhov_2002, Aharony:2015pla}. To begin, note that in $2+1$ dimensions, there exists a conserved current
\begin{equation}
    J = *F,
\end{equation}
where \( F \) is the \( U(1) \) field strength. This current is identically conserved due to the Bianchi identity, \( d * J = dF = 0 \). The corresponding charge is referred to as the vortex charge or topological charge.

A monopole operator of topological charge \( 1 \) is defined by integrating over field configurations with a Dirac singularity at a point \( p \in \mathbb{R}^3 \), such that a magnetic flux of \( 1 \) passes through any \( \mathbf{S}^2 \) surrounding \( p \). This results in an operator \( O(x) \) with operator product expansion (OPE) with \( J^{\mu} \):
\begin{equation}
    J^{\mu}(x) O(0) \sim \frac{1}{4\pi} \frac{x^{\mu}}{|x|^3} O(0) + \text{less singular terms}.
\end{equation}

The quantum numbers of these monopole operators can be determined by analyzing the fermionic zero modes in the radially quantized theory with flux along $S^2$ \cite{Borokhov_2002,Chester_2018}. 
The contribution from fermionic zero modes  to the gauge charge of the monopole can be encoded in an effective shift of $\tfrac{1}{2}$ to the CS level for each charged fermion. Therefore in a $U(1)_k$ theory with $N_f$ fermions the gauge charge of a monopole with GNO flux +1 is $k+\tfrac{N_f}{2}$, which is equal to the \textit{bare} CS level $k_0$. 
Gauge invariant operators can then be constructed by dressing this bare monopole with (derivatives of) fundamental charged fields.

This analysis can be extended to theories with multiple gauge groups, as an example consider the theory with two gauge nodes \eqref{app:two_node_generic_lag}:
\begin{equation}   
\begin{tikzpicture}[baseline=(current bounding box).center]
	\nodeCS(0,0)(g1,1, $k_a$)
	\nodeCS(2,0)(g2,1, $k_b$)
	\path[fcolor,draw,->-] (g1) -- node[midway,above,BFcolor]  {$_{k_{ab}}$} (g2);
\end{tikzpicture}
\end{equation}
The fermion is charged under $a-b$ and its zero modes contributes as a $\tfrac{1}{2}$ shift to the CS level of $a-b$, which is equivalent to a $\tfrac{1}{2}$ shift for the CS levels $k_a$, $k_b$ and a $-1$ shift to the mixed CS $k_{ab}$.
Consider a generic bare monopole \( \mathfrak{M} \) with fluxes \( n \) under \( U(1)_a \) and \( m \) under \( U(1)_b \), denoted as $\mon^{(n,m)}$. Gauss' law dictates that the monopole carries gauge charges \( (n (k_a+\tfrac{1}{2}) + m \tfrac{(k_{ab}-1)}{2}, m (k_b+\tfrac{1}{2}) + n \tfrac{(k_{ab}-1)}{2} ) \) under \( U(1)_a \times U(1)_b \). 
We write:
\begin{equation}
\begin{split}
Q\left[\mon^{(n,m)}\right] =& n Q\left[\mon^{(+,0)}\right]+ m Q\left[\mon^{(0,+)}\right] 
\\ = &
n\left(k_a+\frac{1}{2}, \frac{k_{ab}-1}{2}\right)
+m\left( \frac{k_{ab}-1}{2},k_b+\frac{1}{2}\right)
\\ = &
\left(n (k_a+\frac{1}{2}) + m \frac{(k_{ab}-1)}{2}, m (k_b+\frac{1}{2}) + n \frac{(k_{ab}-1)}{2}\right)
\end{split}
\end{equation}
Gauge-invariant states can be constructed by dressing with modes of \( \psi \) (or \( \bar{\psi} \)) in very specific cases. Since \( \psi \) has charges \( (+1, -1) \),
a monopole dressed $p$ times with $\psi$ and $q$ times with $\bar\psi$ has charge:
\begin{equation}
Q\left[\mon^{(n,m)} \psi^p \bar\psi^q \right]
=
\left(n (k_a+\frac{1}{2}) + m \frac{(k_{ab}-1)}{2}+p-q,
\;m (k_b+\frac{1}{2}) + n \frac{(k_{ab}-1)}{2} + q-p\right)
\end{equation}
As an example, for $k_a=k_b=\tfrac{1}{2}$ and $k_{ab}=-1$ the monopoles $\mon^{(+,0)} \partial_{\bullet} \bar\psi$ and $\mon^{(0,+)} \partial_{\bullet}\psi$ are gauge invariant\footnote{Here $\partial_{\bullet}$ denotes a generic set of derivatives.}:
\begin{equation}   
\begin{tikzpicture}[baseline=(current bounding box).center]
	\nodeCS(0,0)(g1,1, $\frac{1}{2}$)
	\nodeCS(2,0)(g2,1, $\frac{1}{2}$)
	\path[fcolor,draw,->-] (g1) -- node[midway,above,BFcolor]  {$_{-1}$} (g2);
\end{tikzpicture}
\quad \to \quad
\begin{array}{l}
Q\left[\mon^{(+,0)} \partial_{\bullet} \bar\psi \right] = (0,0),
\\
Q\left[\mon^{(0,+)} \partial_{\bullet} \psi \right] = (0,0)
\end{array}
\end{equation}
Throughout this paper we write dressed monopoles by combining the flux assignment and the dressing in the following way:
\begin{equation}
\mon^{(+,0)}  \bar\psi \;\equiv\; \mon^{\tikz \path[-<-,farrowstyle] (0,0) node[black,left] {+}  -- (.5,0) node[right,black] {0}; }
\end{equation}

\begin{equation}
\mon^{(0,+)}  \psi \;\equiv\; \mon^{\tikz \path[->-,farrowstyle] (0,0) node[black,left] {0}  -- (.5,0) node[right,black] {+}; }
\end{equation}

\subsection{Statistical Transmutation via Flux Attachment}
\label{sec: statt}

In the presence of a monopole background, the spin of fundamental fields in the ground state undergoes a shift — a phenomenon known as \textit{statistical transmutation} \cite{Goldhaber:1976dp} (see also \cite{Wu:1976ge,Wilczek:1981du, Polyakov:1988md,Grundberg:1989pn, Wilczek:1995ma, Dyer:2013fja}). This shift depends on the GNO charge (or flux) of the monopole, the intrinsic spin of the particle, and its gauge charge.

In what follows, we summarize the effects of statistical transmutation for a particle of spin $s$, and examine how this behavior is modified in the presence of multiple $U(1)$ gauge factors, particularly in the context of the quiver gauge theories discussed in the main text.

\subsubsection*{1. Single Gauge Group}
Consider a spin $s$ field $\varphi$ with charge $q$ under a $U(1)$ gauge group. 
On the background of a monopole with flux $m$ the ground state of the field $\varphi$ has the following spin channels:
\begin{equation*}
    |\frac{q m}{2}-s|,|\frac{q m}{2}-s|+1,\ldots,\frac{q m}{2}+s,
\end{equation*} 
For example in the background of a monopole with flux $m=1$ a fermion has a spin-$0$ and a spin-$1$ channel and a scalar has a spin-$\tfrac{1}{2}$ channel. 

\subsubsection*{2. Bifundamental Matter}
 Consider a spin-$s$ field $\varphi$ with charge $(+1,-1) q$ under a $U(1)\times U(1)$ gauge symmetry, shown explicitly in Equation \eqref{eq: bifundamental_statt}:
\begin{center}
  \begin{equation}  \label{eq: bifundamental_statt}
    \begin{tikzpicture}
        \node at (0,0) (a1) [gauge,black] {$1$};
        \node at (1.5,0) (a2) [gauge,black] {$1$};
      \draw[black,->-] (a1)--(a2);
        
        \draw[black] node at (.75,.2) {$^{\varphi}$};
    \end{tikzpicture}
    \end{equation}
    \end{center}
The dressed operator $\mon^{(m,n)}\varphi$ will have the following spin channels:
\begin{equation*}
    |\frac{q}{2}(m -n)-s|,|\frac{q}{2}(m-n)-s|+1,\ldots, \frac{q}{2}(m-n)+s,
\end{equation*}
which depends on the difference of the GNO fluxes of the two gauge groups. 
In particular in the background of a monopole $\mathfrak{M}^{(+,+)}$ the bifundamental field $\varphi$ “sees" a vanishing effective flux, and its spin is not transmuted.

\subsubsection*{3. Absence of Transmutation from Distant Fluxes}
Finally, we consider the quiver gauge theory shown in Equation \eqref{eq: nobifundamental_statt}:
\begin{center}
  \begin{equation}  \label{eq: nobifundamental_statt}
    \begin{tikzpicture}
        \node at (-1.5,0) (a0) [gauge,black] {$1$};
        \node at (0,0) (a1) [gauge,black] {$1$};
        \node at (1.5,0) (a2) [gauge,black] {$1$};
      \draw[black,->-] (a1)--(a2); \draw[black,->-] (a0)--(a1);
        \draw[black] node at (.75,.2) {$^{\varphi_2}$};
        \draw[black] node at (-.75,.2) {$^{\varphi_1}$};
    \end{tikzpicture}
    \end{equation}
    \end{center}
where $\varphi_i$ are bifundamental spin-$s_i$ fields. 
On the background of a monopole $\mon^{[0,0,m]}$ the field $\varphi_1$ does not “see" the magnetic flux, and its spin remains $s_1$.
On the other hand the field $\varphi_2$ “sees" an effective flux $-m$, and has spin channels shifted by $-m$:
\begin{equation}
    \begin{array}{l}
        \varphi_1:\; \left| -s_1 \right|,\;\left| -s_1 \right|+1,\;\dots,\;\left| s_1 \right|
        \\
        \varphi_2:\; \left| -s_2 -\frac{m}{2} \right|,\;\left| -s_2-\frac{m}{2} \right|+1,\;\dots,\;\left| s_2-\frac{m}{2} \right|
    \end{array}
    \qquad \text{on }
    \mon^{[0,0,m]}
\end{equation}

Generally, consider a field $\varphi$ of spin-$s$ and charges $q_i$ under some $U(1)_i$ gauge groups, with $i=1,\dots, M$. Then in the background of a monopole with GNO fluxes $m_i$ the field $\varphi$ has the same spin channels as a spin-$s$ field with charge $1$ in the background of an “effective" GNO flux $m_{eff}$:
\begin{equation}
m_{eff} = \sum_{i=1}^{M} q_i m_i
\end{equation}

As an example we consider the following theory, which may be thought of as part of one of the quiver theories considered in the main body of the paper:
\begin{equation}
\begin{tikzpicture}
\foreach \x/\y in {0/3,0/2,1/2,2/2,0/1,1/1,2/1,2/0}
{
	\node at (1.5*\x,1.5*\y) (g\x\y) [gauge,black] {$1$};
}
\draw[red] (g12)++(.2,-.5) node {$\scriptstyle 2$};
\draw[red] (g11)++(.2,-.5) node {$\scriptstyle 2$};

\foreach \from/\to in {02/12,12/22,01/11,11/21,12/03,21/12,11/02,20/11}
{
	\draw[->-,fcolor] (g\from) -- node[midway,above,BFcolor] {$\scriptstyle -1$} (g\to);
}

\draw[->-,bcolor] (g12)++(0,1.5) -- node[midway,left,BFcolor] {$\scriptstyle 0$} (g12);
\draw[->-,bcolor] (g12) 		-- node[midway,left,BFcolor] {$\scriptstyle 0$} (g11);
\draw[-<-,bcolor] (g11)++(0,-1.5) -- node[midway,left,BFcolor] {$\scriptstyle 0$} (g11);

\end{tikzpicture}
\end{equation}
where we only reported CS and mixed CS terms relevant for the computations that follow.
We can consider the monopole with GNO flux $+1/-1$ under the gauge nodes in the central column. Its gauge charges are:
\begin{equation}
Q\left[ \mon{
\left(
	\begin{array}{ccc}
		0 &  & \\
		0 & + & 0 \\
		0 & - & 0 \\
		 & & 0
	\end{array}
\right)
}
\right]
=
\left(
	\begin{array}{ccc}
		-1 &  & \\
		0 & 4 & -1 \\
		1 & -4 & 0 \\
		 & & 1
	\end{array}
\right)
\end{equation}
therefore it can be made gauge invariant with the following dressing:
\begin{equation}
\mon^{
\begin{tikzpicture}
\matrix (A) [matrix of nodes,left delimiter=(,right delimiter={)}]
{
		0 &[3mm]  &[3mm] \\
		0 & + & 0 \\
		0 & $-$ & 0 \\
		 & & 0 \\
};

\draw[->-,farrowstyle,draw] (A-1-1) -- (A-2-2);
\draw[->-,farrowstyle] (A-2-3) -- (A-2-2);
\draw[->-,farrowstyle] (A-2-1) -- (A-2-2);
\draw[->-,farrowstyle] (A-3-3) -- (A-2-2);

\draw[->-,farrowstyle] (A-3-2) -- (A-2-1);
\draw[->-,farrowstyle] (A-3-2) -- (A-3-1);
\draw[->-,farrowstyle] (A-3-2) -- (A-3-3);
\draw[->-,farrowstyle] (A-3-2) -- (A-4-3);

\end{tikzpicture}
}
\end{equation}
Each fermion involved in this dressing “sees" a GNO flux $\pm1$, therefore it has a spin-0 zero mode. Therefore we can dress this monopole to obtain a gauge invariant spin-0 monopole operator. In the theories considered in the main paper this is one of the monopoles entering the $\mathcal{V}_{monopole}$ interaction. 
Similarly, we can consider the monopole operators involved in the $\mathcal{V}_{monopole}$ interaction that have GNO flux in the “bosonic" part of the quivers discussed in Section \ref{sec: qcd_b&f}. There we find that the following monopole can be dressed to be gauge invariant and a Lorentz scalar:
\begin{equation}
\begin{tikzpicture}[baseline=(current bounding box).center]
\foreach \x/\y in {0/3,0/2,1/2,2/2,0/1,1/1,2/1,2/0}
{
	\node at (1.5*\x,1.5*\y) (g\x\y) [gauge,black] {$1$};
}

\draw[red] (g12)++(.2,-.5) node {$\scriptstyle 2$};
\draw[red] (g11)++(.2,-.5) node {$\scriptstyle 2$};

\foreach \from/\to in {02/12,12/22,01/11,11/21}
{
	\draw[->-,bcolor] (g\from) -- node[midway,above,BFcolor] {$\scriptstyle 0$} (g\to);
}

\foreach \from/\to in {12/03,21/12,11/02,20/11}
{
	\draw[->-,bcolor] (g\from) -- node[midway,above,BFcolor] {$\scriptstyle -2$} (g\to);
}

\draw[->-,bcolor] (g12)++(0,1.5) -- node[midway,left,BFcolor] {$\scriptstyle 0$} (g12);
\draw[->-,bcolor] (g12) 		-- node[midway,left,BFcolor] {$\scriptstyle 0$} (g11);
\draw[-<-,bcolor] (g11)++(0,-1.5) -- node[midway,left,BFcolor] {$\scriptstyle 0$} (g11);
\draw[->-,bcolor] (g03) --  (g02);
\draw[->-,bcolor] (g21) --  (g20);

\end{tikzpicture}
\qquad \to \qquad
\mon{
\begin{tikzpicture}[baseline=(current bounding box).center]
\matrix (A) [matrix of nodes,left delimiter=(,right delimiter={)},row sep=4mm]
{
		0 &[3mm]  &[3mm] \\
		0 & + & 0 \\
		0 & $-$ & 0 \\
		 & & 0 \\
};

\draw[->-,barrowstyle] (A-1-1) -- (A-2-1);
\draw[->-,barrowstyle] (A-3-3) -- (A-4-3);
\draw[->-,barrowstyle] (A-3-2.80) -- (A-2-2.280);
\draw[->-,barrowstyle] (A-3-2.100) -- (A-2-2.260);

\end{tikzpicture}
}
\end{equation}
Notice in particular that the middle scalar involved (twice) in the dressing “sees" a GNO flux 2, while the other two scalars “see" no flux, therefore all scalars have a spin-0 mode in the background of the monopole and we obtain a spin-0 gauge invariant operator.
Similarly, one can show that all the monopoles involved in the interactions $\mathcal{V}_{monopole}$ throughout this paper can be dressed to obtain a gauge invariant operator which is a Lorentz scalar.

\section{Abelian Dualities with Two Flavors  }
\label{sec: Abelian-technical}

In this Section we consider in detail the limiting case of the duality in Figure \ref{fig:Abelian_dual_bosons_fermions} with either two fermions or two scalars and $k=0$. 
In these cases both the electric and magnetic theories are QED theories with two charged fields.
We start with the bosonization duality:
\begin{equation}    \label{app_duality_U1_2psi}
\fQED(2,-1)
\dualto
\begin{tikzpicture}[baseline=(current bounding box).center]
	\flavorCS(0,0)(f1, 1,)
	\nodeCS(1.5,0)(g1,1, 2)
	\flavorCS(3,0)(f2, 1,)
	
	\path[bcolor] (f1) edge[->-] node[midway,above] {\tiny$\phi_1$} (g1);
	\path[bcolor] (g1) edge[->-] node[midway,above] {\tiny$\phi_2$} (f2);
	
	\node (V) at (1.5,-1) {$\mathcal{V}=|\phi_1|^4 + |\phi_2|^4$};
\end{tikzpicture}
\end{equation}
This duality was analyzed in detail in \cite{Benini:2017aed}, here we report part of the analysis for completeness.
The potential can be written in terms of the ``easy-plane" potential:
\begin{equation}
\begin{array}{l}
	\mathcal{V} = |\phi_1|^4 + |\phi_2|^4 = \mathcal{V}_{SU(2)} - 2 \mathcal{V}_{EP}
	\\
	\mathcal{V}_{SU(2)} = (|\phi_1|^2 + |\phi_2|^2)^2,
	\qquad
	\mathcal{V}_{EP} = |\phi_1|^2 |\phi_2|^2
\end{array}
\end{equation}
which breaks a possible $SU(2)$ symmetry rotating the two scalars.
The $U(1)_X$ topological symmetry of the bosonic side is believed to enhance in the IR, together with a $\zz_2$ discrete symmetry, to $SO(3)_X$, and is mapped to the flavor symmetry on the fermionic side. The enhancement is due to spin-1 monopole operators with conformal dimension $2$, which play the role of the additional conserved charges $J_{\pm}$ of $SU(2)$. In particular consider the gauge invariant dressed monopoles:
\begin{equation}    \label{app:monopoles_J_U1_2psi}
\mathcal{M}^+ \phi_1 \phi_2^*, 
\qquad
 \mathcal{M}^- \phi_1^* \phi_2, 
\end{equation}
Here, $\mathcal{M}^\pm$ are the bare monopole operators, which have gauge charge $\pm 2$ under the dynamical gauge field. On the background of the bare monopole, the scalar operators have a spin-$\frac{1}{2}$ ground state, therefore the dressed monopoles \eqref{app:monopoles_J_U1_2psi} have a spin-0 and a spin-1 component. The spin-0 component is mapped to the off-diagonal mesons of the fermionic QED, while the spin-1 components are the conserved currents $J_{\pm}$ that induce the symmetry enhancement, as described above.
The mapping of some simple operators is reported in Table \ref{tab:mapping_U1_2psi}.

\begin{table}[t]
\begin{equation}
\begin{array}{c|cc|c|cc|c}
U(1)_{-1} \text { with } 2 \psi & U(1)_X & \mathbb{Z}_2^X & S O(3)_X & U(1)_Y & \mathbb{Z}_2^Y & U(1)_2 \text { with } 2 \phi \text { and } \mathcal{V}_{\mathrm{EP}} \\
\hline \bar{\psi}_1 \psi_1+\bar{\psi}_2 \psi_2 & 0 & + & \mathbf{1} & 0 & + & -\left|\phi_1\right|^2-\left|\phi_2\right|^2 \\
\bar{\psi}_1 \psi_1-\bar{\psi}_2 \psi_2 & 0 & - & \mathbf{3} & 0 & - & \left|\phi_1\right|^2-\left|\phi_2\right|^2 \\
\bar{\psi}_1 \gamma_\mu \psi_1-\bar{\psi}_2 \gamma_\mu \psi_2 & 0 & - & \mathbf{3}^{\prime} & 0 & + & \epsilon_{\mu \nu \rho} F^{\nu \rho} \\
\bar{\psi}_2 \psi_1 \oplus \bar{\psi}_2 \gamma_\mu \psi_1 & 1 & \tikznode{X1}{} & \mathbf{3} \oplus \mathbf{3}^{\prime} & 0 & -/+ & \mathcal{M}^+ \phi_1 \phi_2^* \\
\bar{\psi}_1 \psi_2 \oplus \bar{\psi}_1 \gamma_\mu \psi_2 & -1 & \tikznode{X2}{} & \mathbf{3} \oplus \mathbf{3}^{\prime} & 0 & -/+ & \mathcal{M}^- \phi_1^* \phi_2 \\
\mathcal{N}^+ & 0 & + & \mathbf{1} & 1 & \tikznode{Y1}{} & \phi_1 \phi_2 \\
\mathcal{N}^- & 0 & + & \mathbf{1} & -1 & \tikznode{Y2}{} & \phi_1^* \phi_2^*
\end{array}
\begin{tikzpicture}[overlay,remember picture,line width=.7pt,transform canvas={yshift=1mm}]
	\path (X1) edge[<->,transform canvas={xshift=0mm}]  (X2);
    \path (Y1) edge[<->,transform canvas={xshift=0mm}]  (Y2);
\end{tikzpicture}
\end{equation}
\caption{Mapping of operators across the duality \eqref{app_duality_U1_2psi}, reported from \cite{Benini:2017aed}. The operators group into three singlets of $SO(3)_X$ and two triplets denoted as $\mathbf{3}$ (with spin 0) and $\mathbf{3}'$ (with spin 1). Arrows between different rows indicate that the corresponding $\zz_2$ symmetry exchanges the two operators.}
\label{tab:mapping_U1_2psi}
\end{table}


Let us now discuss the fermionization duality for $N_s=2$, which is the limiting case of \ref{fig:Abelian_dual_bosons_fermions} with two bosons in the electric theory and $k=0$:
\begin{equation}    \label{app:duality_U1_2phi}
\begin{tikzpicture}[baseline=(current bounding box).center]
    \nodeCS(0,0)(g,1,$-2$)
    \node[flavor] (f) at (1.5,0)  {$2$};
    \barrow(g,f,left)
    \node (V) at (0.75,-1)  {$\mathcal{V} = \left(|\elephi_1 |^2+|\elephi_2 |^2\right)^2$};
\end{tikzpicture}
\dualto
\begin{tikzpicture}[baseline=(current bounding box).center]
	\flavorCS(0,0)(f1, 1,)
	\nodeCS(1.5,0)(g1,1, 1)
	\flavorCS(3,0)(f2, 1,)
	
	\path[fcolor] (f1) edge[->-] node[midway,above] {\tiny$\magpsi_1$} (g1);
	\path[fcolor] (g1) edge[->-] node[midway,above] {\tiny$\magpsi_2$} (f2);
	
	\node (V) at (1.5,-1) {$\mathcal{V}=\sigma(\bar\psi_2 \psi_2 - \bar\psi_1 \psi_1)$};
\end{tikzpicture}
\end{equation}
With the inclusion of background fields, this duality reads:
\begin{equation}
\begin{array}{l}
|D_{b-Y} \elephi_1|^2 + |D_{b} \elephi_2|^2 + 
\\
\frac{b d (Y+X)}{2\pi} - 2\frac{bdb}{4\pi}
\end{array}
\dualto
\begin{array}{l}
i \bar\psi_1 \cancel{D}_{-a} \psi_1 + 
i \bar\psi_2 \cancel{D}_{a+X} \psi_2 + 
\sigma(\bar\psi_1 \psi_1 - \bar\psi_2 \psi_2)
\\
+2\frac{ada}{4\pi}+\frac{ad(Y+X)}{2\pi}
+4\CSg +\frac{(X+Y)d(X+Y)}{4\pi}
\end{array}
\end{equation}
The fermionwic theory is a version of the so-called QED-GNY$_-$ theory \cite{Benvenuti:2018cwd,Benvenuti:2019ujm} with a CS interaction.
On the LHS there is an $SO(3)_Y \times O(2)_X$ global symmetry, and we do not expect any enhancement in the IR. On the fermionic side the flavor symmetry is $U(1)_X$, due to the interaction, and there is a topological $U(1)_Y$ symmetry which is conjectured to enhance to $SO(3)_Y$.
Let us consider the action of the symmetries on the operators of the fermionic QED more closely. The gauge invariant operators of the theory are given by gauge invariant polynomials in the fundamental fields as well as dressed monopole operators.
The spectrum of monopole operators can be analyzed via the state operator correspondence by quantizing the fermion fields on $S^2$ with some units of GNO flux for the gauge field. 

The charges of fundamental fields and bare monopoles under the global and gauge symmetries are reported in Table \ref{tab:charges_bare}.

\begin{equation}	\label{tab:charges_bare}
\begin{array}{c|c|cc}
	& U(1)_{a} & U(1)_X & U(1)_Y 
	\\ \hline
	\psi_1 & -1 & 0 & 0  
	\\
	\psi_2 & 1 & 1 & 0 
	\\ \hline
	\mathcal{M}^{\pm}_{bare} & \pm 2 &\pm 1 &\pm 1
\end{array}
\end{equation}
Furthermore there are two $\mathbb{Z}_2$ symmetries, denoted as $\zz_2^X$ and $\zz_2^Y$ acting as:
\begin{equation}
\zz_2^X:  X \to -X, \; \psi_1 \to \psi_2^c,\; \psi_2 \to \psi_1^c, \; Y \to Y, \; \sigma \to -\sigma,\; a\to a + X
\end{equation}

\begin{equation}
\zz_2^Y:  Y \to -Y,\; \psi_1 \to \psi_2,\; \psi_2 \to \psi_1,\; X\to X,\; \sigma \to -\sigma,\; a \to -a-X
\end{equation}
with:
\begin{equation}
	\psi^c = C \bar\psi^{T},
	\qquad
	\bar\psi^c = - \psi^{T} C^{-1}
	\qquad
	C=\left( \begin{array}{cc} 0 & -i \\ i & 0 \end{array} \right)
\end{equation}
The two $\zz_2$ symmetries leave the Lagrangian invariant up to a background CS term, therefore they are anomalous:
\begin{equation}
\zz_2^X: \; \mathcal{L}_f \to \mathcal{L}_f - \frac{X d Y}{2\pi}
\end{equation}

\begin{equation}
\zz_2^Y: \; \mathcal{L}_f \to \mathcal{L}_f - \frac{X d Y}{2\pi}
\end{equation}
while the composition of the two transformations is not anomalous.

In Table \ref{tab:charges_invariant} we report the action of the symmetries on various gauge invariant operators. One can check that the action of $\zz_2^X$ commutes with $U(1)_Y$ on gauge invariants, and the action of $\zz_2^Y$ commutes with $U(1)_X$, therefore the symmetries combine into $O(2)_X \times O(2)_Y$. 

\begin{table}
\begin{equation} 
\begin{array}{c|ccccc|c}
    U(1)_{-2} \text{ with } 2\elephi & U(1)_X & U(1)_Y & \zz_2^X & \zz_2^Y & spin & U(1)_{1} \text { with } 2 \magpsi \text{ and } \mathcal{V}_{GNY}
    \\ \hline 
    \mathcal{N}^{+} \elephi_2^2 & 1 & 1 & \tikznode{X1}{} & \tikznode{Y1}{} & 1 & \mathcal{M}^{+} \magpsi_1^2
    \\
    \mathcal{N}^{-} (\elephi_1^*)^2 & -1 & 1 & \tikznode{X2}{} & \tikznode{Y2}{} & 1 & \mathcal{M}^{+} \bar\magpsi_2^2
    \\
    \elephi_1^* \elephi_2 \oplus \elephi_1^* \partial_{\mu} \elephi_2 & 0 & 1 & \tikznode{X3}{} +/- & \tikznode{Y3}{} & 0,1 & \mathcal{M}^{+} \bar\magpsi_2 \magpsi_1
    \\ \hline 
    \mathcal{N}^{+} \elephi_1^2 & 1 & -1 & \tikznode{X4}{} & \tikznode{Y4}{} & 1 & \mathcal{M}^{-} \magpsi_2^2
    \\
    \mathcal{N}^{-} (\elephi_2^*)^2 & -1 & -1 & \tikznode{X5}{} & \tikznode{Y5}{} & 1 & \mathcal{M}^{-} \bar\magpsi_1^2
    \\
    \elephi_2^* \elephi_1 \oplus \elephi_2^* \partial_{\mu} \elephi_1 & 0 & -1 & \tikznode{X6}{} +/- & \tikznode{Y6}{} & 0,1 & \mathcal{M}^{-} \bar\magpsi_1 \magpsi_2
    \\ \hline  
    \mathcal{N}^{+} \elephi_1 \elephi_2 & 1 & 0 &  \tikznode{X7}{} & -/+ & 0,1 & \magpsi_2^{T} \magpsi_1\oplus \magpsi_2^{T} \gamma_\mu \magpsi_1
    \\
    \mathcal{N}^{-} \elephi_1^* \elephi_2^* & -1 & 0 &  \tikznode{X8}{} &  -/+ & 0,1 & \bar\magpsi_1 \bar\magpsi_2^{T} \oplus \bar\magpsi_1 \gamma_\mu \bar\magpsi_2^{T}
    \\
    |\elephi_1|^2 - |\elephi_2|^2 & 0 & 0 & - & - & 0 & \bar\magpsi_1 \magpsi_1 - \bar\magpsi_2 \magpsi_2
    \\ \hline
    \begin{cases}|\elephi_1|^2 + |\elephi_2|^2 \\ (|\elephi_1|^2 - |\elephi_2|^2)^2 \end{cases} 
    & 0 & 0 & + & + & 0 &
    \begin{cases}\bar\magpsi_1 \magpsi_1 + \bar\magpsi_2 \magpsi_2 \\ \sigma^2 \end{cases}
\end{array}
\begin{tikzpicture}[overlay,remember picture,line width=.7pt,transform canvas={yshift=1mm}]
    \path (X1) edge[<->,transform canvas={xshift=0mm}]  (X2);
    \path (X4) edge[<->,transform canvas={xshift=0mm}]  (X5);
    \path (Y1) edge[<->,transform canvas={xshift=-2mm}]  (Y4);
    \path (Y2) edge[<->,transform canvas={xshift=0mm}]  (Y5);
    \path (Y3) edge[<->,transform canvas={xshift=2mm}]  (Y6);
    \path (X7) edge[<->,transform canvas={xshift=0mm}]  (X8);
\end{tikzpicture}
\end{equation}
\caption{Mapping of operators across the duality \eqref{app:duality_U1_2phi}. The $U(1)_Y \rtimes \zz_2^{Y}$ is a subgroup of an $SO(3)_Y$ symmetry which is manifest in the bosonic Lagrangian and is enhanced in the IR for the fermionic Lagrangian. 
 Arrows between different rows indicate that the corresponding $\zz_2$ symmetry exchanges the two operators.
The operators of the fermionic QED in the bottom-most two rows are expected to mix, and non-trivial linear combinations of the two are expected to map across the duality.}
\label{tab:charges_invariant}
\end{table}

Notice that on the background of a monopole with GNO flux $\pm 1$ the fermions have a spin-0 and a spin-1 ground state. 
When a monopole is dressed with two different fermions we can use two spin-0 mode, one spin-1 and one spin-0 mode and two spin-1 mode. This is the case for the third and sixth entries in \eqref{tab:charges_invariant}. On the other hand if we dress two times with the same fermion we must choose one spin-0 and one spin-1 mode, due to the anticommutativity of the fermion modes. This is the case for the other entries in \eqref{tab:charges_invariant}.
We claim that the spin-1 components of the  monopoles $\mathcal{M}^+ \bar\magpsi_2 \magpsi_1$ and $\mathcal{M}^{-} \bar\magpsi_1 \magpsi_2$  combine with $*F$ to form the conserved current of an enhanced $SO(3)_Y$ symmetry.
The $U(1)_Y$ symmetry is embedded as a maximal torus of $SO(3)_Y$ and $\zz_2^Y$ is the corresponding Weyl group. 

On the LHS there is one relevant deformation that preserves all the symmetries, which corresponds to a uniform mass for all the bosons:
\begin{equation}
\mathcal{O}_1 = |\elephi_1|^2 + |\elephi_2|^2
\end{equation}
With a positive mass, both bosons are gapped and the theory flows to $U(1)_{-2}$ CS theory, while for negative mass the gauge group is Higgsed and the $SO(3)_Y$ flavor group is spontaneously broken to $O(2)_Y$, resulting in an $S^2$ NLSM.
It is also interesting to consider the $V_{EP}$ deformation, triggered by the operator:
\begin{equation}
\mathcal{O}_2 = ( |\elephi_1|^2 - |\elephi_2|^2)^2
\end{equation}
which belongs to a isospin-2 representation of $SO(3)_Y$ and breaks the global symmetry to $U(1)_Y \rtimes \zz_2^Y$. 
When we deform the bosonic theory with $\lambda \mathcal{O}_2$ with $\lambda<0$, the theory is believed to flow to an interacting CFT, which corresponds to the $\mathcal{T}$-tranform of bosonic theory in the bosonization duality \eqref{app_duality_U1_2psi}.

In the fermionic theory we can consider two deformations that are invariant under all the UV global symmetries:
\begin{equation}
	\widetilde{\mathcal{O}}_1 = \sigma^2,
	\qquad
	\widetilde{\mathcal{O}}_2 = -\bar\magpsi_1 \magpsi_1 - \bar\magpsi_2 \magpsi_2
\end{equation}
We expect that only a particular linear combination of $\widetilde{\mathcal{O}}_1$ and $\widetilde{\mathcal{O}}_2$ will preserve the enhanced IR global symmetry.
If we turn on $\lambda_1 \widetilde{\mathcal{O}}_1$ with large and positive $\lambda_1$ the real scalar $\sigma$ is gapped and we can integrating it out, producing quartic fermionic interactions. These interactions are believed to be irrelevant in the IR, therefore the theory flows to the fixed point of $U(1)_{1}$ with two fermions. 
This is the $\mathcal{T}$-reversal of the theory appearing on the fermionic side of the duality \eqref{app_duality_U1_2psi}, matching the bosonic side.
Therefore there is a flow connecting the duality  \eqref{app:duality_U1_2phi} to the $\mathcal{T}$-reversal of  \eqref{app_duality_U1_2psi}, schematically:
\begin{equation}
\begin{array}{ccc|c}
U(1)_{-2} \text{ with } 2\elephi & \leftrightarrow & U(1)_{1} \text{ with } 2\magpsi \text{ and } \mathcal{V}_{GNY} & SO(3)_Y \times O(2)_X
\\
+ \mathcal{V}_{EP} & \Bigg\downarrow  & + \sigma^2
& O(2)_Y \times O(2)_X
\\
\mathcal{T}\Big[
U(1)_{2} \text{ with } 2\elephi \text{ and } \mathcal{V}_{EP} & \leftrightarrow & U(1)_{-1} \text{ with } 2\magpsi
\Big]
& O(2)_Y \times SO(3)_X
\end{array}
\end{equation}
where on the right column we reported the global symmetries of the fixed point of \eqref{app:duality_U1_2phi}, the symmetry preserved along the flow triggered by $+\mathcal{V}_{EP}/+\sigma^2$ and the symmetries of the fixed point of \eqref{app_duality_U1_2psi}.

Turning on  $\lambda_2 \widetilde{\mathcal{O}}_2$ with $\lambda_2>0$, the fermions take negative mass. Integrating them out results in a $U(1)_0$ gauge theory, and generates a quadratic term for $\sigma$. If the effective mass for $\sigma$ is positive, also $\sigma$ is gapped and we obtain an $S^1$ NLSM. 
We conjecture that there is a combination of positive $\lambda_1$ and $\lambda_2$ such that $\sigma$ has vanishing effective mass
and combine with the dual photon resulting in a NLSM with a 2-dimensional target space. 
This can be understood as a symmetry broken phase, associated to the spontaneous breaking $SO(3)_Y \to U(1)_Y$. 
Therefore, we further expect the NLSM to have target space $S^2$.
For $\lambda_2<0$ the two fermions can be integrated out, resulting in a $U(1)_{2}$ CS theory. 
For $\lambda_1<0$ the real scalar condenses, resulting in two vacua each supporting a $U(1)_1$ CS theory. Comparing with the phases of the bosonic theory, analyzed for example in \cite{Benini:2017aed}, we expect that the transition between the $U(1)_2$ phase and the $U(1)_1 \times U(1)_1$ phase, occuring along a line in the $\lambda_1<0, \lambda_2<0$ quadrant, is first order.
It would be interesting to study this transition directly from the fermionic QED point of view, but we leave this to future work.


\bibliographystyle{JHEP}
\bibliography{References}

\providecommand{\href}[2]{#2}\begingroup\raggedright\begin{thebibliography}{10}

\bibitem{Giombi:2011kc}
S.~Giombi, S.~Minwalla, S.~Prakash, S.P.~Trivedi, S.R.~Wadia and X.~Yin,
  \emph{{Chern-Simons Theory with Vector Fermion Matter}},
  \href{https://doi.org/10.1140/epjc/s10052-012-2112-0}{\emph{Eur. Phys. J. C}
  {\bfseries 72} (2012) 2112}
  [\href{https://arxiv.org/abs/1110.4386}{{\ttfamily 1110.4386}}].

\bibitem{Aharony:2011jz}
O.~Aharony, G.~Gur-Ari and R.~Yacoby, \emph{{d=3 Bosonic Vector Models Coupled
  to Chern-Simons Gauge Theories}},
  \href{https://doi.org/10.1007/JHEP03(2012)037}{\emph{JHEP} {\bfseries 03}
  (2012) 037} [\href{https://arxiv.org/abs/1110.4382}{{\ttfamily 1110.4382}}].

\bibitem{Minwalla:2015sca}
S.~Minwalla and S.~Yokoyama, \emph{{Chern Simons Bosonization along RG Flows}},
  \href{https://doi.org/10.1007/JHEP02(2016)103}{\emph{JHEP} {\bfseries 02}
  (2016) 103} [\href{https://arxiv.org/abs/1507.04546}{{\ttfamily
  1507.04546}}].

\bibitem{Aharony:2015mjs}
O.~Aharony, \emph{{Baryons, monopoles and dualities in Chern-Simons-matter
  theories}}, \href{https://doi.org/10.1007/JHEP02(2016)093}{\emph{JHEP}
  {\bfseries 02} (2016) 093}
  [\href{https://arxiv.org/abs/1512.00161}{{\ttfamily 1512.00161}}].

\bibitem{Son:2015xqa}
D.T.~Son, \emph{{Is the Composite Fermion a Dirac Particle?}},
  \href{https://doi.org/10.1103/PhysRevX.5.031027}{\emph{Phys. Rev. X}
  {\bfseries 5} (2015) 031027}
  [\href{https://arxiv.org/abs/1502.03446}{{\ttfamily 1502.03446}}].

\bibitem{Metlitski:2015eka}
M.A.~Metlitski and A.~Vishwanath, \emph{{Particle-vortex duality of
  two-dimensional Dirac fermion from electric-magnetic duality of
  three-dimensional topological insulators}},
  \href{https://doi.org/10.1103/PhysRevB.93.245151}{\emph{Phys. Rev. B}
  {\bfseries 93} (2016) 245151}
  [\href{https://arxiv.org/abs/1505.05142}{{\ttfamily 1505.05142}}].

\bibitem{Seiberg:2016gmd}
N.~Seiberg, T.~Senthil, C.~Wang and E.~Witten, \emph{{A Duality Web in 2+1
  Dimensions and Condensed Matter Physics}},
  \href{https://doi.org/10.1016/j.aop.2016.08.007}{\emph{Annals Phys.}
  {\bfseries 374} (2016) 395}
  [\href{https://arxiv.org/abs/1606.01989}{{\ttfamily 1606.01989}}].

\bibitem{Benini:2017aed}
F.~Benini, \emph{{Three-dimensional dualities with bosons and fermions}},
  \href{https://doi.org/10.1007/JHEP02(2018)068}{\emph{JHEP} {\bfseries 02}
  (2018) 068} [\href{https://arxiv.org/abs/1712.00020}{{\ttfamily
  1712.00020}}].

\bibitem{Senthil:2018cru}
T.~Senthil, D.T.~Son, C.~Wang and C.~Xu, \emph{{Duality between $(2+1)d$
  Quantum Critical Points}},
  \href{https://doi.org/10.1016/j.physrep.2019.09.001}{\emph{Phys. Rept.}
  {\bfseries 827} (2019) 1} [\href{https://arxiv.org/abs/1810.05174}{{\ttfamily
  1810.05174}}].

\bibitem{Turner:2019wnh}
C.~Turner, \emph{{Dualities in 2+1 Dimensions}},
  \href{https://doi.org/10.22323/1.349.0001}{\emph{PoS} {\bfseries Modave2018}
  (2019) 001} [\href{https://arxiv.org/abs/1905.12656}{{\ttfamily
  1905.12656}}].

\bibitem{Jain:2013gza}
S.~Jain, S.~Minwalla and S.~Yokoyama, \emph{{Chern Simons duality with a
  fundamental boson and fermion}},
  \href{https://doi.org/10.1007/JHEP11(2013)037}{\emph{JHEP} {\bfseries 11}
  (2013) 037} [\href{https://arxiv.org/abs/1305.7235}{{\ttfamily 1305.7235}}].

\bibitem{Gur-Ari:2015pca}
G.~Gur-Ari and R.~Yacoby, \emph{{Three Dimensional Bosonization From
  Supersymmetry}}, \href{https://doi.org/10.1007/JHEP11(2015)013}{\emph{JHEP}
  {\bfseries 11} (2015) 013}
  [\href{https://arxiv.org/abs/1507.04378}{{\ttfamily 1507.04378}}].

\bibitem{Karch:2016sxi}
A.~Karch and D.~Tong, \emph{{Particle-Vortex Duality from 3d Bosonization}},
  \href{https://doi.org/10.1103/PhysRevX.6.031043}{\emph{Phys. Rev. X}
  {\bfseries 6} (2016) 031043}
  [\href{https://arxiv.org/abs/1606.01893}{{\ttfamily 1606.01893}}].

\bibitem{Kachru:2016rui}
S.~Kachru, M.~Mulligan, G.~Torroba and H.~Wang, \emph{{Bosonization and Mirror
  Symmetry}}, \href{https://doi.org/10.1103/PhysRevD.94.085009}{\emph{Phys.
  Rev. D} {\bfseries 94} (2016) 085009}
  [\href{https://arxiv.org/abs/1608.05077}{{\ttfamily 1608.05077}}].

\bibitem{Kachru:2016aon}
S.~Kachru, M.~Mulligan, G.~Torroba and H.~Wang, \emph{{Nonsupersymmetric
  dualities from mirror symmetry}},
  \href{https://doi.org/10.1103/PhysRevLett.118.011602}{\emph{Phys. Rev. Lett.}
  {\bfseries 118} (2017) 011602}
  [\href{https://arxiv.org/abs/1609.02149}{{\ttfamily 1609.02149}}].

\bibitem{Karch:2018mer}
A.~Karch, D.~Tong and C.~Turner, \emph{{Mirror Symmetry and Bosonization in 2d
  and 3d}}, \href{https://doi.org/10.1007/JHEP07(2018)059}{\emph{JHEP}
  {\bfseries 07} (2018) 059}
  [\href{https://arxiv.org/abs/1805.00941}{{\ttfamily 1805.00941}}].

\bibitem{Hanany:2005ve}
A.~Hanany and K.D.~Kennaway, \emph{{Dimer models and toric diagrams}},
  \href{https://arxiv.org/abs/hep-th/0503149}{{\ttfamily hep-th/0503149}}.

\bibitem{Franco:2005rj}
S.~Franco, A.~Hanany, K.D.~Kennaway, D.~Vegh and B.~Wecht, \emph{{Brane dimers
  and quiver gauge theories}},
  \href{https://doi.org/10.1088/1126-6708/2006/01/096}{\emph{JHEP} {\bfseries
  01} (2006) 096} [\href{https://arxiv.org/abs/hep-th/0504110}{{\ttfamily
  hep-th/0504110}}].

\bibitem{Benvenuti:2005cz}
S.~Benvenuti and M.~Kruczenski, \emph{{Semiclassical strings in Sasaki-Einstein
  manifolds and long operators in $\mathcal{N}=1$ gauge theories}},
  \href{https://doi.org/10.1088/1126-6708/2006/10/051}{\emph{JHEP} {\bfseries
  10} (2006) 051} [\href{https://arxiv.org/abs/hep-th/0505046}{{\ttfamily
  hep-th/0505046}}].

\bibitem{Benvenuti:2005ja}
S.~Benvenuti and M.~Kruczenski, \emph{{From Sasaki-Einstein spaces to quivers
  via BPS geodesics: $L^{p,q|r}$}},
  \href{https://doi.org/10.1088/1126-6708/2006/04/033}{\emph{JHEP} {\bfseries
  04} (2006) 033} [\href{https://arxiv.org/abs/hep-th/0505206}{{\ttfamily
  hep-th/0505206}}].

\bibitem{Franco:2005sm}
S.~Franco, A.~Hanany, D.~Martelli, J.~Sparks, D.~Vegh and B.~Wecht,
  \emph{{Gauge theories from toric geometry and brane tilings}},
  \href{https://doi.org/10.1088/1126-6708/2006/01/128}{\emph{JHEP} {\bfseries
  01} (2006) 128} [\href{https://arxiv.org/abs/hep-th/0505211}{{\ttfamily
  hep-th/0505211}}].

\bibitem{Butti:2005sw}
A.~Butti, D.~Forcella and A.~Zaffaroni, \emph{{The Dual superconformal theory
  for L**pqr manifolds}},
  \href{https://doi.org/10.1088/1126-6708/2005/09/018}{\emph{JHEP} {\bfseries
  09} (2005) 018} [\href{https://arxiv.org/abs/hep-th/0505220}{{\ttfamily
  hep-th/0505220}}].

\bibitem{Benvenuti:2024seb}
S.~Benvenuti, R.~Comi, S.~Pasquetti, G.~Pedde~Ungureanu, S.~Rota and A.~Shri,
  \emph{{Planar Abelian Mirror Duals of $\mathcal{N}=2$ SQCD$_3$}},
  \href{https://arxiv.org/abs/2411.05620}{{\ttfamily 2411.05620}}.

\bibitem{Benvenuti:2025a}
S.~Benvenuti, R.~Comi, S.~Pasquetti, G.~Pedde~Ungureanu, S.~Rota and A.~Shri,
  \emph{{A Chiral-Planar dualization algorithm for $3d$ $\mathcal{N}=2$
  Chern-Simons-matter theories}},
  \href{https://arxiv.org/abs/2505.02913}{{\ttfamily 2505.02913}}.

\bibitem{Benvenuti:2025b}
S.~Benvenuti, R.~Comi, S.~Pasquetti, G.~Pedde~Ungureanu, S.~Rota and A.~Shri,
  \emph{{Phases of $3d$ $\mathcal{N}=2$ Chern-Simons-Matter Theories}},
  {\emph{to appear} (2025) }.

\bibitem{Intriligator:1996ex}
K.A.~Intriligator and N.~Seiberg, \emph{{Mirror symmetry in three-dimensional
  gauge theories}},
  \href{https://doi.org/10.1016/0370-2693(96)01088-X}{\emph{Phys. Lett. B}
  {\bfseries 387} (1996) 513}
  [\href{https://arxiv.org/abs/hep-th/9607207}{{\ttfamily hep-th/9607207}}].

\bibitem{Hanany:1996ie}
A.~Hanany and E.~Witten, \emph{{Type IIB superstrings, BPS monopoles, and
  three-dimensional gauge dynamics}},
  \href{https://doi.org/10.1016/S0550-3213(97)00157-0}{\emph{Nucl. Phys. B}
  {\bfseries 492} (1997) 152}
  [\href{https://arxiv.org/abs/hep-th/9611230}{{\ttfamily hep-th/9611230}}].

\bibitem{Shaji:1990is}
N.~Shaji, R.~Shankar and M.~Sivakumar, \emph{{On Bose-fermi Equivalence in a
  U(1) Gauge Theory With Chern-Simons Action}},
  \href{https://doi.org/10.1142/S0217732390000664}{\emph{Mod. Phys. Lett. A}
  {\bfseries 5} (1990) 593}.

\bibitem{Chen:1993cd}
W.~Chen, M.P.A.~Fisher and Y.-S.~Wu, \emph{{Mott transition in an anyon gas}},
  \href{https://doi.org/10.1103/PhysRevB.48.13749}{\emph{Phys. Rev. B}
  {\bfseries 48} (1993) 13749}
  [\href{https://arxiv.org/abs/cond-mat/9301037}{{\ttfamily
  cond-mat/9301037}}].

\bibitem{Fradkin:1994tt}
E.H.~Fradkin and F.A.~Schaposnik, \emph{{The Fermion - boson mapping in
  three-dimensional quantum field theory}},
  \href{https://doi.org/10.1016/0370-2693(94)91374-9}{\emph{Phys. Lett. B}
  {\bfseries 338} (1994) 253}
  [\href{https://arxiv.org/abs/hep-th/9407182}{{\ttfamily hep-th/9407182}}].

\bibitem{Karch:2016aux}
A.~Karch, B.~Robinson and D.~Tong, \emph{{More Abelian Dualities in 2+1
  Dimensions}}, \href{https://doi.org/10.1007/JHEP01(2017)017}{\emph{JHEP}
  {\bfseries 01} (2017) 017}
  [\href{https://arxiv.org/abs/1609.04012}{{\ttfamily 1609.04012}}].

\bibitem{Benvenuti:2023qtv}
S.~Benvenuti, R.~Comi and S.~Pasquetti, \emph{{Mirror dualities with four
  supercharges}}, \href{https://doi.org/10.1007/JHEP10(2024)234}{\emph{JHEP}
  {\bfseries 10} (2024) 234}.

\bibitem{Comi:2022aqo}
R.~Comi, C.~Hwang, F.~Marino, S.~Pasquetti and M.~Sacchi, \emph{{The SL(2,
  \ensuremath{\mathbb{Z}}) dualization algorithm at work}},
  \href{https://doi.org/10.1007/JHEP06(2023)119}{\emph{JHEP} {\bfseries 06}
  (2023) 119} [\href{https://arxiv.org/abs/2212.10571}{{\ttfamily
  2212.10571}}].

\bibitem{Benini:2017dus}
F.~Benini, P.-S.~Hsin and N.~Seiberg, \emph{{Comments on global symmetries,
  anomalies, and duality in (2 + 1)d}},
  \href{https://doi.org/10.1007/JHEP04(2017)135}{\emph{JHEP} {\bfseries 04}
  (2017) 135} [\href{https://arxiv.org/abs/1702.07035}{{\ttfamily
  1702.07035}}].

\bibitem{Bando:1987br}
M.~Bando, T.~Kugo and K.~Yamawaki, \emph{{Nonlinear Realization and Hidden
  Local Symmetries}},
  \href{https://doi.org/10.1016/0370-1573(88)90019-1}{\emph{Phys. Rept.}
  {\bfseries 164} (1988) 217}.

\bibitem{Nakahara:2003nw}
M.~Nakahara, \emph{{Geometry, topology and physics}}, Institute of Physics
  Publishing, The Institute of Physics, London (2003).

\bibitem{Borokhov_2002}
V.~Borokhov, A.~Kapustin and X.~Wu, \emph{Topological disorder operators in
  three-dimensional conformal field theory},
  \href{https://doi.org/10.1088/1126-6708/2002/11/049}{\emph{Journal of High
  Energy Physics} {\bfseries 2002} (2002) 049–049}.

\bibitem{Pufu:2013vpa}
S.S.~Pufu, \emph{{Anomalous dimensions of monopole operators in
  three-dimensional quantum electrodynamics}},
  \href{https://doi.org/10.1103/PhysRevD.89.065016}{\emph{Phys. Rev. D}
  {\bfseries 89} (2014) 065016}
  [\href{https://arxiv.org/abs/1303.6125}{{\ttfamily 1303.6125}}].

\bibitem{Chester_2018}
S.M.~Chester, L.V.~Iliesiu, M.~Mezei and S.S.~Pufu, \emph{{Monopole operators
  in U(1) Chern-Simons-matter theories}},
  \href{https://doi.org/10.1007/jhep05(2018)157}{\emph{Journal of High Energy
  Physics} {\bfseries 2018} (2018) }.

\bibitem{witten2003sl2zactionthreedimensionalconformal}
{Edward Witten}, \emph{{$SL(2,\mathbb{Z})$ Action On Three-Dimensional
  Conformal Field Theories With Abelian Symmetry}},
  \href{https://arxiv.org/abs/hep-th/0307041}{{\ttfamily hep-th/0307041}}.

\bibitem{Einhorn:1977qv}
M.B.~Einhorn and R.~Savit, \emph{{Topological Excitations in the Abelian Higgs
  Model}}, \href{https://doi.org/10.1103/PhysRevD.17.2583}{\emph{Phys. Rev. D}
  {\bfseries 17} (1978) 2583}.

\bibitem{Peskin:1977kp}
M.E.~Peskin, \emph{{Mandelstam 't Hooft Duality in Abelian Lattice Models}},
  \href{https://doi.org/10.1016/0003-4916(78)90252-X}{\emph{Annals Phys.}
  {\bfseries 113} (1978) 122}.

\bibitem{Fisher:1989dnp}
M.P.A.~Fisher and D.H.~Lee, \emph{{Correspondence between two-dimensional
  bosons and a bulk superconductor in a magnetic field}},
  \href{https://doi.org/10.1103/PhysRevB.39.2756}{\emph{Phys. Rev. B}
  {\bfseries 39} (1989) 2756}.

\bibitem{Savit:1979ny}
R.~Savit, \emph{{Duality in Field Theory and Statistical Systems}},
  \href{https://doi.org/10.1103/RevModPhys.52.453}{\emph{Rev. Mod. Phys.}
  {\bfseries 52} (1980) 453}.

\bibitem{Delmastro:2019vnj}
D.~Delmastro and J.~Gomis, \emph{{Symmetries of Abelian Chern-Simons Theories
  and Arithmetic}}, \href{https://doi.org/10.1007/JHEP03(2021)006}{\emph{JHEP}
  {\bfseries 03} (2021) 006}
  [\href{https://arxiv.org/abs/1904.12884}{{\ttfamily 1904.12884}}].

\bibitem{Ellis_2017}
J.P.~Ellis, \emph{{TikZ-Feynman: Feynman diagrams with TikZ}},
  \href{https://doi.org/10.1016/j.cpc.2016.08.019}{\emph{Computer Physics
  Communications} {\bfseries 210} (2017) 103–123}.

\bibitem{Witten_2016}
E.~Witten, \emph{Fermion path integrals and topological phases},
  \href{https://doi.org/10.1103/revmodphys.88.035001}{\emph{Reviews of Modern
  Physics} {\bfseries 88} (2016) }.

\bibitem{seiberg2016gappedboundaryphasestopological}
N.~Seiberg and E.~Witten, \emph{Gapped boundary phases of topological
  insulators via weak coupling},
  \href{https://arxiv.org/abs/1602.04251}{{\ttfamily 1602.04251}}.

\bibitem{Aharony:2015pla}
O.~Aharony, P.~Narayan and T.~Sharma, \emph{{On monopole operators in
  supersymmetric Chern-Simons-matter theories}},
  \href{https://doi.org/10.1007/JHEP05(2015)117}{\emph{JHEP} {\bfseries 05}
  (2015) 117} [\href{https://arxiv.org/abs/1502.00945}{{\ttfamily
  1502.00945}}].

\bibitem{Goldhaber:1976dp}
A.S.~Goldhaber, \emph{{Spin and Statistics Connection for Charge-Monopole
  Composites}}, \href{https://doi.org/10.1103/PhysRevLett.36.1122}{\emph{Phys.
  Rev. Lett.} {\bfseries 36} (1976) 1122}.

\bibitem{Wu:1976ge}
T.T.~Wu and C.N.~Yang, \emph{{Dirac Monopole Without Strings: Monopole
  Harmonics}}, \href{https://doi.org/10.1016/0550-3213(76)90143-7}{\emph{Nucl.
  Phys. B} {\bfseries 107} (1976) 365}.

\bibitem{Wilczek:1981du}
F.~Wilczek, \emph{{Magnetic Flux, Angular Momentum, and Statistics}},
  \href{https://doi.org/10.1103/PhysRevLett.48.1144}{\emph{Phys. Rev. Lett.}
  {\bfseries 48} (1982) 1144}.

\bibitem{Polyakov:1988md}
A.M.~Polyakov, \emph{{Fermi-Bose Transmutations Induced by Gauge Fields}},
  \href{https://doi.org/10.1142/S0217732388000398}{\emph{Mod. Phys. Lett. A}
  {\bfseries 3} (1988) 325}.

\bibitem{Grundberg:1989pn}
J.~Grundberg, T.H.~Hansson, A.~Karlhede and U.~Lindstrom, \emph{{Spin,
  Statistics and Linked Loops}},
  \href{https://doi.org/10.1016/0370-2693(89)91589-X}{\emph{Phys. Lett. B}
  {\bfseries 218} (1989) 321}.

\bibitem{Wilczek:1995ma}
F.~Wilczek, \emph{{Statistical transmutation and phases of two-dimensional
  quantum matter}},  \href{https://arxiv.org/abs/cond-mat/9509085}{{\ttfamily
  cond-mat/9509085}}.

\bibitem{Dyer:2013fja}
E.~Dyer, M.~Mezei and S.S.~Pufu, \emph{{Monopole Taxonomy in Three-Dimensional
  Conformal Field Theories}},
  \href{https://arxiv.org/abs/1309.1160}{{\ttfamily 1309.1160}}.

\bibitem{Benvenuti:2018cwd}
S.~Benvenuti and H.~Khachatryan, \emph{{QED's in $2{+}1$ dimensions: complex
  fixed points and dualities}},
  \href{https://arxiv.org/abs/1812.01544}{{\ttfamily 1812.01544}}.

\bibitem{Benvenuti:2019ujm}
S.~Benvenuti and H.~Khachatryan, \emph{{Easy-plane QED$_{3}$\textquoteright{}s
  in the large N$_{f}$ limit}},
  \href{https://doi.org/10.1007/JHEP05(2019)214}{\emph{JHEP} {\bfseries 05}
  (2019) 214} [\href{https://arxiv.org/abs/1902.05767}{{\ttfamily
  1902.05767}}].

\end{thebibliography}\endgroup

\end{document}